\documentclass[11pt,preprint]{aastex}

\usepackage{amsmath}

\newcommand\be{\begin{equation}}
\newcommand\en{\end{equation}}

\shorttitle{Structure of evolved disks}
\shortauthors{Sicilia-Aguilar et al.}

\begin{document}

\title{Dust properties and disk structure of evolved protoplanetary
disks in Cep OB2: Grain growth, settling, gas and dust mass, and inside-out evolution}

\author{Aurora Sicilia-Aguilar\altaffilmark{1,2}, Thomas Henning\altaffilmark{1}, Cornelis P. Dullemond\altaffilmark{1,3}, Nimesh Patel\altaffilmark{4}}
\author{Attila Juh\'{a}sz\altaffilmark{5}, Jeroen Bouwman\altaffilmark{1}, Bernhard Sturm\altaffilmark{1}}

\altaffiltext{1}{Max-Planck-Institut f\"{u}r Astronomie, K\"{o}nigstuhl 17, 69117 Heidelberg, Germany}
\altaffiltext{2}{Departamento de F\'{\i}sica Te\'{o}rica, Universidad Aut\'{o}noma de Madrid, Cantoblanco 28049, Madrid, Spain}
\altaffiltext{3}{Institut f\"{u}r Theoretische Astrophysik, Zentrum f\"{u}r Astronomie, Universit\"{a}t Heidelberg, Albert-Ueberle-Str. 2, 69120 Heidelberg, Germany}
\altaffiltext{4}{Harvard-Smithsonian Center for Astrophysics, 60 Garden Street, Cambridge, MA 02138, USA}
\altaffiltext{5}{Leiden Observatory, Niels Bohrweg 2, NL-2333 CA Leiden, The Netherlands}

\email{sicilia@mpia.de, aurora.sicilia@uam.es}

\begin{abstract}
We present $Spitzer$/IRS spectra of 31 TTS
and IRAM/1.3mm observations for 34 low- and intermediate-mass stars in
the Cep~OB2 region. Including our previously published data, we analyze 
56 TTS and the 3 intermediate-mass stars with silicate features in 
Tr~37 ($\sim$4~Myr) and NGC~7160 ($\sim$12~Myr). 
The silicate emission features are well reproduced with
a mixture of amorphous (with olivine, forsterite, and silica stoichiometry) 
and crystalline grains (forsterite, enstatite). We explore grain size 
and disk structure using radiative transfer disk models, finding that most objects 
have suffered substantial evolution (grain growth, settling). About 
half of the disks show inside-out evolution,
with either dust-cleared inner holes or
a radially-dependent dust distribution, typically with larger grains 
and more settling in the innermost disk. The typical strong silicate features 
require nevertheless
the presence of small dust grains, and could be explained by differential settling
according to grain size, anomalous dust distributions, 
and/or optically thin dust populations within disk gaps. M-type stars tend to have
weaker silicate emission and steeper SEDs than K-type objects. 
The inferred low dust masses 
are in a strong contrast with the relatively high gas accretion rates, suggesting
global grain growth and/or an anomalous gas to dust ratio.
Transition disks (TD)  in the Cep OB2 region display strongly processed grains, 
suggesting that they are dominated by dust evolution and
settling. Finally, the 
presence of rare but remarkable disks with strong accretion 
at old ages reveals that some very massive disks may still survive to 
grain growth, gravitational instabilities, and planet formation.
\end{abstract}


\section{Introduction \label{intro}}

Evolved protoplanetary disks are among the most interesting objects
to understand how (and if) planet formation occurs in the disks around
solar-type stars. Although there is well-accepted evidence 
suggesting that most IR excesses from protoplanetary disks 
have dissipated by $\sim$6-10 Myr (e.g. Haisch et al. 2001; Sicilia-Aguilar et al.
2006a, from now on SA06; Hern\'{a}ndez et al. 2007), the way disks become optically 
thin and disperse after a (brief or long) transition
phase is not clear. Different disk dispersal processes have been proposed,
and observational evidence suggests that they all happen to some extent,
but the relative importance and the interplay between them is unknown.

Grain growth/settling to the midplane without fragmentation would render the
disk optically thin in a timescale too short to be consistent
with observations (Dullemond \& Dominik 2005; Brauer et al. 2008). Nevertheless, evidence for 
grain growth to millimeter-sized particles is found in (sub)millimeter observations 
as flatter dust emissivity curves and reduced disk masses compared 
to gas observations (Andrews \& Williams 2005, 2007; Rodmann et al.
 2006; Natta et al. 2007). High accretion rates
observed in objects with negligible near-IR excess also
point to strong grain growth/settling (Sicilia-Aguilar et al. 2010).
Planet formation would substantially affect the disk structure, 
removing large amounts of mass and affecting the
viscous evolution and accretion (Najita et al. 2007). 
UV/X-ray photoevaporation by the host star may not 
be able to fully disperse the disk in a reasonable time
if it is too massive or too strongly accreting, but it can remove a large
amount of mass, leading to inner holes and fast disk dispersal 
with the help of viscous evolution (Clarke et al. 2001; Alexander et al. 2006; 
Gorti et al. 2009; Ercolano et al. 2009). In addition, the presence of stellar companions
may contribute to the disk clearing (Bouwman et al. 2006;
Ireland \& Kraus 2008), although
it may not be the leading cause of inner holes (Pott et al. 2010).

The observations of objects with no near-IR excess 
and strong mid-IR fluxes and thus optically thin/clean inner disks a 
few AU in size and optically thick outer disks, also called transition
disks (TD; Skrutskie et al. 1990), suggests inside-out evolution to be 
an important disk dispersal mechanism, as predicted by Hayashi et al. (1985).
Nevertheless, $Spitzer$ revealed systems where the disk dispersal processes 
seem to operate over larger radial distances, resulting in disks with global 
dust depletion (Currie et al. 2009).
The connection between the different disk dispersal mechanisms is
complex. Photoevaporation is efficient at dispersing the full disk only once the disk mass
and accretion rate have dropped below certain levels, and it is
also favored by grain growth, which decreases the shielding of the
gas to the FUV radiation (Gorti et al. 2009). Grain growth
and settling are suppressed by gas turbulence (Schr\"{a}pler \& Henning 2004), and may thus 
be important in the disk dead zones, but dead zones may disappear
once the disk mass decreases, or never exist if the initial disk
mass is too low (Hartmann et al. 2006). 
In addition, the relative importance of the various dispersal processes is expected to 
vary with time (Alexander \& Armitage 2009).

Although even the youngest star-forming regions contain 
evolved and transitional protoplanetary disks
(e.g. Hartmann et al. 2005; Fang et al. 2009), such objects
with reduced IR excess seem to
become increasingly frequent with cluster age (Megeath et al. 2005; SA06;
Lada et al. 2006; Hern\'{a}ndez et al. 2007; 
Sicilia-Aguilar et al. 2009; Currie et al. 2009;
Muzerolle et al. 2010). Nevertheless, there is still substantial 
debate regarding the definition of transition and evolved disks and 
where to place the limits to distinguish them from primordial ones, as
quantifying mass depletion and the presence of inner holes is not
easy without complete multiwavelength data and spatially resolved
observations (Luhman et al. 2010; Currie \& Sicilia-Aguilar 2011).
 The changes in the spectral energy distribution (SED) and thus,
in the dusty disk component, observed with age are accompanied
by a decrease in the accretion rate (Hartmann et al. 1998;
Sicilia-Aguilar et al. 2006b, 2010), which even drops to undetectable
or zero levels for roughly half of the transition objects 
(Sicilia-Aguilar et al. 2006b, 2010; Muzerolle et al. 2010).
The optically thin emission of silicate grains (Calvet et al. 1992; 
Natta et al. 2000) also indicates dust evolution from ISM-like to larger and 
crystalline grains (Meeus et al. 2001; Bouwman et al. 2001;
van Boekel et al. 2003; Kessler-Silacci et al. 2005; Furlan et al. 2005;
Watson et al. 2009). Global disk structure (Lommen et al. 2010) and 
accretion/turbulence in the disk (Sicilia-Aguilar et al. 2007; from now on SA07) are
thought to affect the grains we observe in the disk atmosphere
(Dullemond \& Dominik 2008; Zsom et al. 2011).
Therefore, dust mineralogy can be used as a tracer of the energetic processes in
the disk and the effect of turbulence and transport in the innermost regions. 

Evolved clusters and star-forming regions constitute 
an ideal laboratory to explore the characteristics
and effects of disk dispersal, studying
dust properties and disk structure
for statistically significant samples of objects.
One of the most interesting and well studied evolved regions is the
Cep OB2 association (Patel et al. 1995; Patel et al. 1998),
located at 900 pc distance (Contreras et al. 2002) and containing 
the clusters Tr 37 and NGC 7160 (Platais et al. 1998). 
These clusters have been extensively studied at optical wavelengths in 
order to determine the membership, spectral types, extinction, 
the presence of accretion, and the accretion rate for a 
large number of solar-type members. The approximate median ages derived from the solar-type 
members are $\sim$4 and $\sim$12 Myr for Tr 37 and NGC 7160, respectively
(Sicilia-Aguilar et al. 2004, 2005).
Our $Spitzer$ observations (SA06; SA07) revealed
strong disk evolution. The disk fractions for the solar-type stars are  $\sim$45\% in Tr 37 
and $\sim$4\% in NGC 7160, and a large fraction of the objects present IRAC 
fluxes systematically lower than younger regions (e.g. Taurus;
Hartmann et al. 2005) that suggest important grain growth/dust settling.
Although among the objects with MIPS detections the number of evolved disks
in Tr 37 is not much different than what is seen in Taurus (Luhman et al. 2010),
the cluster lacks very massive and strongly accreting
objects (with GM Cep being the only one in the region, Sicilia-Aguilar et al. 2008a) 
and a substantial fraction of objects have very low IRAC excesses
and MIPS fluxes below the detection limits, characteristics of transitional or
very evolved (strongly settled and likely dust depleted) disks.
Up to 20\% of the disks have no or very small near-IR excesses at $\lambda \leq$6$\mu$m
([3.6]-[4.5]$<$0.2), being thus empirically classified as TD\footnote{Note that
there is still lack of consensus regarding the definition of TD. We apply here 
the definition we used previously in SA06, while some authors include as TD
those showing global or homologous dust depletion (e.g. Currie et al. 2009), or
use stronger near-IR constrains to ensure the presence of a cleared inner
hole (e.g. Luhman et al. 2010). We also put our focus on
discussing the different disk structures and the physical processes to which they
can be related, rather than on the naming conventions.}.
Accretion rates or strong constraints on the accretion (via high-resolution
H$\alpha$ spectroscopy) are available for a large number of members
(Sicilia-Aguilar et al. 2005, 2006b, 2010), and previous IRS observations 
revealed a large variety in the shape of the silicate emission 
and a weak correlation between grain size and accretion (SA07). 

Here we present 31 new IR spectra of T Tauri stars (TTS) in the Cep OB2 region
obtained with $Spitzer$ IRS, and 1.3 millimeter observations obtained for 
34 objects with the IRAM 30m telescope, both intended
to study and constrain the structure of evolved
disks. In Section \ref{data}, we introduce the new observations and
also reanalyze our previous IRS/IRAC/MIPS $Spitzer$ results. 
In Section \ref{analysis}, we model the properties
of the silicate features and the global SEDs in order to derive the
mineralogy and disk structure. We then discuss the different implications
of the observed dust mineralogy and disk structure
for disk dispersal evolution in Section \ref{discussion}, summarizing the results in 
Section \ref{conclu}.

\section{Observations and data reduction\label{data}}

\subsection{IRS data \label{irs}}

We obtained $5.4-35$~$\mu$m low-resolution spectra ($R = 60-120$) of 31 members
of the clusters Tr 37 and NGC 7160, within the Cep OB2 association, 
with the Infrared Spectrograph (IRS, Houck et al. 2004) on-board of $Spitzer$.
Our data were obtained within our $Spitzer$ GO program 30523, observed between 2006 and
2008. To improve the statistics, adding up to 59 objects (56 of them solar-type stars), 
we also include here the data from our 
program 8 (already published; SA07), and five objects
from program 3137, already public, (14-141, 72-875, 72-1427, 21364762, 
21362507\footnote{The ID of objects listed in this work correspond to the numbering
used in our previous optical papers (see Sicilia-Aguilar et al. 2005; SA06)}). 
The objects in program 30523 were selected among the more than 100 disked
objects detected with IRAC and/or MIPS by SA06, avoiding
those that presented a complicated background or nearby bright sources.
We did not intend to obtain a complete collection, but to sample
the different types of disk morphology (from flared, Taurus-like disks, to 
TD with inner holes) and accretion rates (ranging from $\sim$10$^{-7}$ to few 
times 10$^{-10}$ M$_{\odot}$/yr, and including some non-accreting
TD) revealed by our Spitzer photometry and optical study. 
The complete sample with the basic properties of the objects
is listed in Table \ref{obs-table}. Many of the
disks are substantially evolved, with IR fluxes well below the Taurus median,
although we also include a few very active disks, one of
which is the variable star 13-277 (also known as GM Cep). GM Cep, together with
the brightest among the TD, 21392541, were also observed with the high resolution 
IRS modules. The targets for programs 8 and 
30523 were positioned in the slit using the high accuracy PCRS peak-up before the
observations. 

Our reduced low-resolution spectra are based on the {\tt droopres} products processed 
through the S15.3.0 (program 30523) and S18.6.0 (program 3137) version of the 
$Spitzer$ data pipeline. 
For consistency, we also re-reduced the data from program 8 using a
more recent pipeline, S16.1.0, finding no significant differences with our previous
data reduction\footnote{Reduced spectra can be provided upon request 
by contacting the authors.}.  Pixels flagged by the data pipeline or by visual inspection of the images as 
``bad'' were replaced with a value interpolated from an 8 pixel perimeter surrounding the errant 
pixel. Our data were further processed using spectral extraction tools 
developed for the ``Formation and Evolution 
of Planetary Systems" (FEPS) {\it $Spitzer$\,} science legacy project
(see also Bouwman et al. 2008), based on the {\tt SMART} software package (Higdon et al. 2004).
The spectra were extracted using a 6.0 pixel and 5.0 pixel fixed-width aperture in the spatial 
dimension for the observations with the first order of the short- ($5.4-14$~$\mu$m) and the 
long-wavelength ($14-35$~$\mu$m) modules, respectively.  
The low-level fringing at wavelengths  $>20$~$\mu$m was removed using the
{\tt irsfinge} package (Lahuis \& Boogert 2003). To remove any effect of pointing offsets, we 
matched orders based on the point spread function of the IRS instrument, correcting for possible 
flux losses (see Swain et al. 2008 for further details). This was especially 
important for the objects from program 3137, which systematically presented offsets  
up to 2" from the real position. The background was subtracted using 
associated pairs of imaged spectra from the two nodded positions along the slit for most of the
objects, also eliminating stray light contamination and anomalous dark currents.
A few stars are embedded in nebulosity (73-472, 14-141, 21-998, 21364762), 
having a background that suffers from 
variations on small spatial scales. For them, the standard method (subtracting the nodded 
spectra from each other) could not be used for the background correction, and we applied a customized 
routine for background subtraction. After bad pixel correction and flatfielding, the two--dimensional 
spectra for each of the nod--positions were extracted. The source positions in both nods 
were matched (they are typically at 1/3 and 2/3 of the spatial dimension) and the spectra 
combined. The source is then in the center of the spatial axis of the combined spectrum. 
For each wavelength resolution element a second- or third-oder polynomial
was fitted to the regions to the left and right of the source spectrum, carefully 
avoiding the edges of the detector. These polynomials are evaluated in the central (source) 
region and subtracted as background. While this procedure results in increased
uncertainty for the faintest sources, the extracted SEDs are in excellent agreement with the
IRAC and MIPS photometry. 

The spectra were calibrated using a 
spectral response function derived from multiple IRS spectra of the calibration star 
$\eta^1$~Doradus and a {\sc marcs} stellar model provided by the $Spitzer$ Science Center. The 
spectra of the calibration target were extracted in an identical way as our science targets.
The relative errors between spectral points within one order are dominated by the noise on each 
individual point and not by the calibration. We estimate a relative flux calibration across an 
order of $\approx 1$~\% and an absolute calibration error between orders/modules of 
$\approx 3$~\%, which is mainly due to uncertainties in the scaling of the {\sc marcs} model.

The high-resolution data for GM Cep and 21392541 were reduced using the standard
pipeline developed for the C2D data (Lahuis et al. 2006, 2007). The C2D pipeline is
based on the {\tt SMART} routines  (Higdon et al. 2004), and includes advanced 
reduction and calibration routines developed for the C2D and FEPS data in order
to perform the full aperture and optimal PSF extraction and pointing
correction. The observations included 
separate sky observations, in order to improve the background subtraction. 
The data reduction starts with the SSC pipeline RSC products, which are corrected
for crosstalk, in addition to the other corrections of the Basic Calibrated Data (BCD).
The spectrum is extracted by using an optimal, wavelength-dependent PSF fitting.
The full aperture spectrum is also extracted separately, in order to check for
extended emission. The flux calibration is based on a large number of high-S/N calibrator stars
and the {\sc marcs} stellar model provided by the $Spitzer$ Science Center. 
The extracted spectra were corrected for fringing (Lahuis \& Boogert 2003),
the orders were matched, and the data were corrected for the flux loses due to 
pointing offsets. The bright star GM Cep is well detected with the 
high-resolution IRS modules, but the object
21392541 is too faint, and its spectrum could not be extracted.

\subsection{IRAM 30m observations \label{iram}}

A total of 34 members of Cep OB2 were observed at 1.3mm with the
bolometer at the IRAM 30m telescope (see Table \ref{iram-table}).
The observations included most of the objects with IRS spectra,
and took place within 3 different programs carried out between
2006 and 2009 (142-06, 132-07, 145-08) as part of the MAMBO Pool
Observations, using the MAMBO2 117 channel bolometer. In addition,
GM Cep had been previously observed with IRAM and the MAMBO1 37 channel
bolometer (Sicilia-Aguilar et al. 2008).

All observations were performed as standard ON-OFF scans, with
each scan including 10 min on-source and 10 min off-source
integrations, divided in 1 min integrations following the ON-OFF-OFF-ON
sequence. For each source, we obtained between 1 and 10 scans,
varying the position (by observing at different times) and distance
(30-70'') of the off-source pointing for repeated observations. 
The beam of the telescope at 1.3mm is 11''. While for most of the
objects this is enough to avoid contamination by nearby sources
or nebular emission, the bolometer observations revealed extended
emission in a couple of cases (see Table \ref{iram-table}), which
are therefore excluded from this study. Skydip observations were
performed every 2-3 hours or more frequently, depending on
the weather, in order to determine the sky opacity, $\tau$. The
pointing and flux calibration were done using the nearby sources LK H$\alpha$ 234
and NGC 7538, from the IRAM Pool catalog, and the planet Uranus. 

The data were reduced using the MOPSIC pipeline, specially developed by
R. Zylka for the reduction of ON-OFF observations. The MOPSIC pipeline uses the  
RSD (Resampled Signal and then the Difference) to subtract the atmospheric
emission from the data while minimizing the sky 
noise\footnote{See http://iram.fr/IRAMFR/ARN/dec05/node9.html},
which is measured from the correlated signal of adjacent pixels.
The total emission is calculated taking into account the $\tau$ 
measured in the skydip observations and the elevation of the source,
and calibrated according to the flux calibrators above mentioned.
The final result is the weighted mean of all the scans.
For each source, we started reducing all individual scans,
in order to detect potential problems. In particular, observations
taken during the winter 2007 (program 132-07) suffered from sporadic
technical problems due to malfunctions in the preamplifier. The
affected scans, together with other scans suffering from poor weather
or spikes, were labeled and removed. Finally, the good scans were
reduced together for each source. The final fluxes (or 3$\sigma$ upper limits
in case of non-detection) are given in Table \ref{iram-table}, together
with the estimated disk mass based on the Beckwith et al. (1990) prescription,
for an opacity $\kappa$$_\nu$=2 cm$^2$ g$^{-1}$ at 1.3mm. Interestingly, we do not find
any correlation between the estimated disk masses and the accretion rates,
except for the fact that the most massive disk corresponds to the
strongest accretor, GM Cep.

\subsection{IRAC/MIPS \label{iracmips}}

The IRAC and MIPS data for the cluster members had been presented in
SA06. Given the improvement of the pipeline
and calibration since then, we have re-reduced the data and 
redone the photometry for the objects analyzed here. In addition to
our IRAC/MIPS data (AORs 3959040 and 4316416 for Tr 37, 3959296 and 4320000
for NGC 7160), we have also analyzed
the archive data from AORs 6051840, which include
a few of the objects to the west of Tr 37 that were out of our original $Spitzer$
field.

The data reduction was done with the Mosaicking and Point-source Extraction
pipeline (MOPEX\footnote{http://ssc.$Spitzer$.caltech.edu/dataanalysistools/tools/mopex/}). The mosaics were
created from the BCD data, using standard parameters for each 
band and including the masks for detector artifacts and associated status
masks. The process included pointing refinement and background matching
to produce uniform mosaics. Short and long exposures
were treated separately. AORs 3959040 and 3959296 contained
5 individual images per pointing in both the short (1s) and
long (26.4s) exposures, while the total exposure time of AOR 6051840
was 4.8s. MIPS AORs 4316416 and 4320000
mapped the region with median exposure times 78s. 
Aperture photometry was done with the APEX multiframe pipeline within
MOPEX for IRAC channels 1,2, and 3. The presence of substantial
extended emission in IRAC channel 4 and MIPS 24$\mu$m in Tr 37 precluded the
use of APEX for the photometry at these wavelengths, which was
then done using IRAF standard tasks within the $phot$ and $apphot$ 
packages. The apertures and sky annuli were 3 and 12-20 pixels for
IRAC, and 13" and 20-32" for MIPS, which are standard values. 
The corresponding aperture corrections were taken from the IRAC and
MIPS Handbook, being 1.112, 1.113, 1.125, 1.218, and 1.167 for the
3.6, 4.5, 5.8, 8.0, and 24$\mu$m bands, respectively.
Given that long-exposure images suffer more often from scattered-light
artifacts and extended nebular emission, we preferentially use the
short-exposure photometry for objects detected in the short-exposure
mosaics with high signal to noise ratio (S/N). Nevertheless, in most cases
the differences for objects detected in both short and long exposure
mosaics are negligible.

The newly reduced data (see Table \ref{spitzer-table}) agree very well with our 
previous results, and the use of smaller apertures improves the observations of
the objects with complicated background within the Tr 37 globule, 
whose IRAC and MIPS magnitudes are now 
in better agreement with the IRS spectra. The new MIPS data for
NGC 7160 also shows a better agreement with the IRS spectrum of 01-580,
due to a better background treatment. The IRAC photometry of two objects
(21364762 and 72-875) is strongly contaminated by nebular emission,
being thus excluded from the analysis.

\section{Analysis \label{analysis}}

\subsection{SEDs and silicate features \label{seds}}

\begin{figure}
\plotone{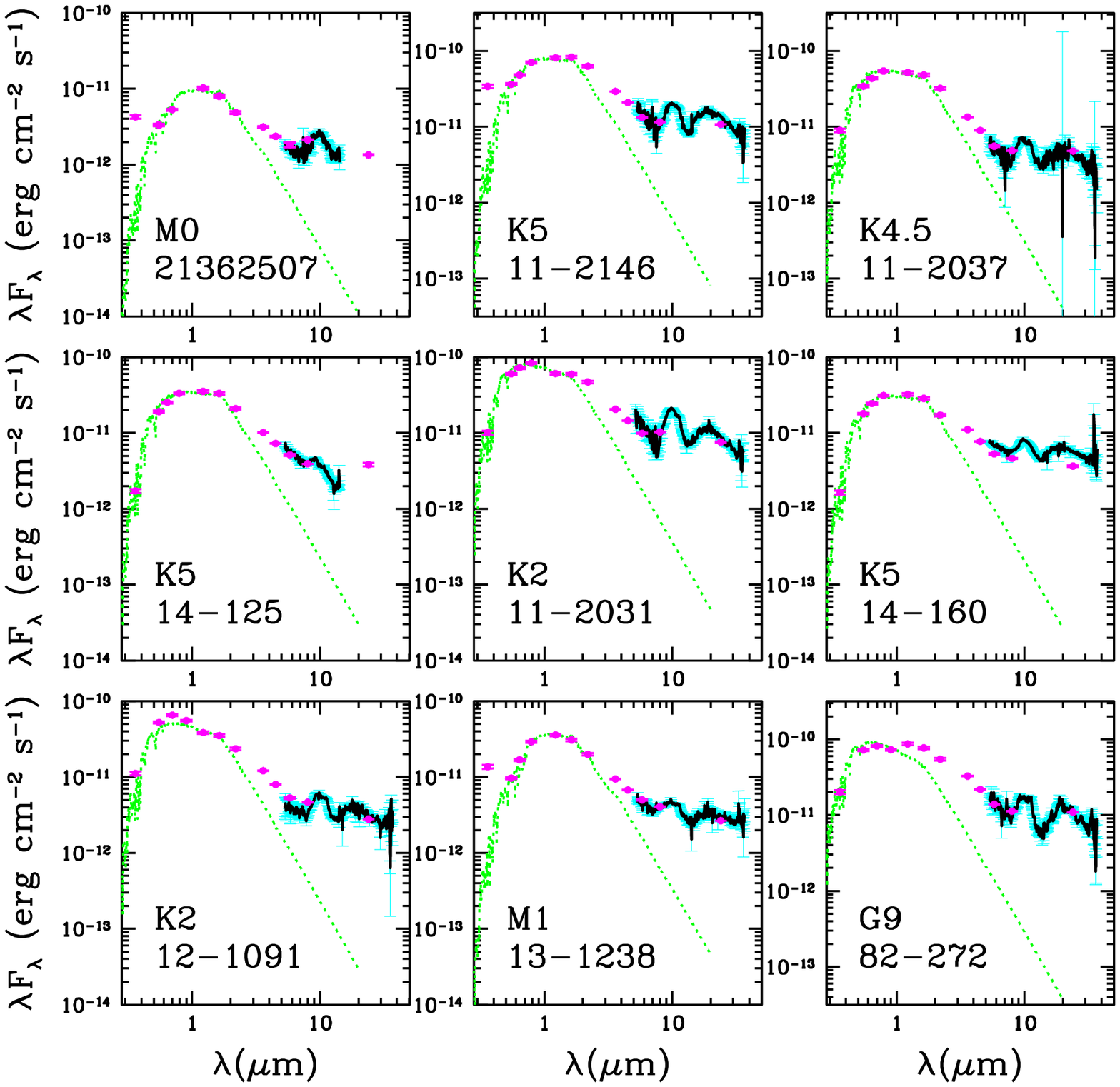}
\caption{SEDs for the objects with SED type \#1, normal CTTS (see text). 
The optical, 2MASS, IRAC, and MIPS
data are represented by magenta dots, with their errors (usually, smaller than
the dots). The IRS spectra are shown in black with errors in cyan. Photospheric
models (Gustafsson et al. 2008; Kenyon \& Hartmann 1995) are shown in green.
The data are corrected for extinction using the A$_V$ derived for each object.
\label{seds1-fig}}
\end{figure}

\begin{figure}
\plotone{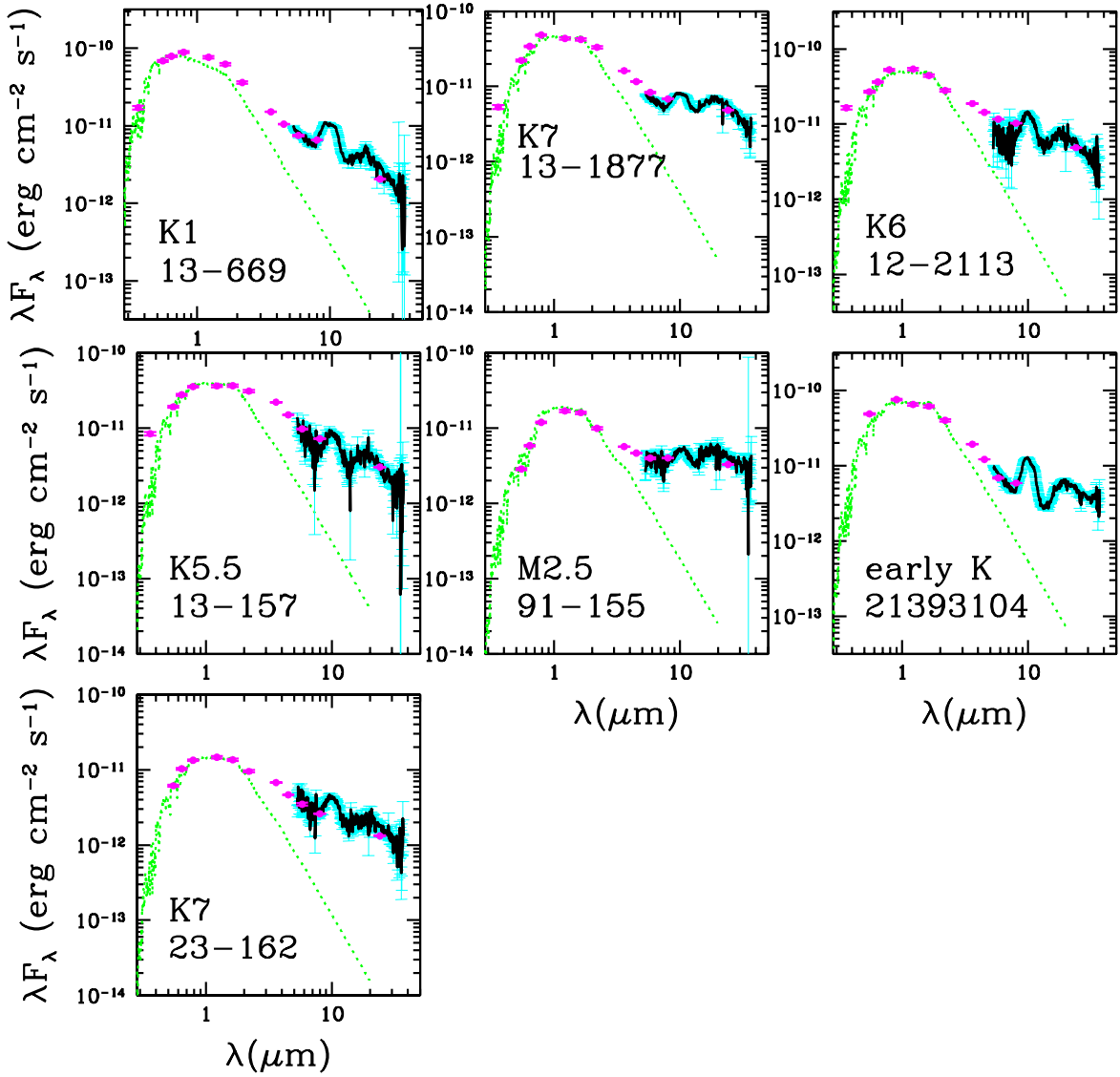}
\caption{SEDs for the objects with sed type \#1, normal CTTS (continued). Same as Figure \ref{seds1-fig}. \label{seds1b-fig}}
\end{figure}

\begin{figure}
\plotone{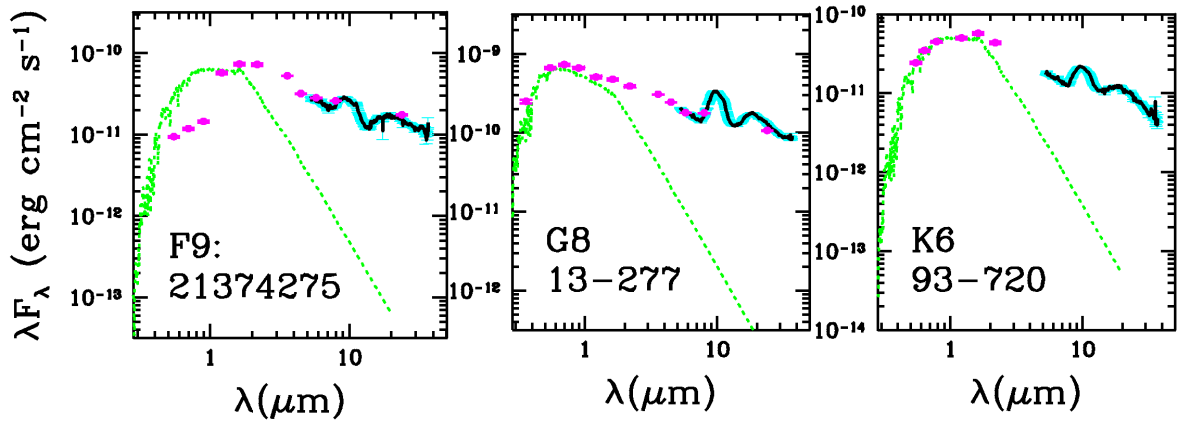}
\caption{SEDs for the objects with sed type \#2, massive CTTS (see text). Same as Figure \ref{seds1-fig}. \label{seds2-fig}}
\end{figure}

\begin{figure}
\plotone{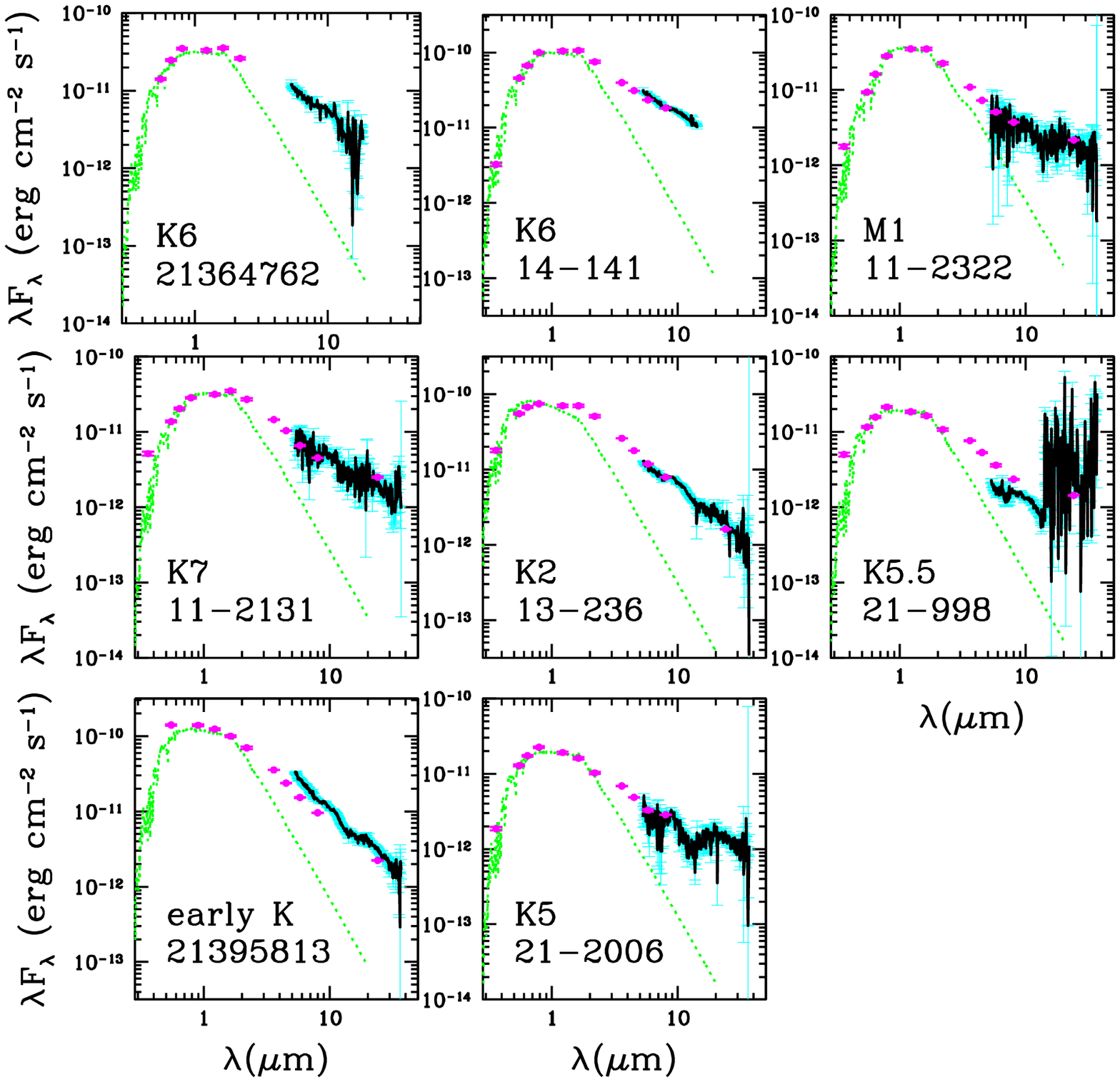}
\caption{SEDs for the objects with sed type \#3, depleted CTTS (see text). Same as Figure \ref{seds1-fig}. \label{seds3-fig}}
\end{figure}

\begin{figure}
\plotone{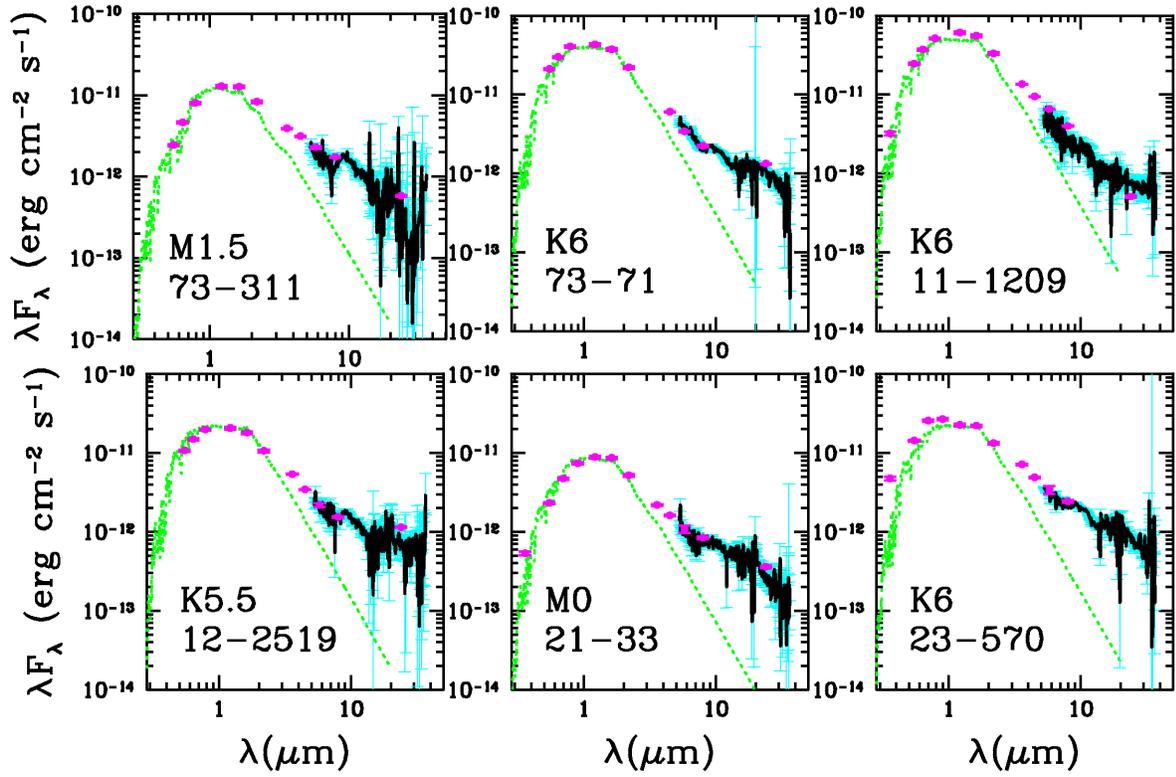}
\caption{SEDs for the objects with sed type \#4, globally settled/depleted 
CTTS (see text). Same as Figure \ref{seds1-fig}. \label{seds4-fig}}
\end{figure}

\begin{figure}
\plotone{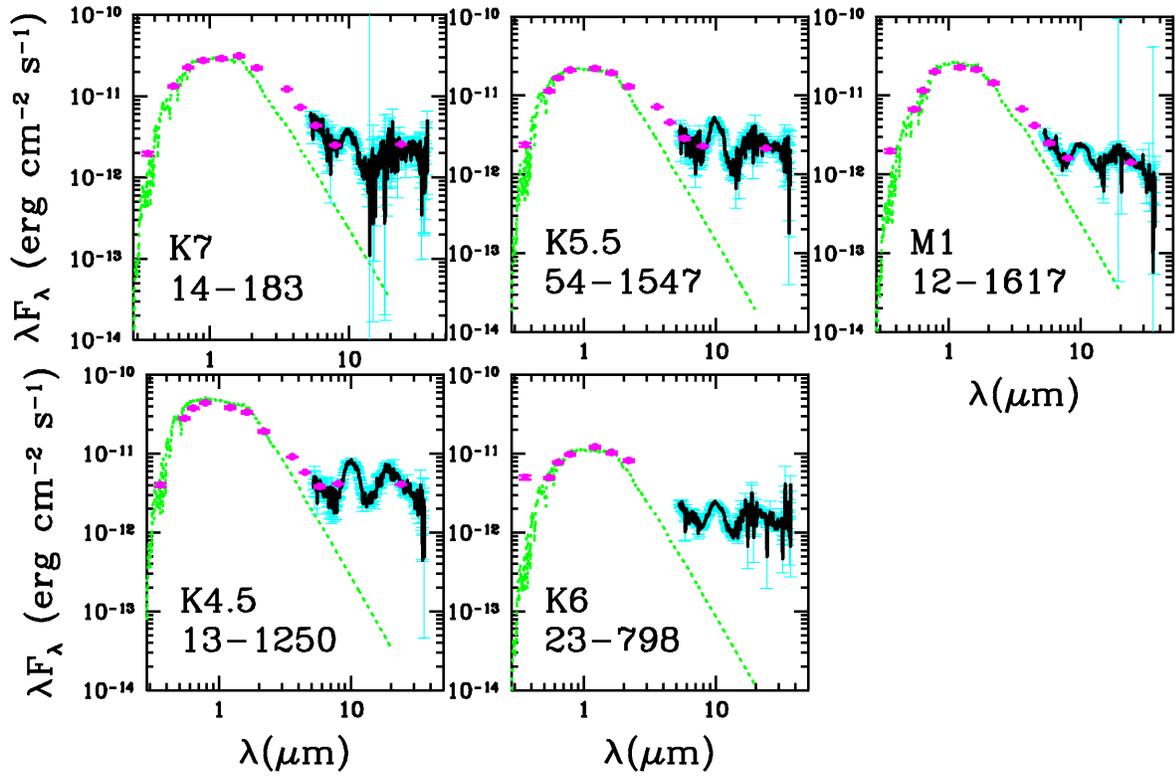}
\caption{SEDs for the objects with sed type \#5, ``kink" disk (see text). Same as Figure \ref{seds1-fig}. \label{seds5-fig}}
\end{figure}

\begin{figure}
\plotone{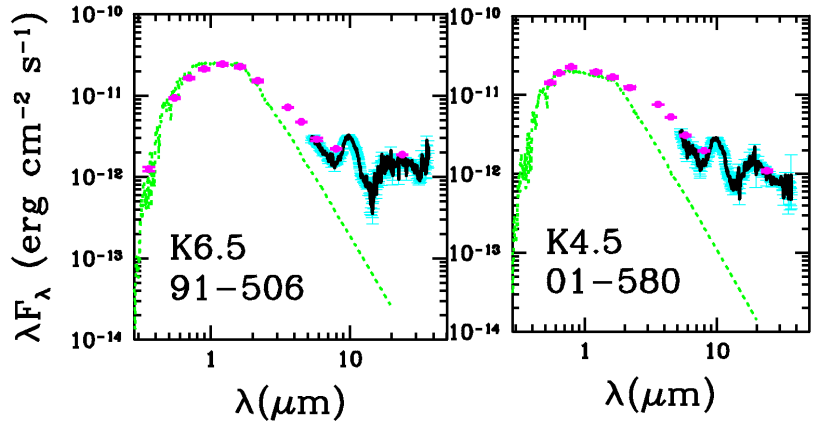}
\caption{SEDs for the objects with sed type \#6, dust-depleted disk with strong
silicate feature (see text). Same as Figure \ref{seds1-fig}. \label{seds6-fig}}
\end{figure}

\begin{figure}
\plotone{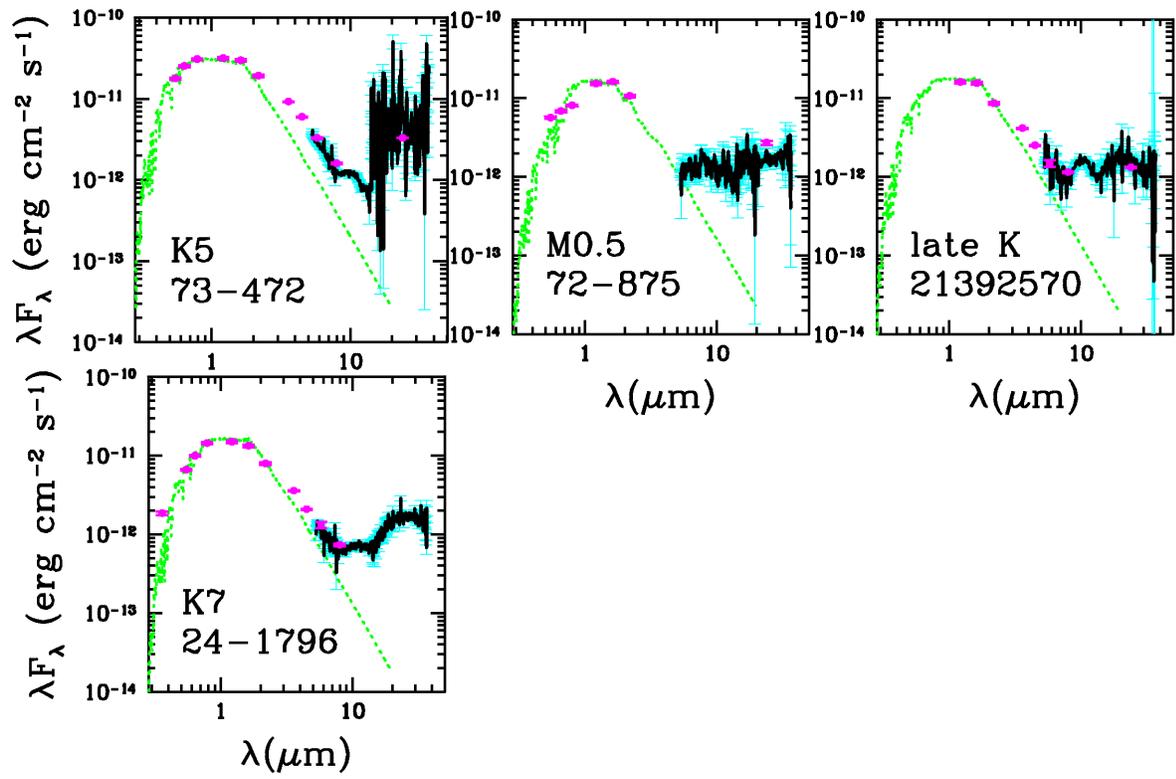}
\caption{SEDs for the objects with sed type \#7, classical TD with weak silicate (see text). Same as Figure \ref{seds1-fig}. \label{seds7-fig}}
\end{figure}

\begin{figure}
\plotone{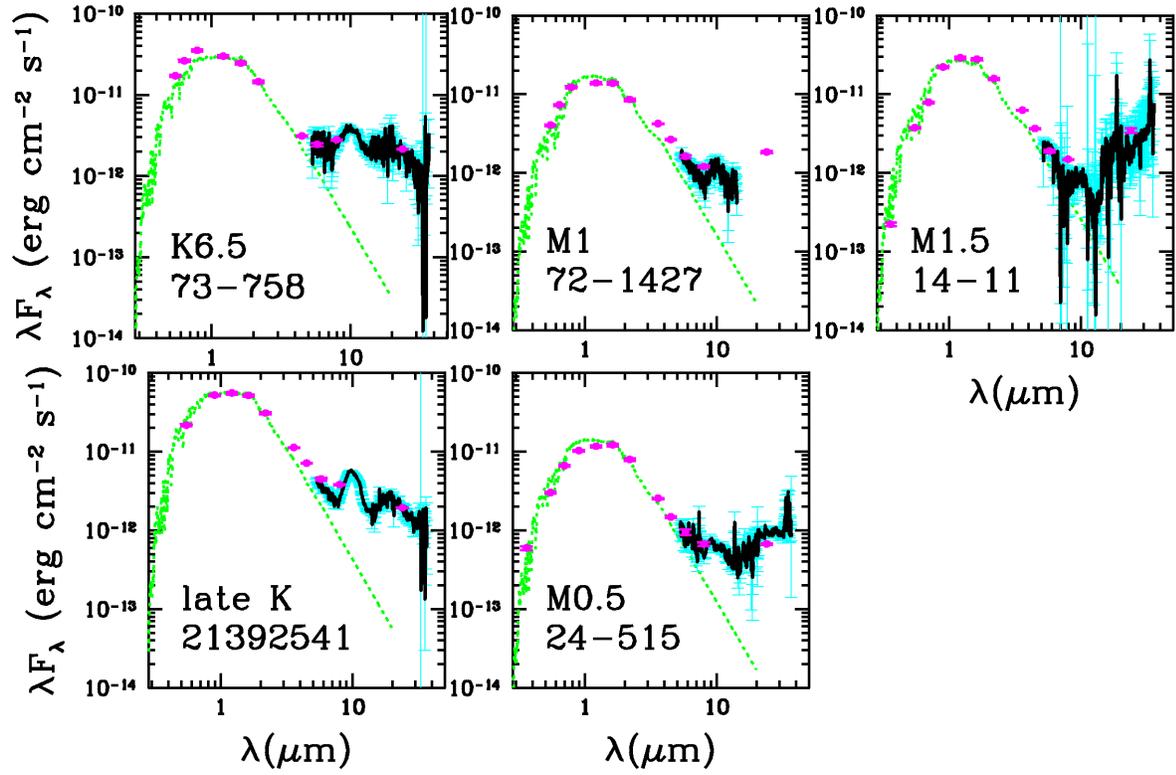}
\caption{SEDs for the objects with sed type \#8, classical TD with strong silicate (see text). Same as Figure \ref{seds1-fig}. \label{seds8-fig}}
\end{figure}

\begin{figure}
\plotone{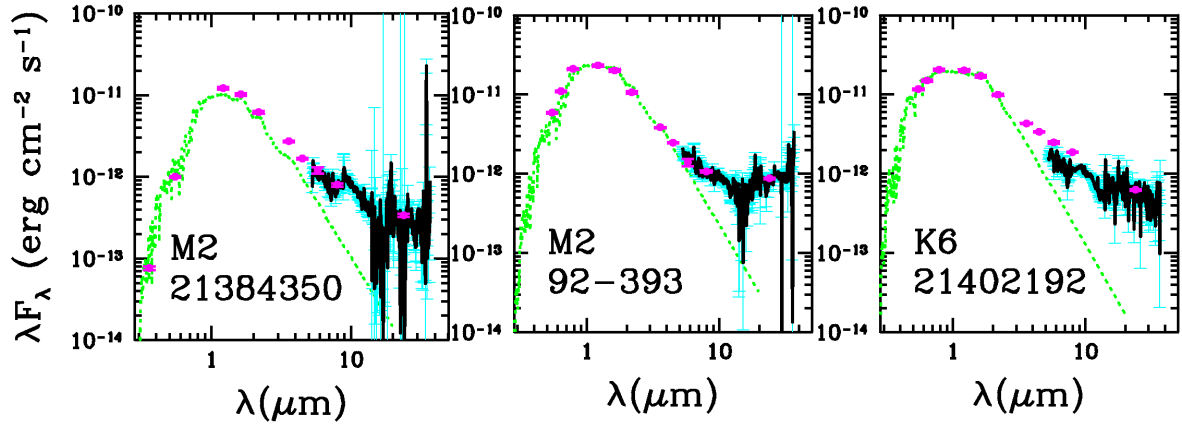}
\caption{SEDs for the objects with sed type \#9, settled TD with weak silicate (see text). Same as Figure \ref{seds1-fig}. \label{seds9-fig}}
\end{figure}

\begin{figure}
\plotone{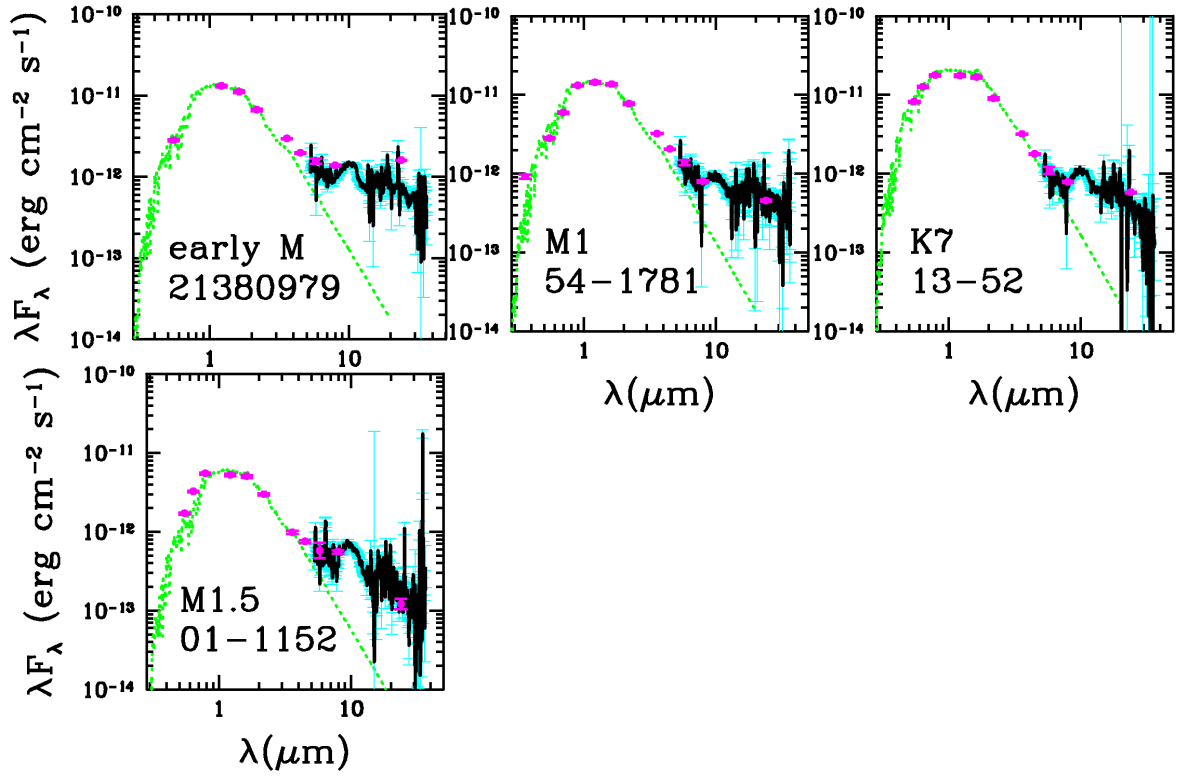}
\caption{SEDs for the objects with sed type \#10, settled TD with strong silicate (see text). Same as Figure \ref{seds1-fig}. \label{seds10-fig}}
\end{figure}

\begin{figure}
\plotone{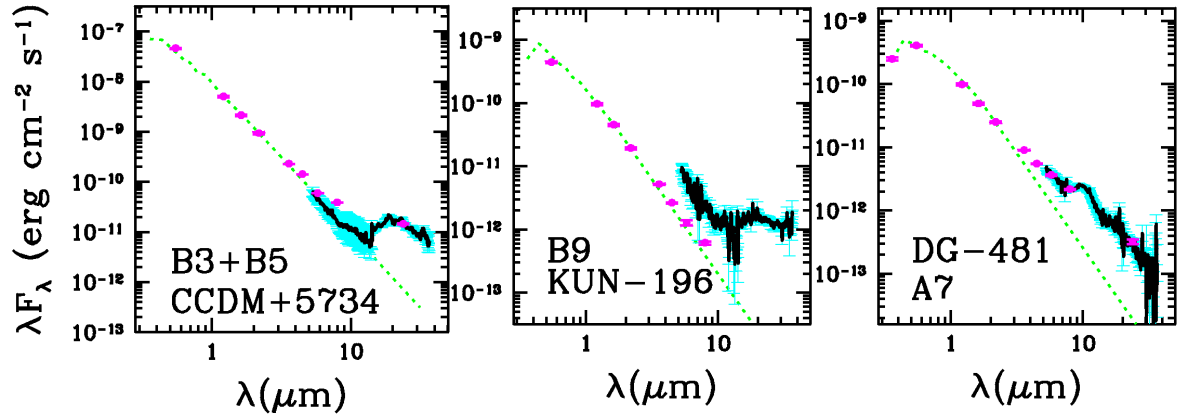}
\caption{SEDs for the intermediate-mass objects, which are most likely dust depleted TD
with large holes or debris disks. Same as Figure \ref{seds1-fig}. \label{seds11-fig}}
\end{figure}

For the solar-type stars that conform most of our sample, the region from $\sim$5.4$\mu$m to $\sim$35 $\mu$m
traces disk material located at $\sim$0.1 to $\sim$30 AU
(Calvet et al. 1992; Chiang \& Goldreich 1997), and 1.3mm observations
are sensitive to the bulk of the disk mass (Beckwith 1990). In order to
complete our picture of the objects, we include the available optical photometry
(see Sicilia-Aguilar et al. 2004, 2005, 2010).
The spectral types and extinctions are known for most of the members (Sicilia-Aguilar et al. 2005). 
In the few cases where this information is not available, we take the initial extinction to be
the cluster average (A$_V$=1.67$\pm$0.45 mag; Sicilia-Aguilar et al. 2005) and
obtain an approximate spectral type and a corrected extinction by fitting the SED to 
different {\sc marcs} models (Gustafsson et al. 2008). The SED fitting also lead to 
better extinction estimates for a few objects (72-875 and 21362507 have slightly higher 
extinctions than estimated before, A$_V \sim$2 and 3.5 mag, respectively). 
The offsets between optical/2MASS/$Spitzer$ data in a few
objects are typically related to known strong variability (e.g. 82-272, GM Cep).
The  SEDs, compared to the photospheric emission given by  the {\sc marcs} models (for the
low-mass stars) and to the stellar photospheres in Kenyon \& Hartmann (1995; for the
intermediate-mass stars) are displayed in Figures \ref{seds1-fig}-\ref{seds11-fig} (arranged
according to their SED types, see Section \ref{radmc}).
Tables \ref{opt-table} and \ref{2mass-table} offer a summary of the previously published optical
and 2MASS JHK data.

\begin{figure}
\plottwo{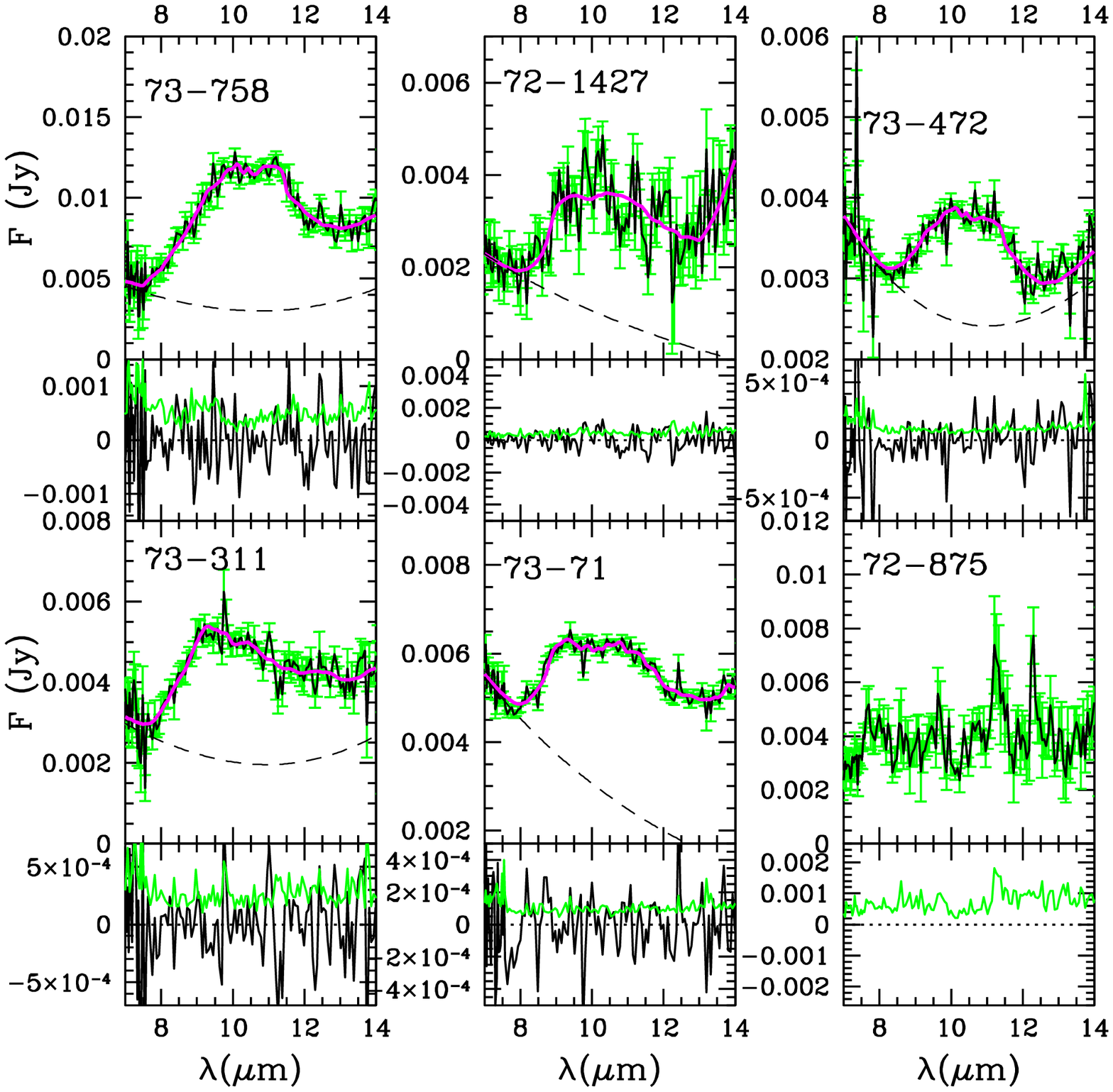}{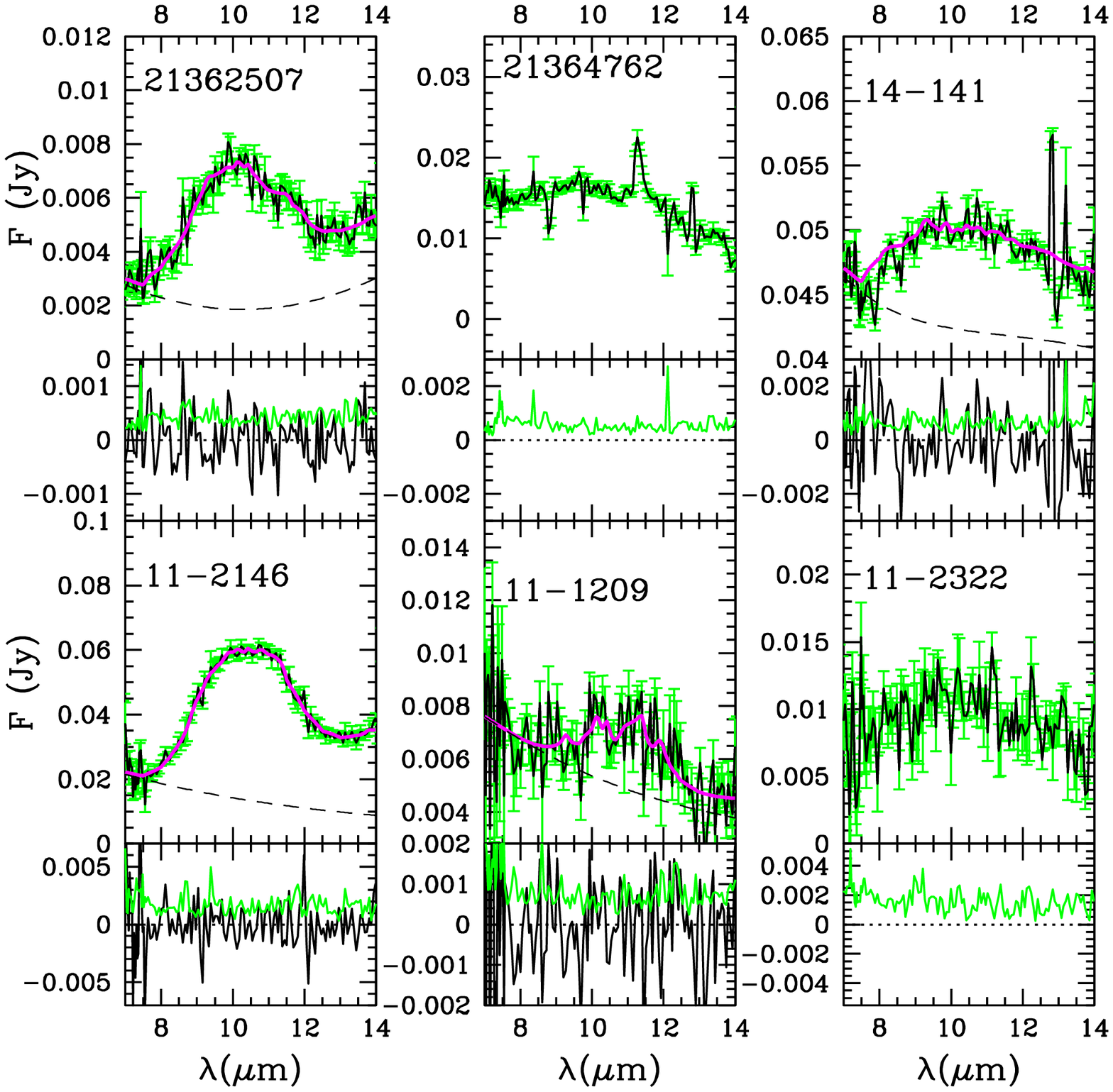}
\caption{Silicate features at in the $\sim$10$\mu$m region for the objects in Tr 37. 
For each object, the upper panel shows the IRS spectrum in black, with errors in 
green. The fitted TLTD model is shown (if appropriate) as a magenta line, and the corresponding
continuum is depicted as a dashed line. The lower panel shows the residua (black)
compared to the errors (green).\label{10um1-fig}}
\end{figure}

\begin{figure}
\plottwo{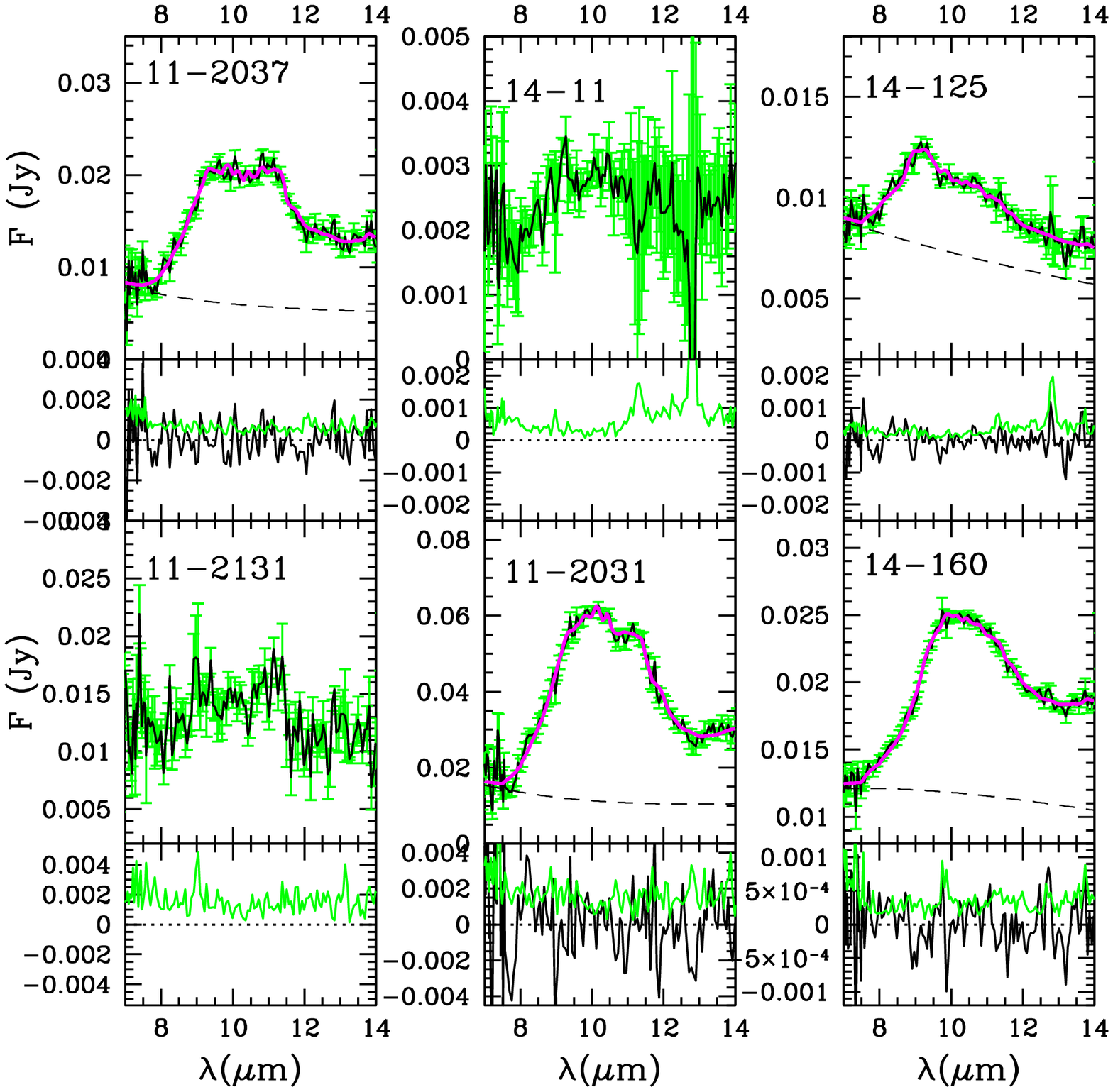}{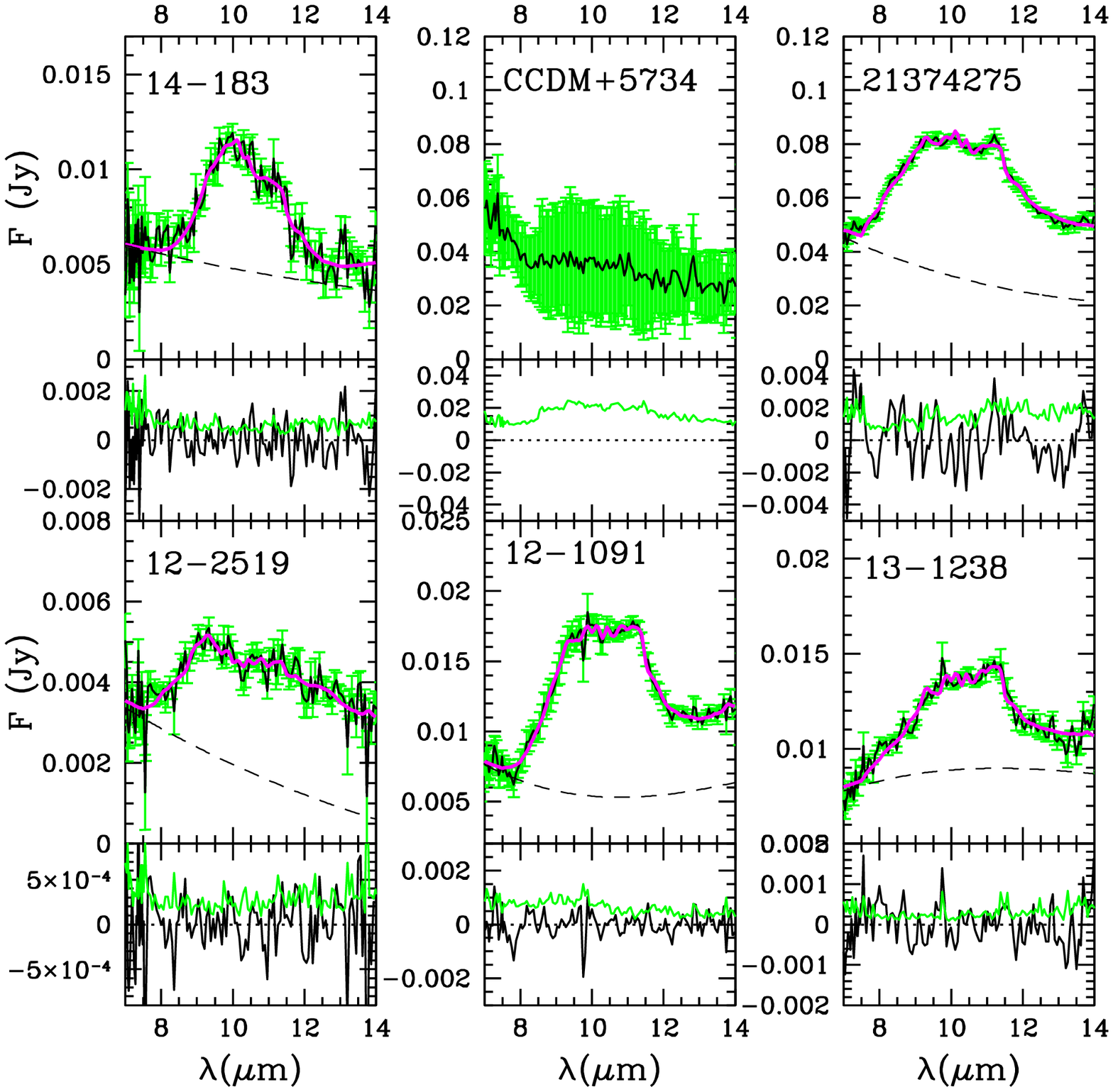}
\caption{Silicate features at in the $\sim$10$\mu$m region for the objects in Tr 37 (continued).
Same as Figure \ref{10um1-fig}. \label{10um2-fig}}
\end{figure}

\begin{figure}
\plottwo{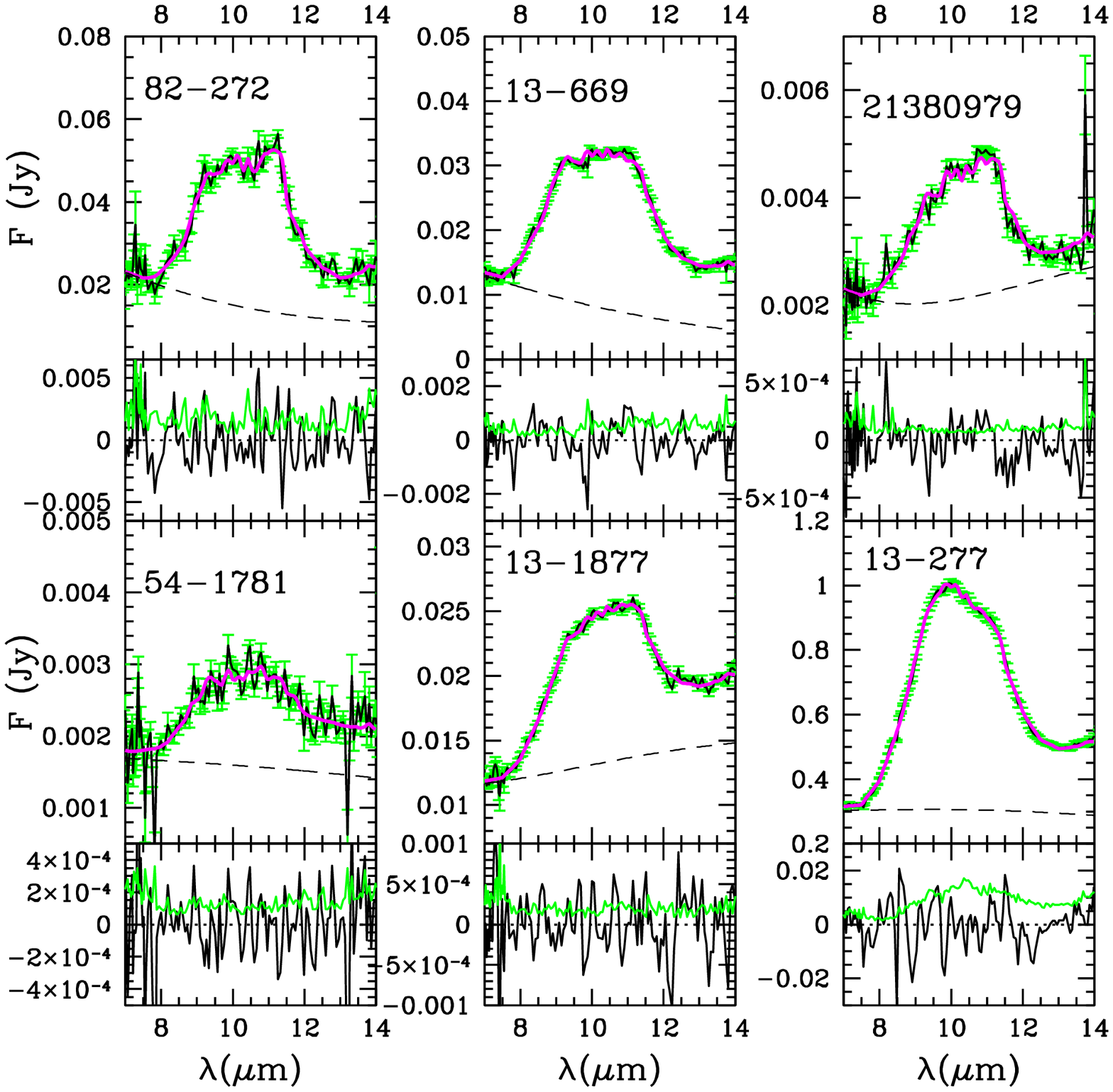}{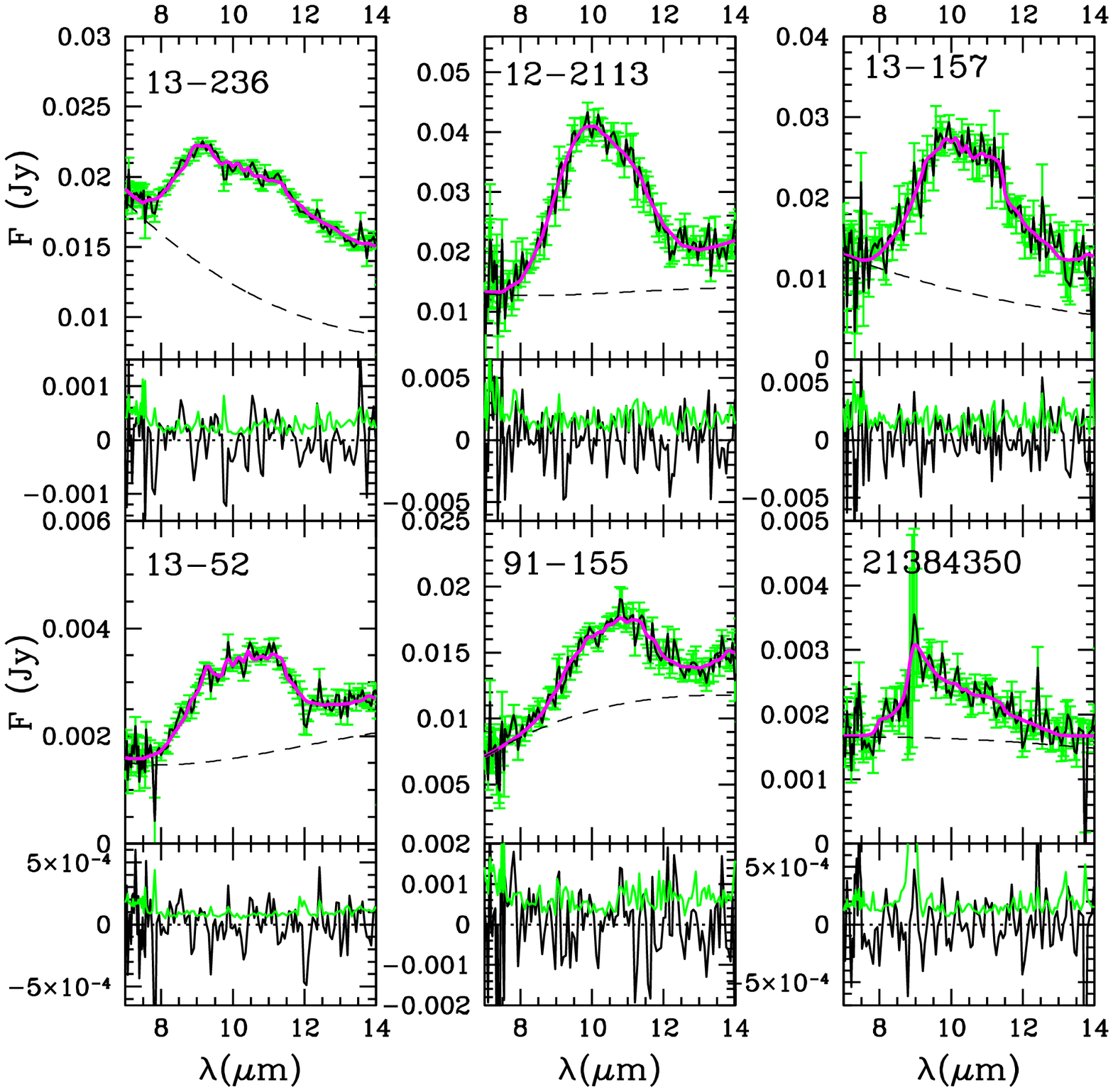}
\caption{Silicate features at in the $\sim$10$\mu$m region for the objects in Tr 37 (continued). 
Same as Figure \ref{10um1-fig}.\label{10um3-fig}}
\end{figure}

\begin{figure}
\plottwo{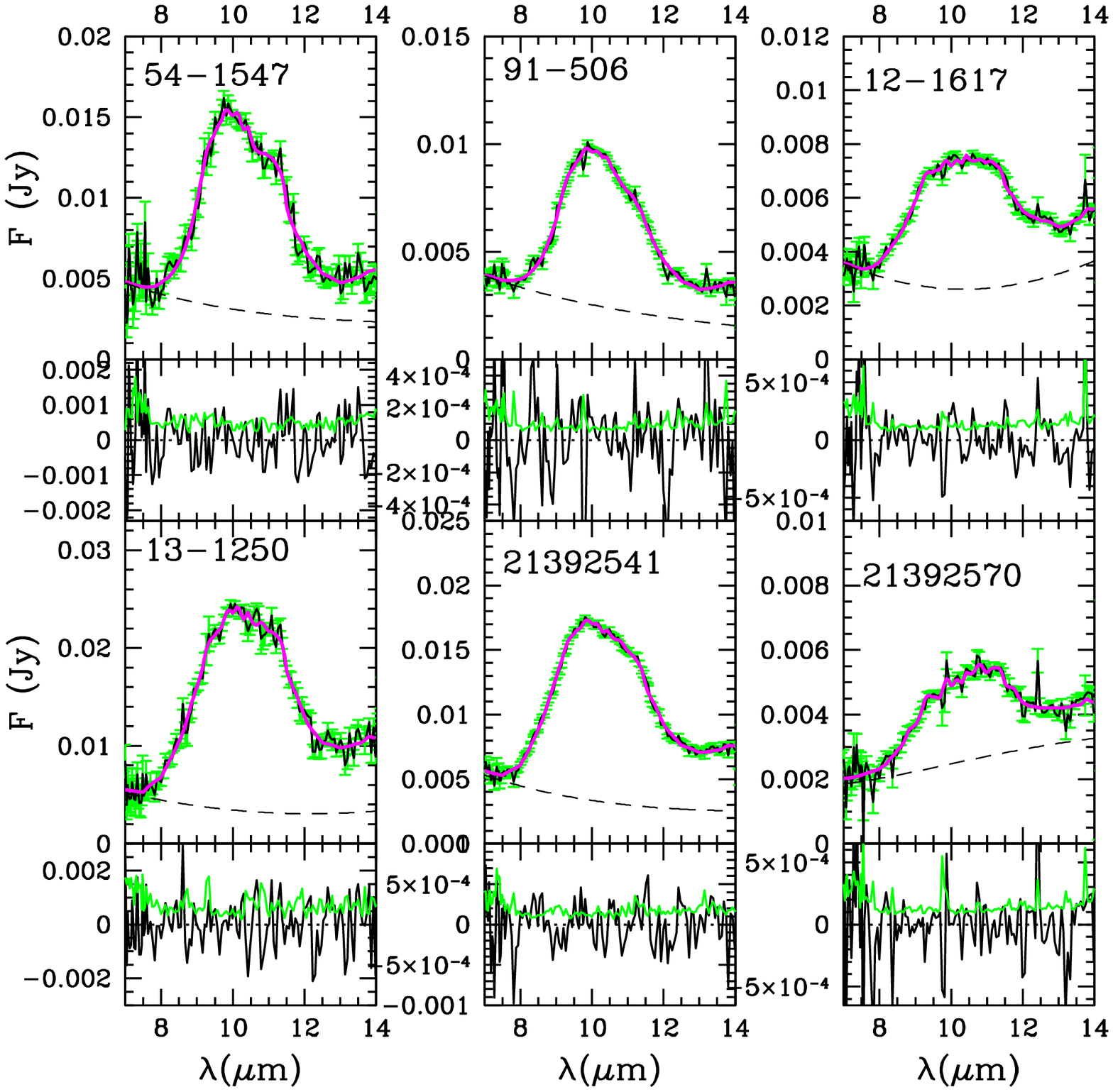}{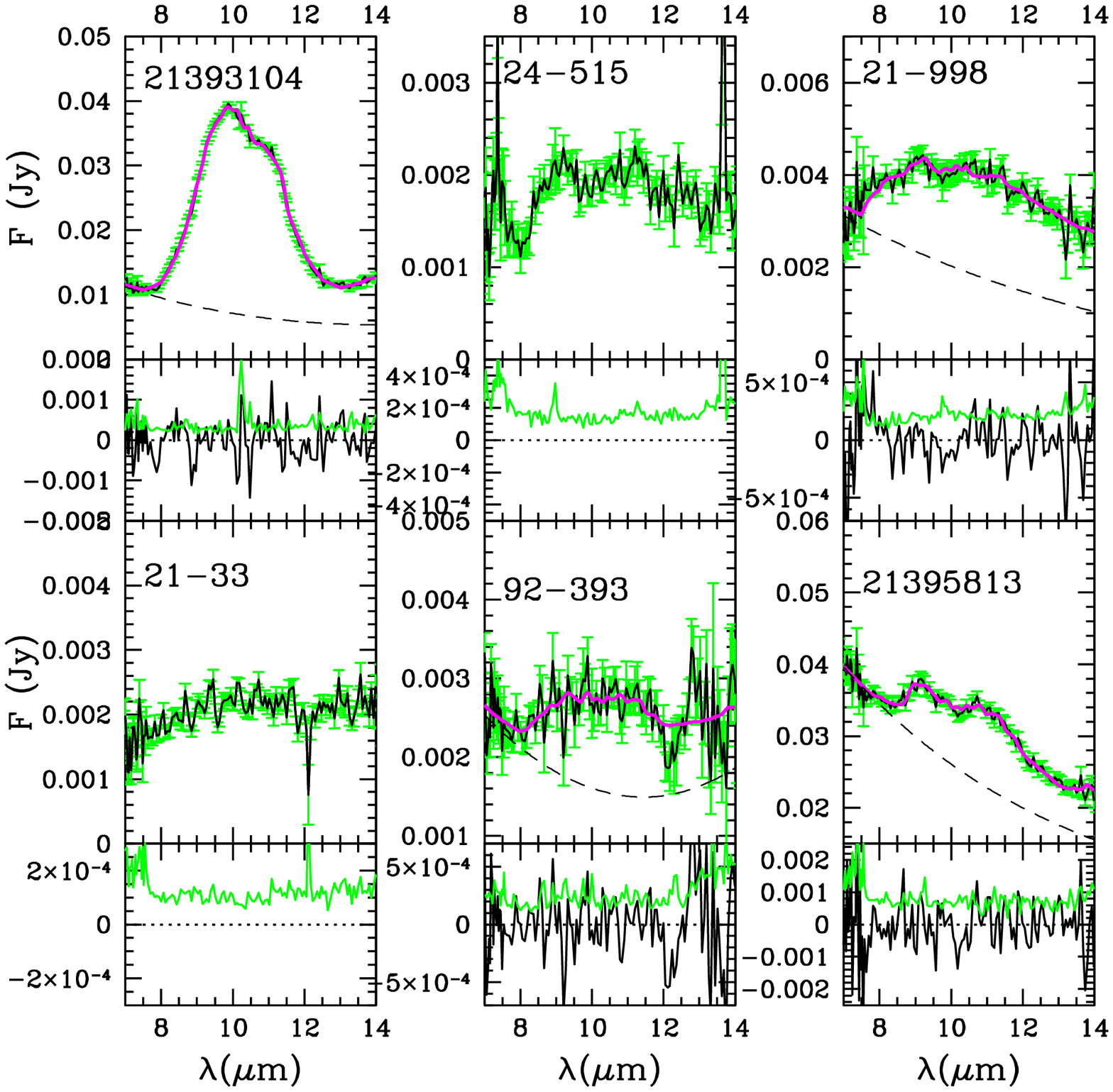}
\caption{Silicate features at in the $\sim$10$\mu$m region for the objects in Tr 37 (continued). 
Same as Figure \ref{10um1-fig}.\label{10um4-fig}}
\end{figure}

\begin{figure}
\plottwo{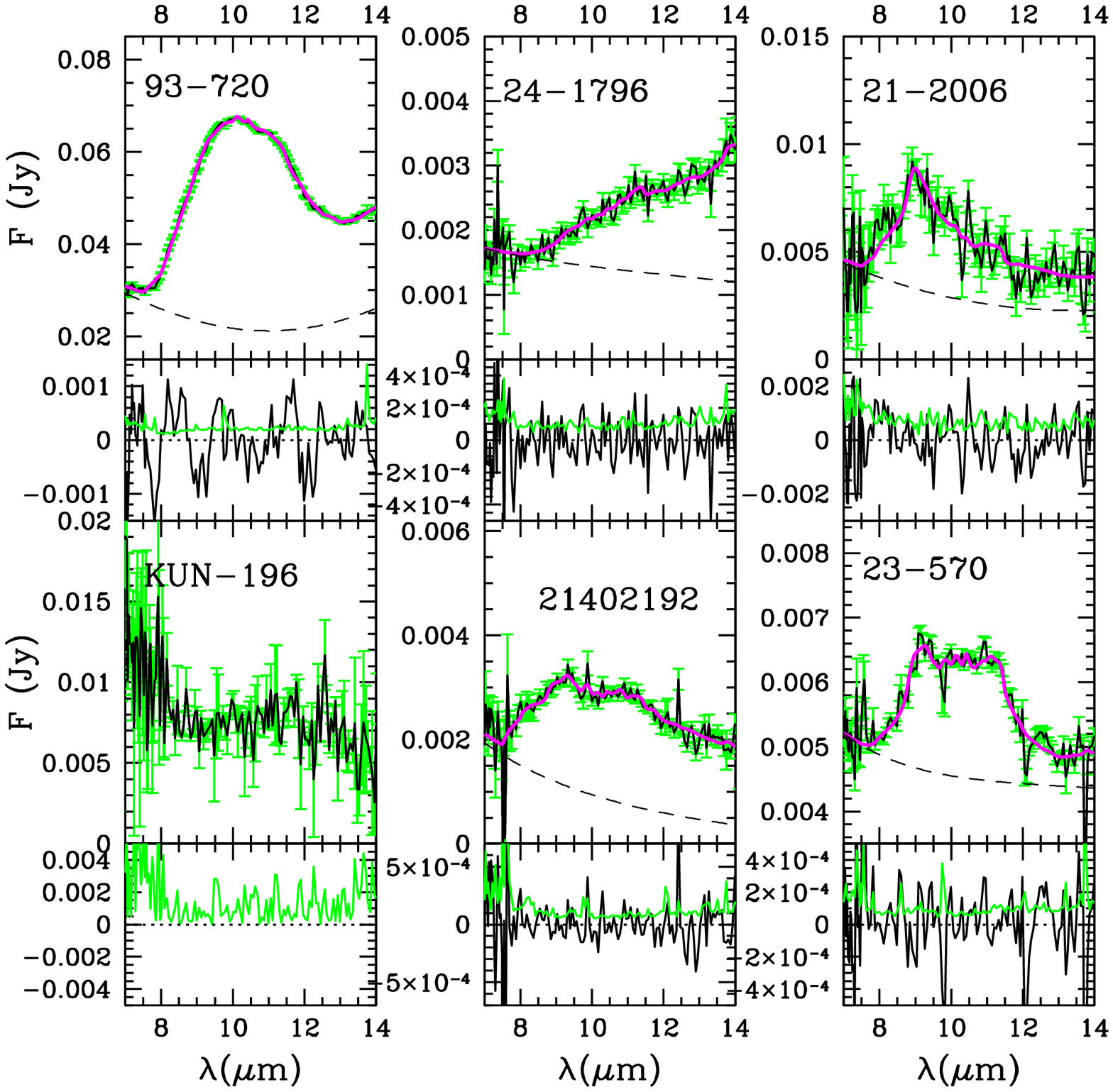}{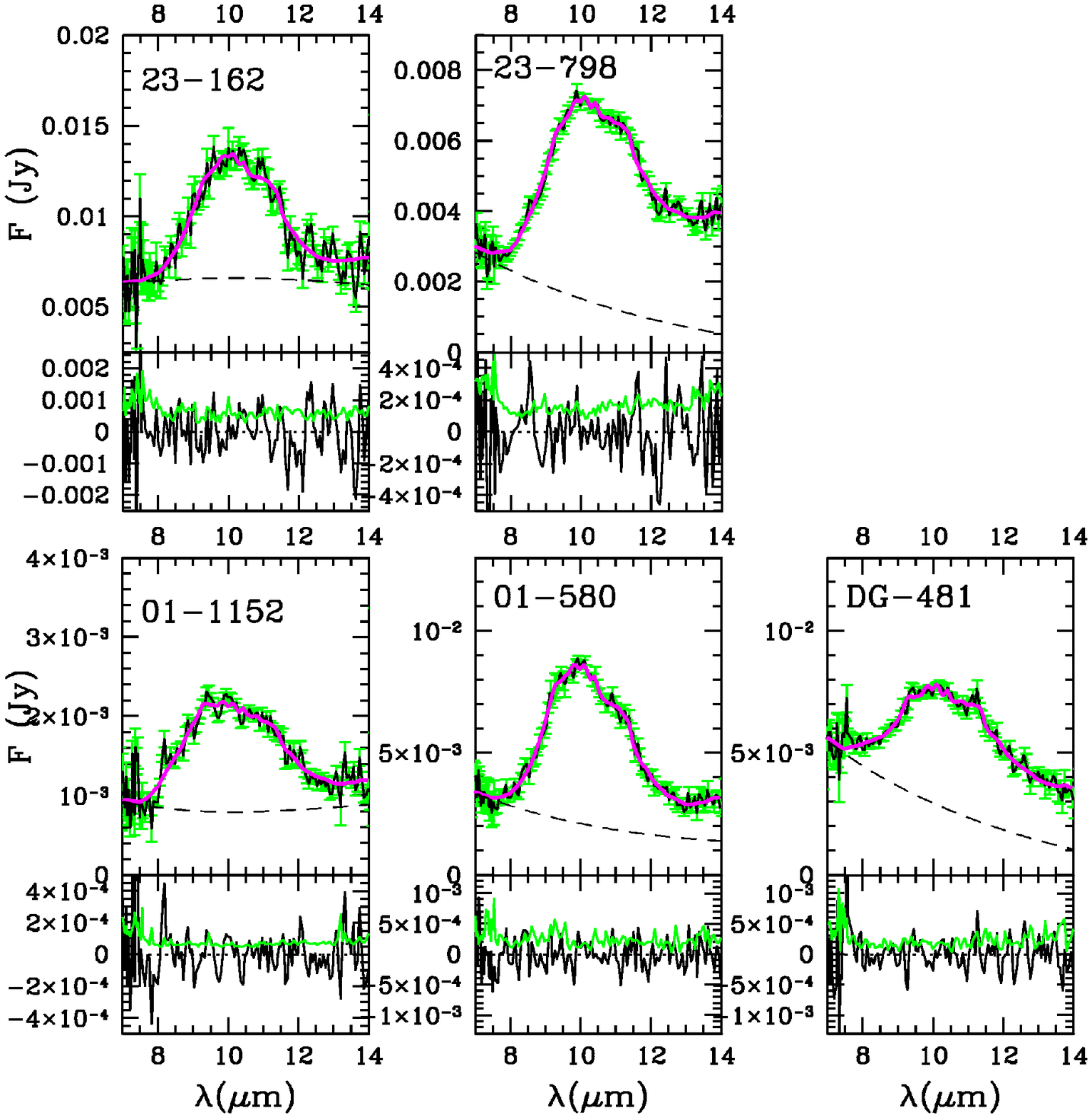}
\caption{Silicate features at in the $\sim$10$\mu$m region for the objects in Tr 37 (continued)
and NGC 7160 (lower right panels including the objects 01-1152, 01-580, and DG-481). 
Same as Figure \ref{10um1-fig}. \label{10um5-fig}}
\end{figure}

Figures \ref{10um1-fig} to \ref{10um5-fig} show the details of the 10$\mu$m silicate features.
Most of the sources present the characteristic silicate emission found in
TTS, due to the presence of a mixture of small (0.1-6$\mu$m), warm (150-450 K) 
silicate grains with amorphous and crystalline components (e.g. Natta et al. 2000;
Bouwman et al. 2001; Meeus et al. 2003).
Although silicate emission is present in most disks, it is not evident in
a few cases. In addition to the intermediate-mass stars CCDM+5734 and KUN-196, 
the 10$\mu$m silicate emission is not visible in 72-875 nor in 11-2322. 
The first object is a TD with very low accretion rate (the lack of IRAC data does not allow to check
the presence of near-IR excess, but the shape of the IRS spectrum suggest photospheric
flux down to $\sim$5-6 $\mu$m), so the lack of silicate emission could be due to 
a real depletion of small grains in the innermost disk. On the other hand, 11-2322 is
a M1 star with IR excess at all wavelengths, so the lack of silicate emission could be due to
strong grain growth and settling (in fact, due to the S/N limitations, we cannot exclude the
presence of some few-micron grains in this object), as has been suggested for the Coronet cluster
(Sicilia-Aguilar et al. 2008b). A few more disks have 
weak (or no) silicate emission, including some TD (92-393, 24-1796)
and normal disks (11-1209, 11-2131, 21-998, 21-33). 

A few other objects show atypical silicate emission. 
One of the TD (24-515) shows only evidence of
crystalline silicates, suggesting that most of the amorphous grains are in
large (few-microns) aggregates. The spectroscopic 
binary 82-272 presents prominent emission at $\sim$11 $\mu$m, suggestive of a 
dominating forsterite dust component,
as has been observed for RECX-5 in the $\eta$ Cha cluster (Bouwman et al. 2010),
and 21380979 could be another forsterite source. 
A few other sources have strong enstatite emission at $\sim$9.3 $\mu$m (e.g. 13-236,
21395813, 23-570). The emission
of 21-2006 and 21384350 is mostly consistent with amorphous silica (note that
21384350 is also partially affected by the strong 
emission of a bright nearby object), and 12-2519 may be another
silica source. Nevertheless, due to the complex shape of the continuum, extracting the 
silicate emission from the spectrum is complicated in TD and in objects with small
near-IR excesses whose SEDs show a change 
in the slope around 6--12um (what we call "kink" disks, see SA06). 
The features observed in 24-1796 and 12-2519 
could be affected by the strong change in the SED slope occurring at
these wavelengths, although the crystalline features in 12-2519 reveal that
at least some small crystalline grains are present in this disk.

\subsection{Disk mineralogy and grain size derived from the silicate feature\label{fit}}

In order to study the disk mineralogy, we fit the optically
thin silicate emission using the Two Layers Temperature Distribution (TLTD) method,
developed by Juh\'{a}sz et al. (2009), from now on J09. We exclude
from the fitting procedure and subsequent mineralogy analysis the 
objects with low S/N (S/N$\leq$20), those with no evidence of silicate emission, and those
suffering from strong contamination by nearby sources or nebular
emission in the 10$\mu$m region (e.g. 21384350). Although the silicate emission at 
$\sim$10$\mu$m and $\sim$20$\mu$m traces only a small part of the dust content in 
the disk (small grains in the optically thin disk atmosphere with temperatures
T$\sim$150-450 K; Calvet et al. 1992; Natta et al. 2000), the presence or absence of crystalline
grains and the grain size can reveal information about heating processes in
the disk, irradiation, transport and turbulence (Meeus et al. 2001; Honda et al. 2003;
Dullemond et al. 2006;
Dullemond \& Dominik 2004, 2008; Watson et al. 2009; Zsom et al. 2011).

The TLTD method (see J09 for details) reproduces the
silicate feature using a multicomponent continuum, including
emission from the stellar photosphere, an inner rim, and the optically thick disk midplane.
It considers a collection of four different dust species with sizes 
0.1, 1.5, and 6.0~$\mu$m: amorphous silicates
with olivine and pyroxene stoichiometry (Dorschner et al.
1995), forsterite (Sogawa et al. 2006), amorphous silica (Henning \& Mutschke 1997);
and, in addition, enstatite grains (J\"{a}ger et al. 1998) with sizes 0.1 and 1.5~$\mu$m
(since the emission of 6$\mu$m enstatite grains is indistinguishable from 
the continuum in cases of low S/N; see J09).
The mass absorption coefficients are calculated from the material optical
constants assuming that the crystalline grains can be assimilated to a
Distribution of Hollow Spheres (Min et al. 2005), and considering
standard Mie theory for spherical particles for the amorphous dust.
The main difference with standard one- or two-temperature fitting methods
is that the TLTD model considers that the 
emission originates in regions characterized by a continuous
distribution of temperatures. Therefore, the flux can be written as (see J09
for further details):

 \begin{eqnarray}
F_\nu = F_{\nu, {\rm cont}} & + & \sum_{i=1}^N\sum_{j=1}^MD_{i,j}\kappa_{i,j}
\int_{\rm{T_{\rm a, max}}}^{\rm{T_{\rm a, min}}}\frac{2\pi}{d^2}B_\nu(T){T}^{\frac{2-qa}{qa}}dT
\label{eq:1}
\end{eqnarray}

where the flux in the continuum is:

\begin{eqnarray}
F_{\nu, {\rm cont}} = D_0 \frac{\pi R_\star^2}{d^2} B_\nu(T_\star)&+& D_1\int_{\rm{T_{\rm r,max}}}^{\rm{T_{\rm r, min}}}\frac{2\pi}{d^2}B_\nu(T){T}^{\frac{2-qr}{qr}}dT \\
&+& D_2\int_{\rm{T_{\rm m,max}}}^{\rm{T_{\rm m, min}}}\frac{2\pi}{d^2}B_\nu(T){T}^{\frac{2-qm}{qm}}dT.
\label{eq:2}
\end{eqnarray}

Here, $R_*$ and $T_*$ are the radius and effective temperature of the
star. The temperatures of the disk atmosphere, rim, and midplane 
($T_{a}$, $T_{r}$, and $T_{m}$) are parameterized as power laws of the
radius (with exponents qa, qr, and qm, respectively) and vary between a minimum
and a maximum value. The exponents of the temperatures (qa, qr, qm) 
and the coefficients of each contribution ($D_0$, $D_1$, $D_2$, and $D_{i,j}$) are 
fitted to the data using a genetic optimization algorithm (PIKAIA, Charbonneau 1995;
see J09). The uncertainties in the IRS spectra are estimated by adding random 
Gaussian noise to the original spectrum at the noise level, and then
repeating the fit 100 times. The final silicate composition, the total crystalline 
fraction, and the average grain sizes are obtained as the average of the whole 
set, and the errors are derived from the standard deviation in  the positive
and negative direction. Since the mineralogy at different radial distances is 
not necessarily the same and the S/N of our spectra is also strongly 
wavelength-dependent, we fitted separately the 7-14 and 17-35$\mu$m ranges, 
as we have done before for similar sets of data
(Sicilia-Aguilar et al. 2008b). In the long-wavelength region (17-35$\mu$m), 
the emission of large amorphous grains strongly resembles
the continuum, so the crystalline fractions may have a tendency to being overestimated.

\begin{figure}
\plotone{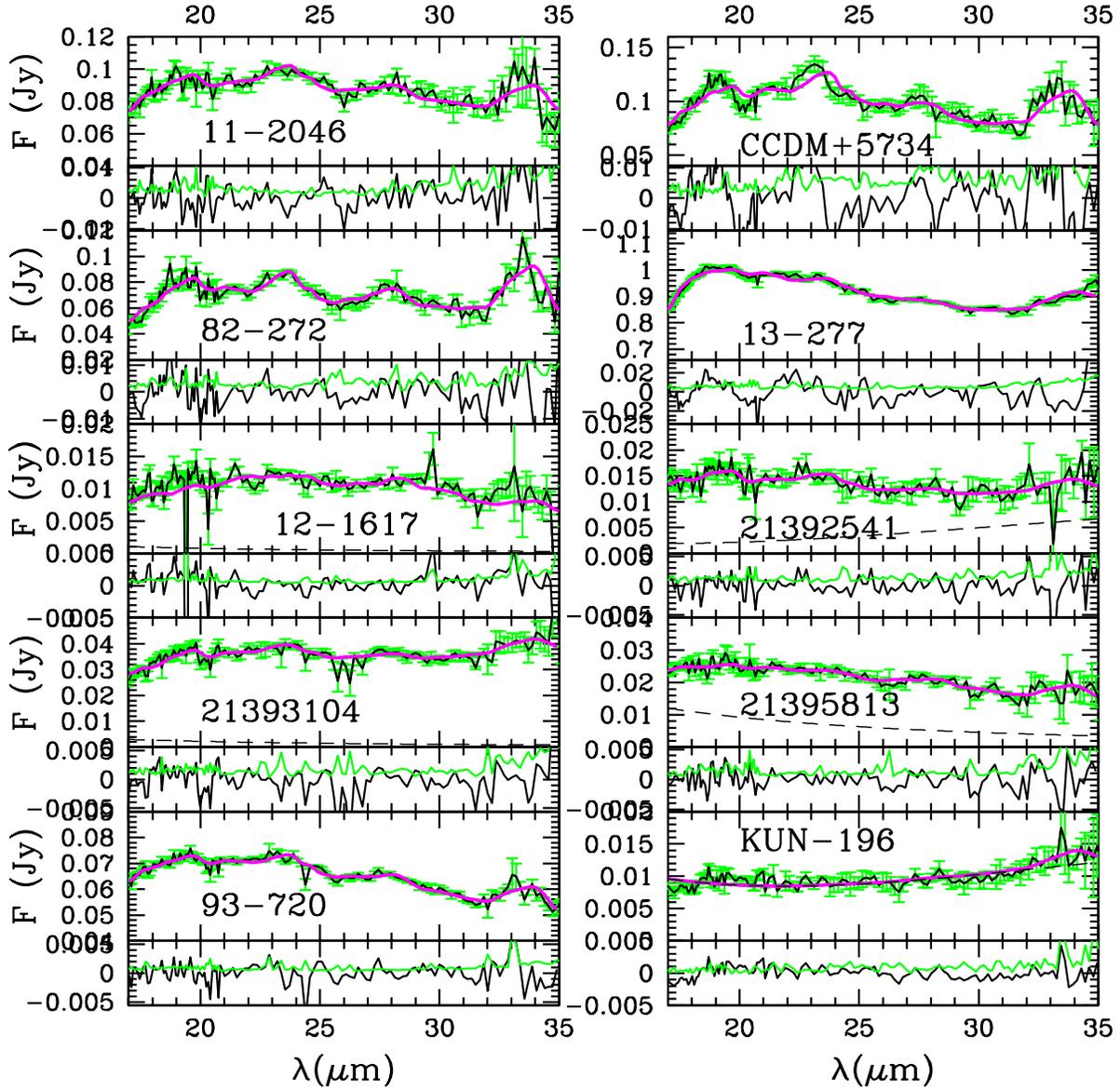}
\caption{Silicate features at in the $\sim$20-30$\mu$m region for the objects in Tr 37 with good S/N at long
wavelengths. The S/N in these spectra allow to identify features (11-2046, CCDM+5734,
82-272, 13-277, 93-720) as well as to exclude the presence of them (21395813, KUN-196).
\label{20um-fig}}
\end{figure}

The results are displayed in Tables \ref{silshort-table}-\ref{sizecrystlong-table}
and the fits to the silicate features are shown in Figures \ref{10um1-fig}-\ref{20um-fig}. 
As in our previous studies, we find that the TLTD model with the assumed grain
compositions and size distribution reproduces very well the observed features
at both short and long wavelengths. The residuals do not suggest the presence 
of other additional components, and the main source of uncertainty in the models
is the S/N of the spectra. The sensitivity to large grains depends strongly 
on the S/N of the spectrum:
The emission from grains larger than $\sim$6$\mu$m is not very different from the continuum 
emission in this wavelength range, so separating the emission of large grains from the
continuum is only possible in objects with high S/N, and the maximum grain
size detectable in noisy spectra (S/N$\sim$20-50) may be underestimated.
Therefore, in order to extend the study of the grain size to all spectra, we also measured 
the peak of the silicate feature by normalizing the spectra to the continuum
obtained by fitting a straight line between the average 
values in the regions 7.2-7.4$\mu$m and 12.6-12.7$\mu$m (the peak
values are listed in Table \ref{obs-table}). This
is a similar procedure to what we used in SA07, and provides a good
estimate of the grain size in objects with low S/N. The
normalized peak is strongly correlated with the grain size (see Bouwman
et al. 2008), as objects with larger grains show shallower
and broader features. For our sample, the correlation between measured
silicate peak and fitted grain size is very strong, with a Spearman rank coefficient
-0.57 (range -0.77 --- -0.26) and a probability p=9$\times$10$^{-5}$ that
the two quantities are uncorrelated, even though the continuum fit used to measure 
the normalized peak is not the same defined by the TLTD models. 
The disk mineralogy derived here
is further analyzed in Sections \ref{mineral} and \ref{transition}, where
we compare the dust characteristics with the stellar and disk properties.

\subsection{Tracing the global disk structure with RADMC\label{radmc}}

In order to explore in more detail not only the disk mineralogy, but
also the global disk structure, we have performed detailed modeling using a full 2D 
radiative transfer code. Our aim with this analysis is 
not to produce accurate fits to the SEDs of the individual objects in
the region, but to understand the types of disk structures 
that reproduce the observed SEDs. Detailed, full-disk models
offer an advantage over simple silicate-feature fitting procedure, which is that the
whole disk structure is taken into account, allowing to explore the global
disk details together with the grain size distributions. This is particularly important
in objects where the SED slope changes at the wavelengths 
where the silicate emission appears, for which defining the continuum level
with a simple model can be uncertain and lead to variations in the strength and width 
of the extracted silicate feature and thus errors in the estimated grain size. Given the
various morphologies of the disks in our sample, up to 20\% of 
our objects could be affected by this continuum uncertainty
(see Figures \ref{10um1-fig}-\ref{10um5-fig} and 
comments in Table \ref{sizecrystshort-table}). 

We explored what are the minimal changes that need
to be performed to a normal CTTS disk structure in order to reproduce
the observed SEDs. For this purpose, we have selected 10 prototype 
disks with different SED morphologies that represent the main types
of objects observed in the Cep~OB2 region (see Table \ref{diskproto-table}
for a detailed summary of the prototypes and their properties). 
The prototype disks include:  a normal TTS disk with silicate
emission (model 1, prototype 11-2037), a very massive and flared
disk (model 2, prototype GM Cep), a dust-depleted CTTS (model 3, prototype
13-236), a strongly depleted and settled disk (model 4, prototype 11-1209), 
a ``kink'' or pre-transitional disk 
(model 5, prototype 13-1250\footnote{This object had been previously 
classified as TD, but the new IRAC photometry produces a color slightly over our
TD cutoff, being rather an example of a pre-transitional (Espaillat
et al. 2010) or ``kink" disk, according to its IR colors.}), 
a depleted but strongly accreting disk
with strong silicate emission
(model 6, prototype the old disk around 01-580 in NGC 7160), a classical
(large 24$\mu$m flux or ``turn-up") TD with weak silicate emission (model 7,
prototype 24-1796), a classical TD with strong silicate emission
(model 8, prototype 14-11), a TD with no silicate
and low 24$\mu$m flux or settled TD (model 9, prototype 92-393), and
a settled TD with strong silicate emission (model 10, prototype 13-52).

\begin{figure}
\plottwo{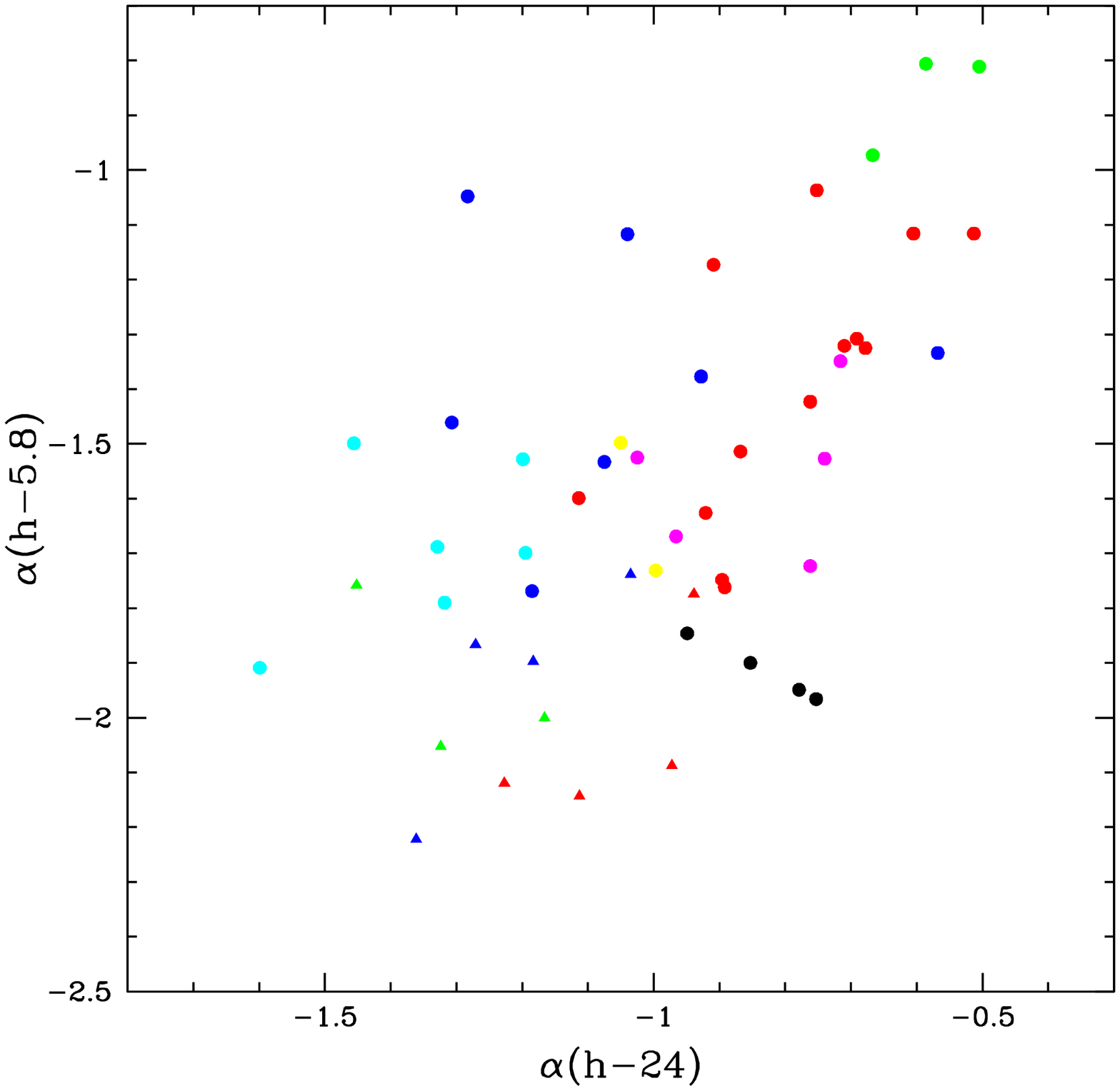}{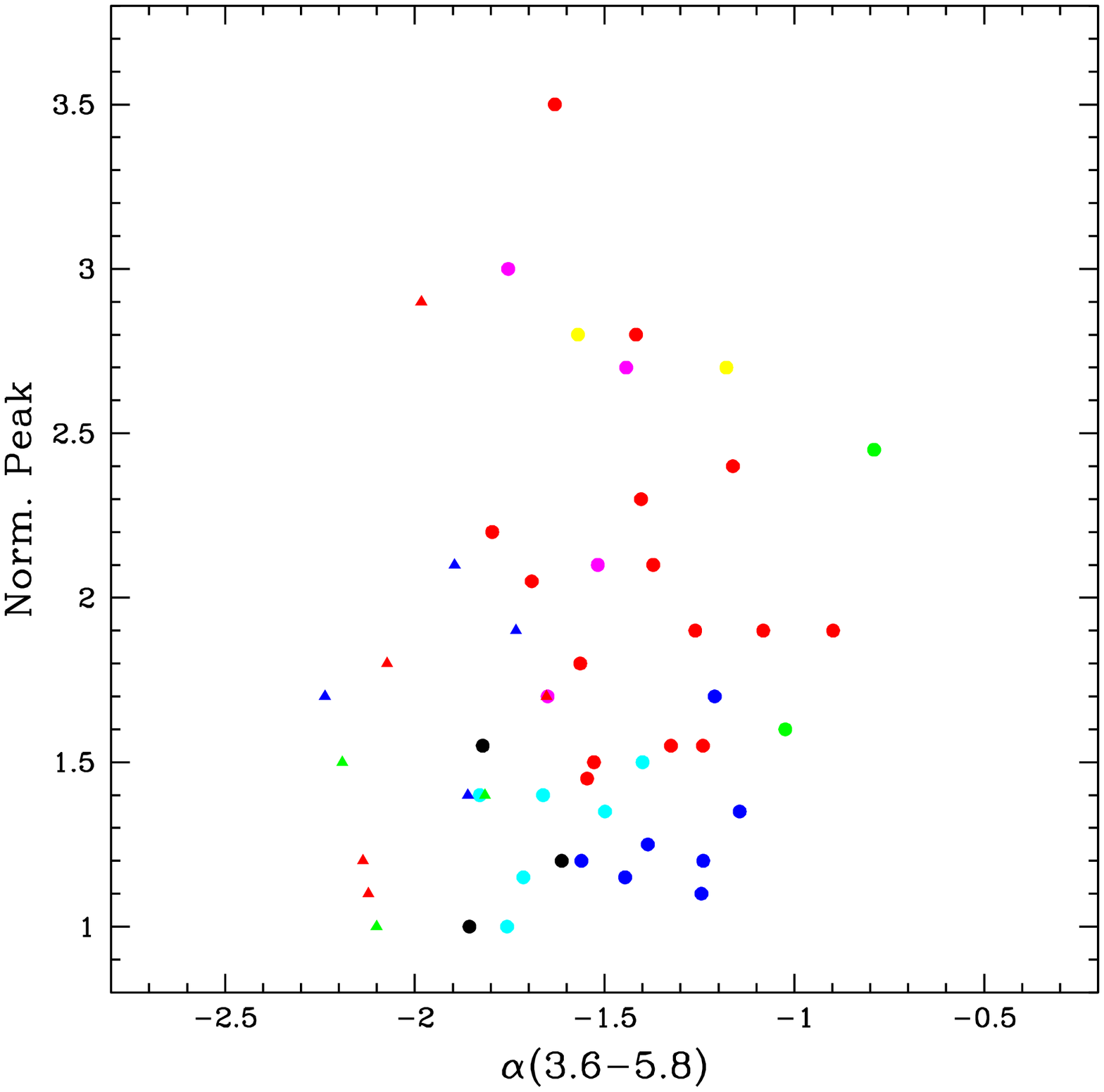}
\caption{SED slope and normalized silicate peak for visualization of the SED classification scheme
(see text). The SED types are determined considering the whole SED and the
relative strength of the 10$\mu$m silicate feature, here we present the 
SED slope at H-[5.8] vs. H-[24] (left) and the normalized silicate peak versus
the inner disk slope at [3.6]-[5.8].
The correspondence of SED class and symbol is as follows: red circles = \# 1, green
circles = \# 2, blue circles = \#3, cyan circles = \#4, magenta circles = \#5,
yellow circles = \#6, black circles = \#7, red triangles = \#8, green triangles = \#9,
blue triangles = \#10.   
\label{sedtype-fig}}
\end{figure}

All disks, with some variations due to differences in stellar luminosity
and mineralogy, can be classified within one of these subtypes. 
Note that we exclude from this classification the 3 intermediate-mass
stars (CCDM+5734, KUN-196, and DG-481) since they are most likely
TD with very large inner holes and low dust masses, or debris disks. The SEDs
in Figures \ref{seds1-fig}-\ref{seds10-fig} have been arranged according to
this classification, and the SED type is also listed in Table \ref{obs-table} for each object.
Figure \ref{sedtype-fig} helps to visualize the SED type, which is based
on the SED shape and the silicate feature. 
We do not attempt to define a unique set of indices for SED classification, 
but to organize the SEDs observed in Cep OB2 in order to explore the underlying
physics and disk structures. Given that the disk slope is far from constant in the IR, especially
for the disks with inside-out evolution, we checked the SED
slope, $\alpha$, at different wavelengths, considering for this 
the 2MASS, IRAC, MIPS, and IRS data. From the IRS data, we derive
the fluxes at 5.8, 8, 12, and 17$\mu$m by integrating the flux 
in a 0.8 $\mu$m interval around these wavelengths. For longer wavelengths,
we also derived the flux by integrating within a 2$\mu$m interval
centered at 23 and 33$\mu$m. Although both the 23 and 33$\mu$m photometric points 
could be affected by forsterite emission, strong crystalline
features are only present in very few sources, so general trends should remain unaffected.
In general, for the plots and the analysis,
we give priority to the IRS-derived slopes, using the IRAC and MIPS values
whenever no IRS data was available. In any case, IRAC, MIPS, and IRS data
are in excellent agreement. The SED slope values are listed
in Table \ref{alpha-table}. 
For the SED classification, we note that we consider as TD those with
[3.6]-[4.5]$<$0.2 and small or negligible near-IR excess, since also a few
normal CTTS (e.g. 14-183) have [3.6]-[4.5]$<$0.2. We also list as TD some objects
with [3.6]-[4.5]$\sim$0.2-0.25 and strong turn-ups (72-1427, 73-472, 21392541, 21392570),
since their SEDs are more similar to our TD prototypes than to other models
(e.g., ``kink" disks or depleted CTTS).

We used the RADMC radiative transfer code 
package\footnote{See http://www.mpia.de/homes/dullemon/radtrans/radmc/} (Dullemond \& Dominik 2004)
for modeling 3-D axisymmetric circumstellar dust configurations.
This package was described and first used in Dullemond
\& Dominik (2004), and has
been extensively used and tested over the last 7 years for continuum radiative
transfer in protoplanetary disks and for computing the disk structure
(see for instance Pascucci et al. 2004; Pinte et al. 2009).
Although the code assumes axially symmetric density distribution (2D), 
photon packages are followed in 3D.
The code uses a variant of the Monte Carlo method of Bjorkman
\& Wood (2001) to compute how the stellar photons
penetrate the disk and determine the dust temperature and scattering
source function everywhere in the disk. Using volume ray-tracing
the spectra and images can then be determined at any inclination.
For evolved disks, which can be very geometrically
thin, special care is taken of the grid, ensuring that the vertical
structure of the disk is always well resolved. RADMC can handle
multiple dust size and/or composition populations, each having its
own spatial distribution. This makes the code well suited for
studying how dust growth/fragmentation/settling affect the SED
of the disk. In case of different dust distributions (e.g.,
for the innermost and outer disk), the temperature of each one is treated 
separately, although for simplification all the grains contained in a 
collisional equilibrium size distribution are assumed to have the same 
temperature, independently of their size.

We run different RADMC models in order to reproduce the full SED of the prototype disks,
using the stellar parameters effective temperature, stellar radius, and stellar mass
T$_{eff}$, R$_*$, M$_*$ derived from the spectral types and optical
photometry  (see Table \ref{diskproto-table}). The disks are modeled in the
simplest way, assuming a pressure scale height (H$_p$) that varies as a power-law with the
radius, H$_p$/R$\propto$R$^{1/7}$ . We observed that models with well-mixed dust and gas 
in hydrostatic equilibrium failed to reproduce the disks, so the scale height 
at the outer disk radius, H$_{rdisk}$/R$_{disk}$, had to be set for each case. 
This procedure is equivalent to including some degree of settling in the model.
Although settling is likely to affect 
differently grains of different sizes, our simple model assumes that all the grains 
have the same scale height and are well mixed. We also assume a negligible accretion 
rate, which is true for most of the objects
in Cep OB2, given their typical \.{M}$\sim$10$^{-8}$--10$^{-9}$M$_\odot$/yr rates.
The disk surface density is estimated from the total disk mass, assuming that
it follows a power law of the radius with exponent -1. 
In all cases, we take the gas-to-dust ratio to be 100, and
the total disk mass is varied to fit the millimeter emission and the
shape of the mid-IR flux as seen in the IRS spectra. 
This method allows to better reproduce the dust content in evolved disks where 
we may expect strong grain growth, anomalous gas to dust ratios, and 
variable viscosity and accretion rate with the radial distance. 
We note that for most objects, the total disk masses inferred from the
millimeter and 20-30$\mu$m data are below the values one would
expect to account for the typical accretion rates of 10$^{-8}$--$^{-9}$M$_\odot$/yr,
which is suggestive of strong grain growth and/or anomalous gas-to-dust ratios.
The outer disk radius is relatively unconstrained, as the main parameter
is the total assumed disk mass.
For each case, we explored different sizes for the inner hole by
setting the temperature of the disk inner rim, from
no hole or dust distribution starting at the dust destruction radius, T$\sim$1500 K,
up to few-AU cleared holes, or inner rim temperature $\sim$150-200 K. We
explored different silicate grain size distributions, assuming
a collisional power-law distribution with an exponent -3.5 and a minimum and
a maximum grain size, which vary between 0.1-10$\mu$m and 100-10000$\mu$m,
respectively. A total of 25\% of amorphous carbon dust was also included in all the models,
assuming the same grain size distribution for carbon as we do for silicate grains.
Since the inclination of the objects is unknown, we assume always a 45 degrees
angle. 

\begin{figure}
\plottwo{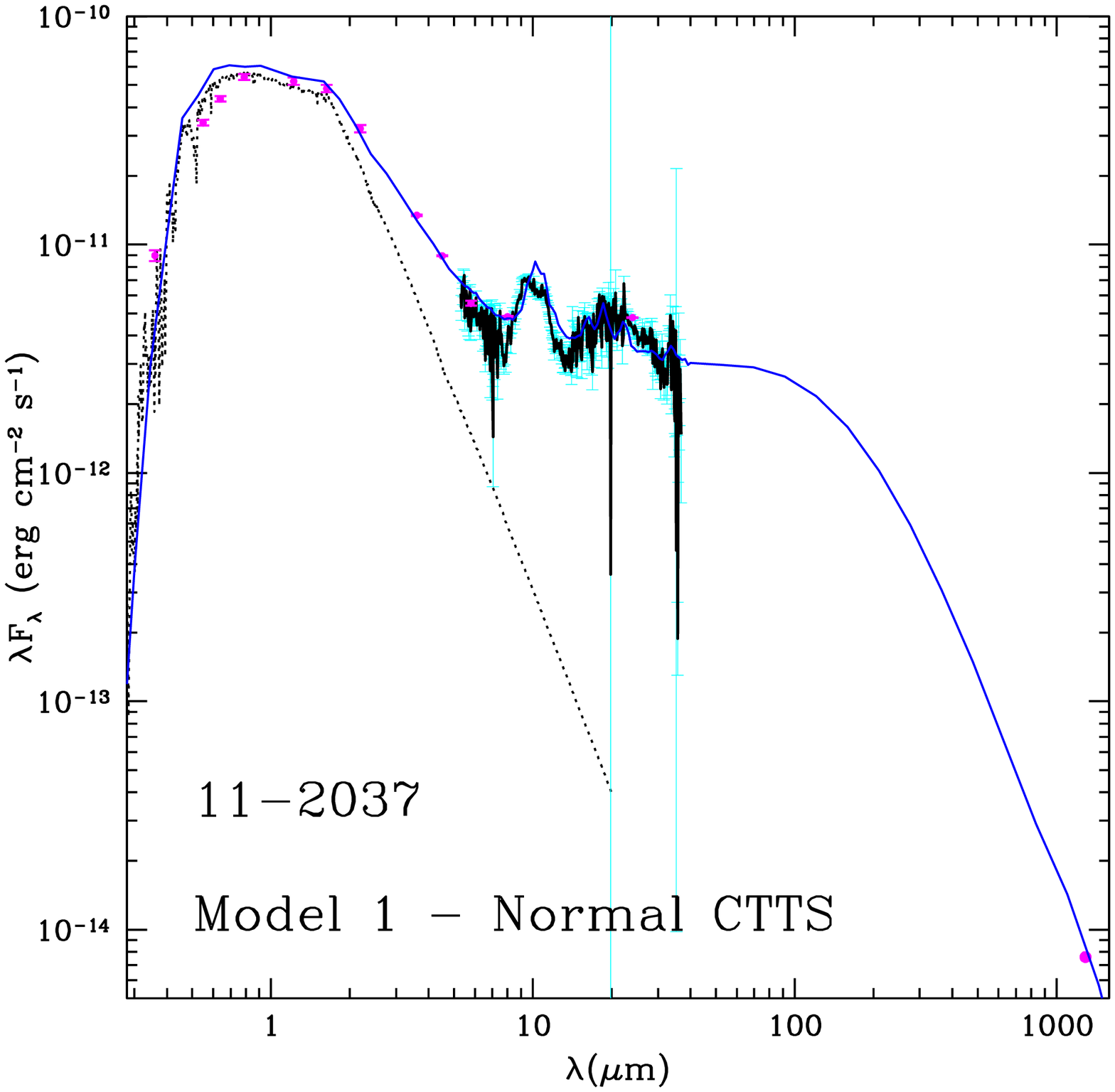}{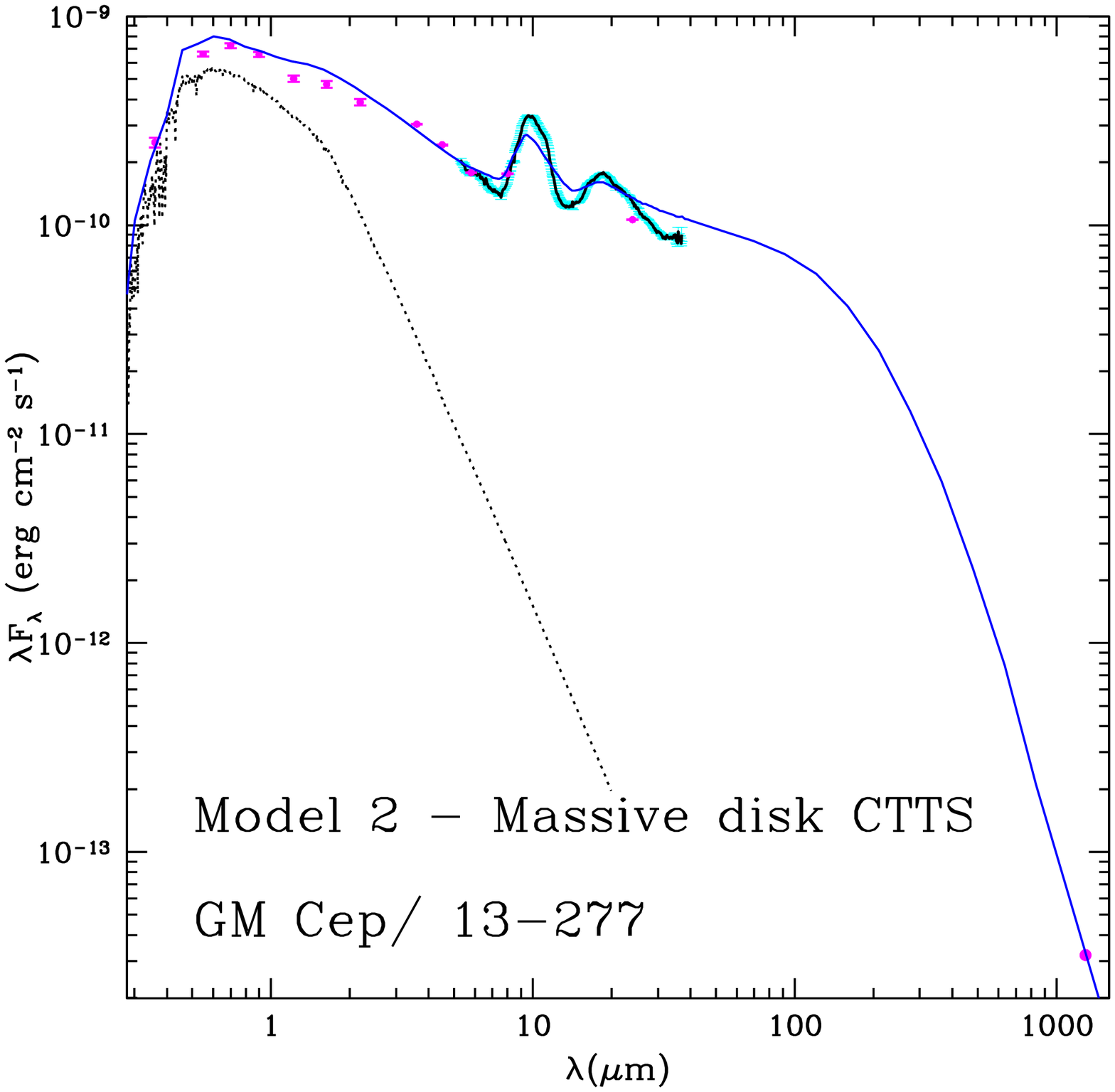}
\caption{Models for normal (left, \#1) and massive(right, \#2) CTTS, compared
to the SEDs and IRS spectra of their prototypes 11-2037 and GM Cep, respectively.\label{modelsA-fig}}
\end{figure}

\begin{figure}
\plottwo{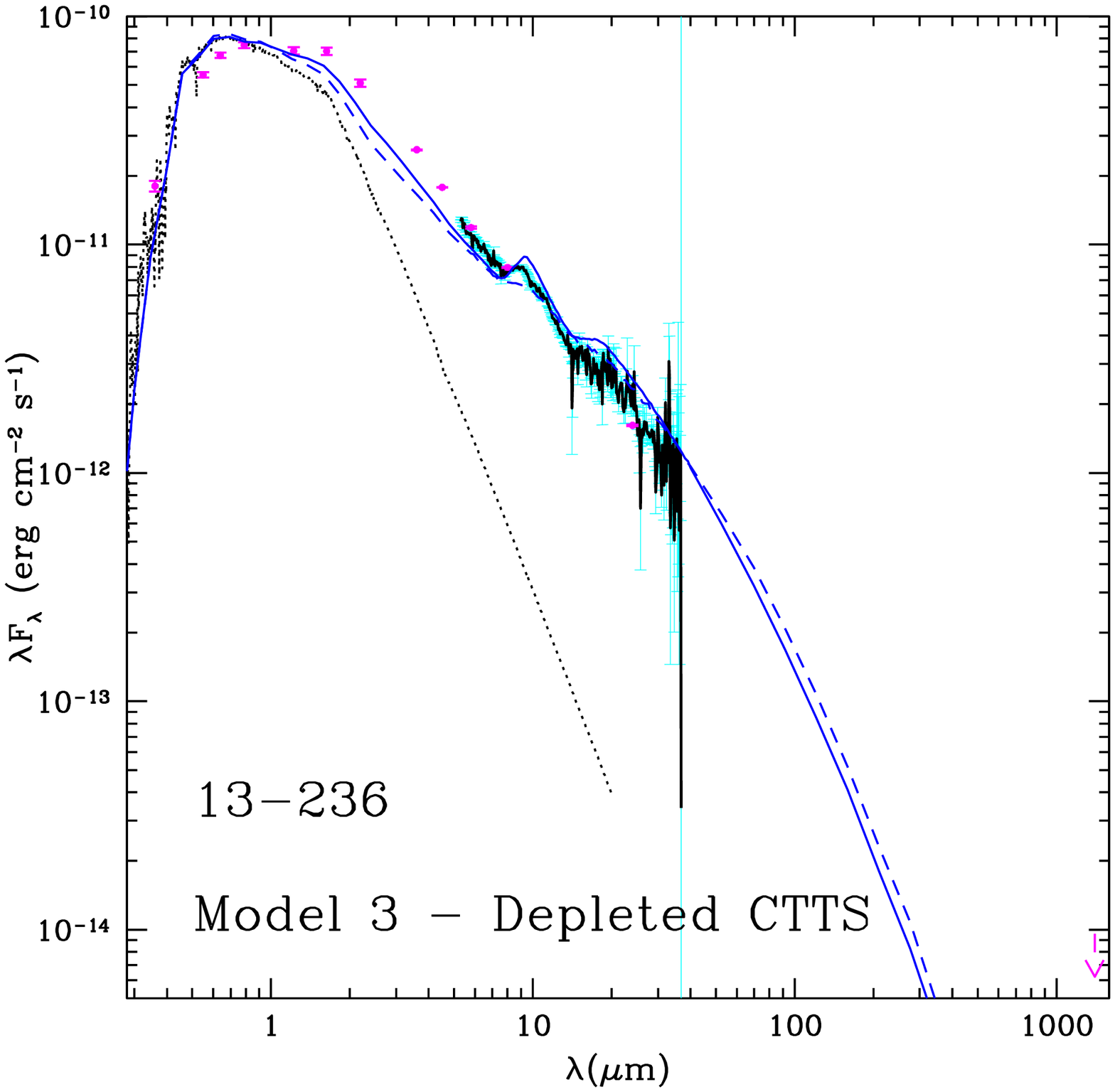}{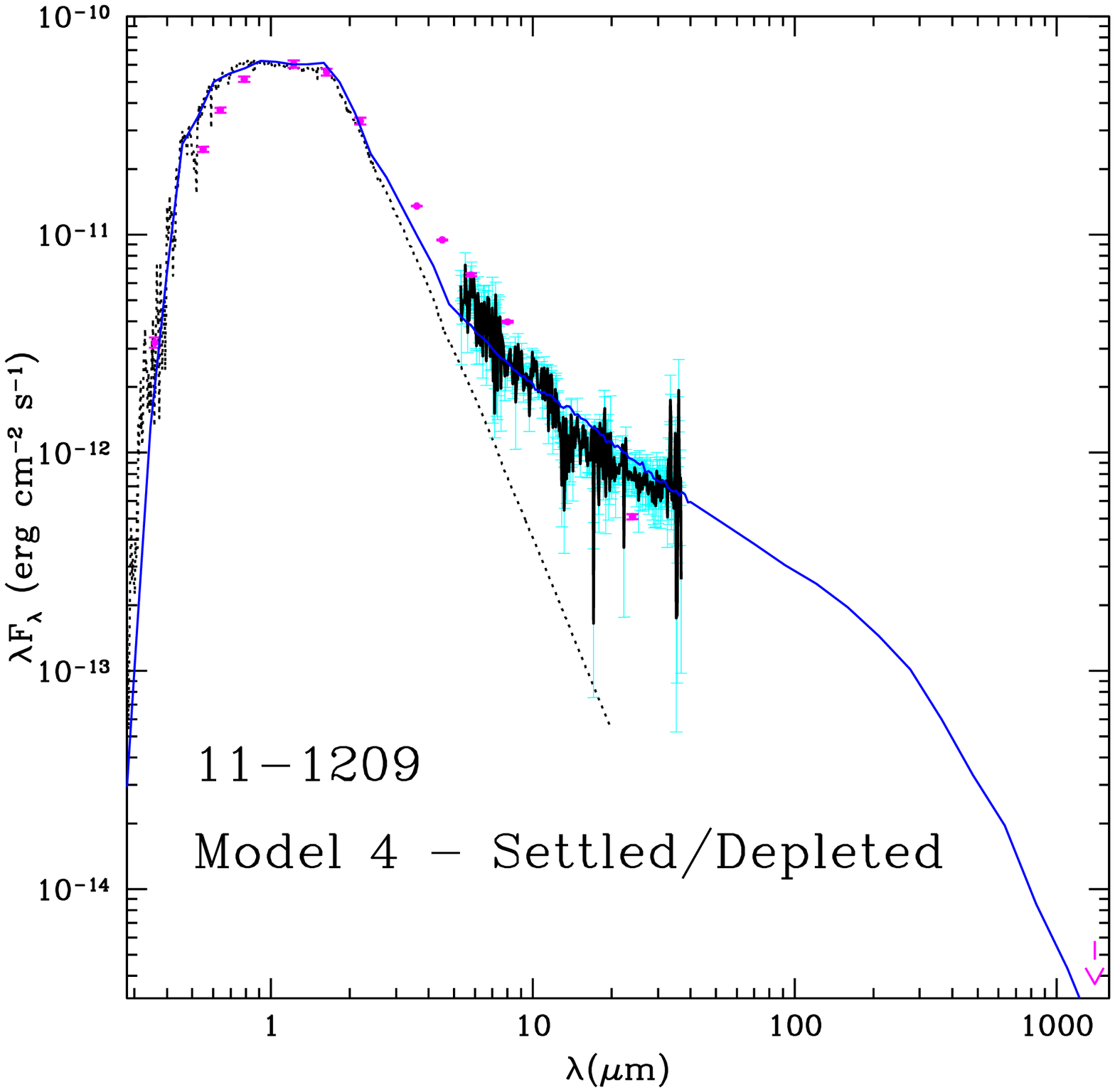}
\caption{Models for depleted CTTS (right, \#3) and globally settled/depleted disk (left, \#4), compared
to the SEDs and IRS spectra of their prototypes 13-236 and 11-1209, respectively.
An upper limit at 1.3mm is shown as a down-pointing arrow.
For Model 3, we display the results of assuming a uniform dust distribution (3A, dashed line) versus
the case with small grains in the innermost disk (3B, bold line; see Table \ref{models-table} for
details). \label{modelsB-fig}}
\end{figure}

\begin{figure}
\plottwo{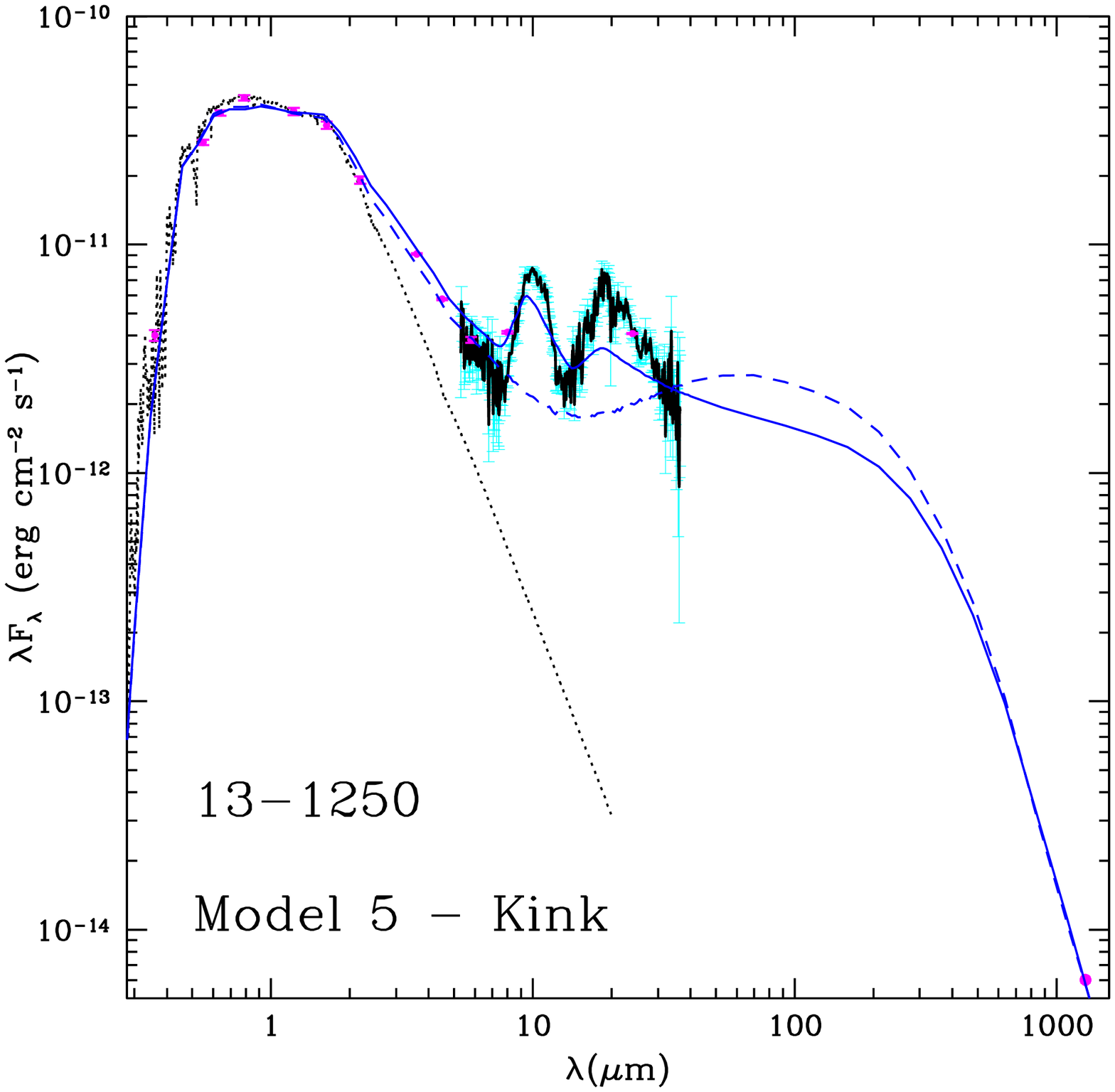}{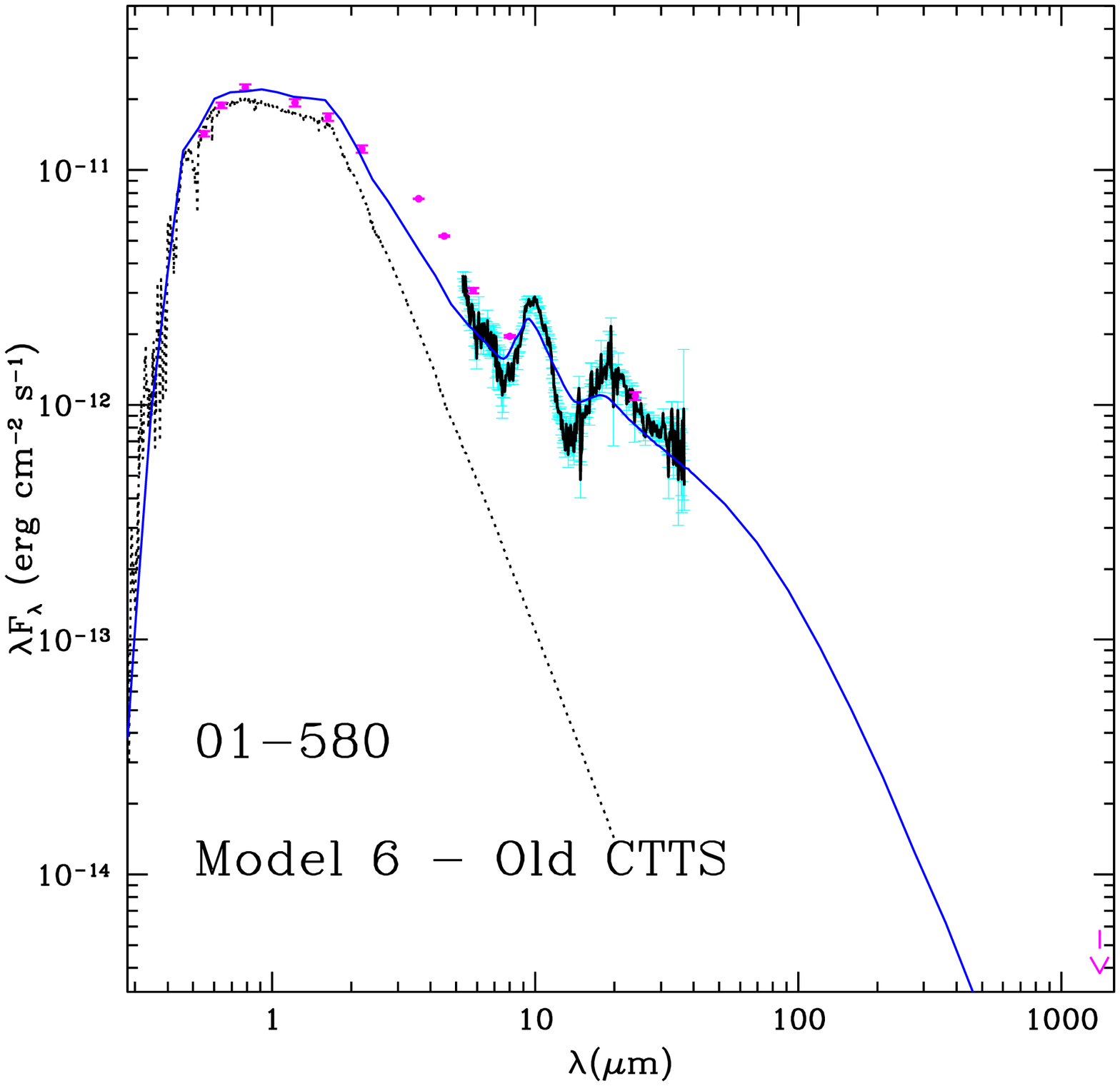}
\caption{Models for ``kink" disk (right, \#5) and long-surviving CTTS disk
(left, \#6), compared to the SEDs and IRS spectra of their prototypes 13-1250 and 01-580, respectively.
An upper limit at 1.3mm is shown as a down-pointing arrow.
For Model 5, we display the two fits: A, with small grains inside, which reproduces
well the silicate emission (bold line), and B, with large grains inside, which fits the
general structure and the slope change of the disk (dashed line; see Table \ref{models-table} for details about
models 5A and 5B).\label{modelsC-fig}}
\end{figure}

\begin{figure}
\plottwo{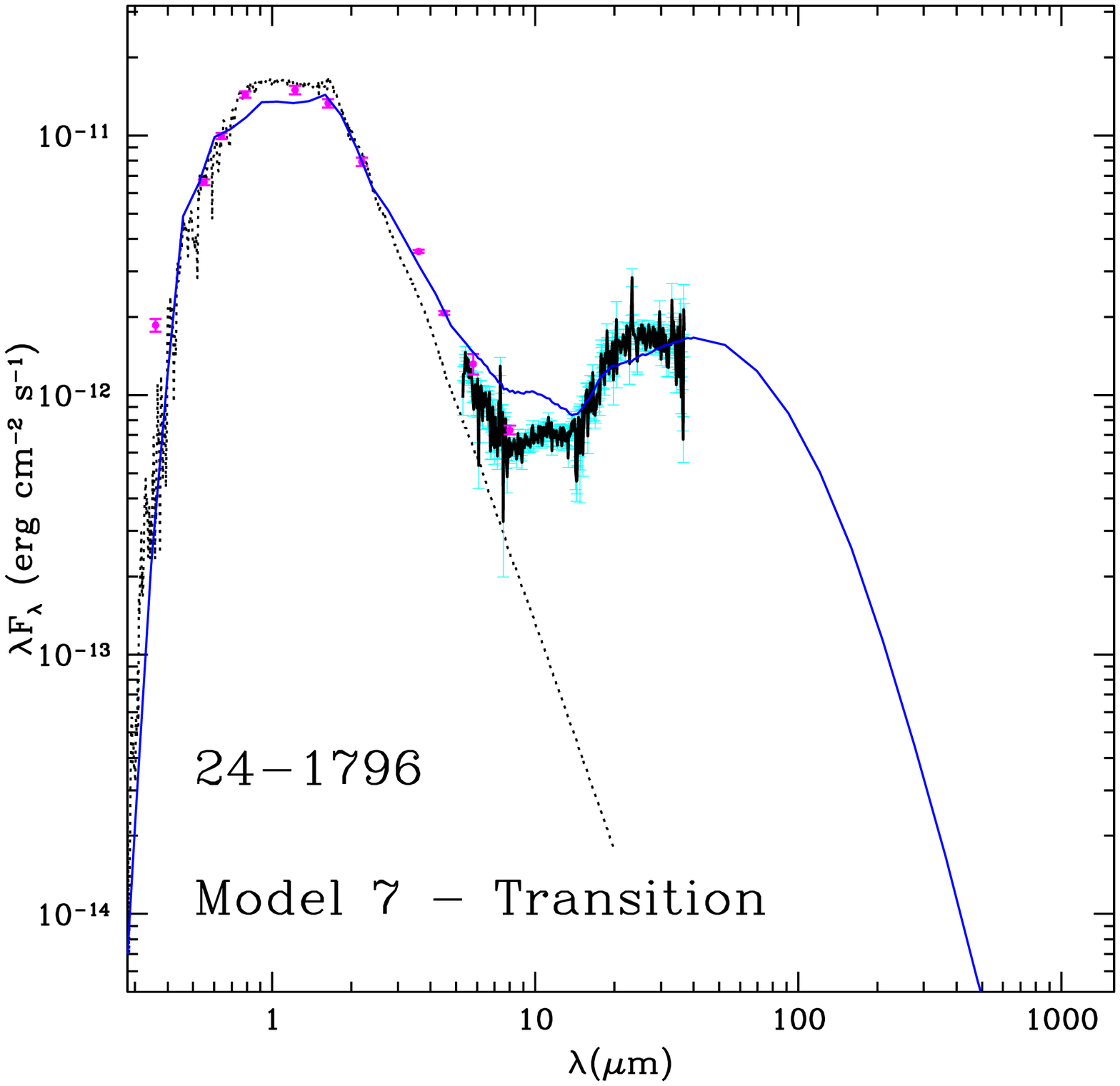}{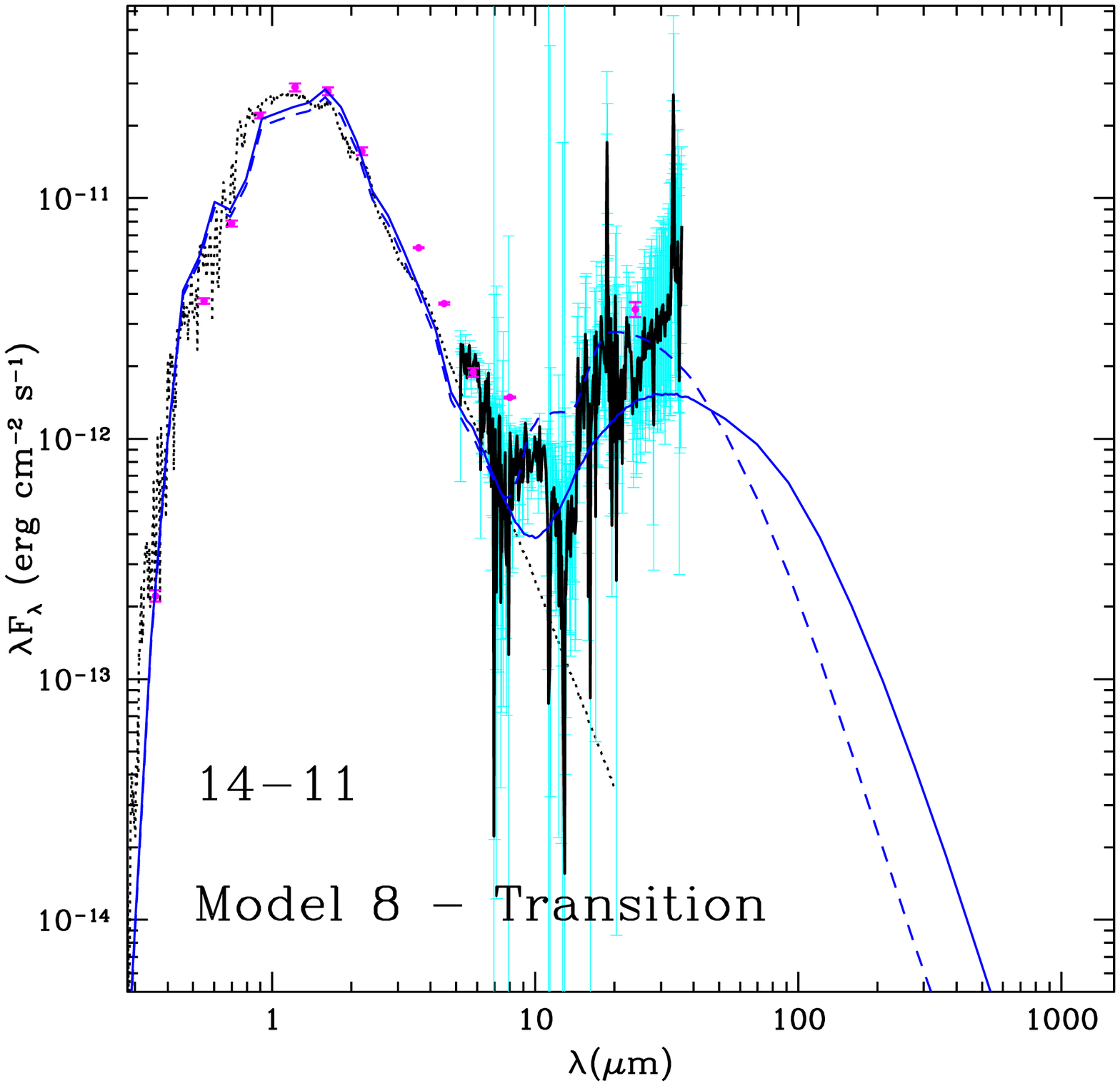}
\caption{Models for transition disks with strong mid-IR excesses, for disks with weak silicate
(left, \#7) and strong silicate (right, \#8) features, compared
to the SEDs and IRS spectra of their prototypes 24-1796 and 14-11, respectively.
For Model 8, we display the models with large grains and large vertical scale height
(A, bold line) and the model with small grains and small vertical scale height (B, dashed
line; see Table \ref{models-table} for more details about the models).\label{modelsD-fig}}
\end{figure}

\begin{figure}
\plottwo{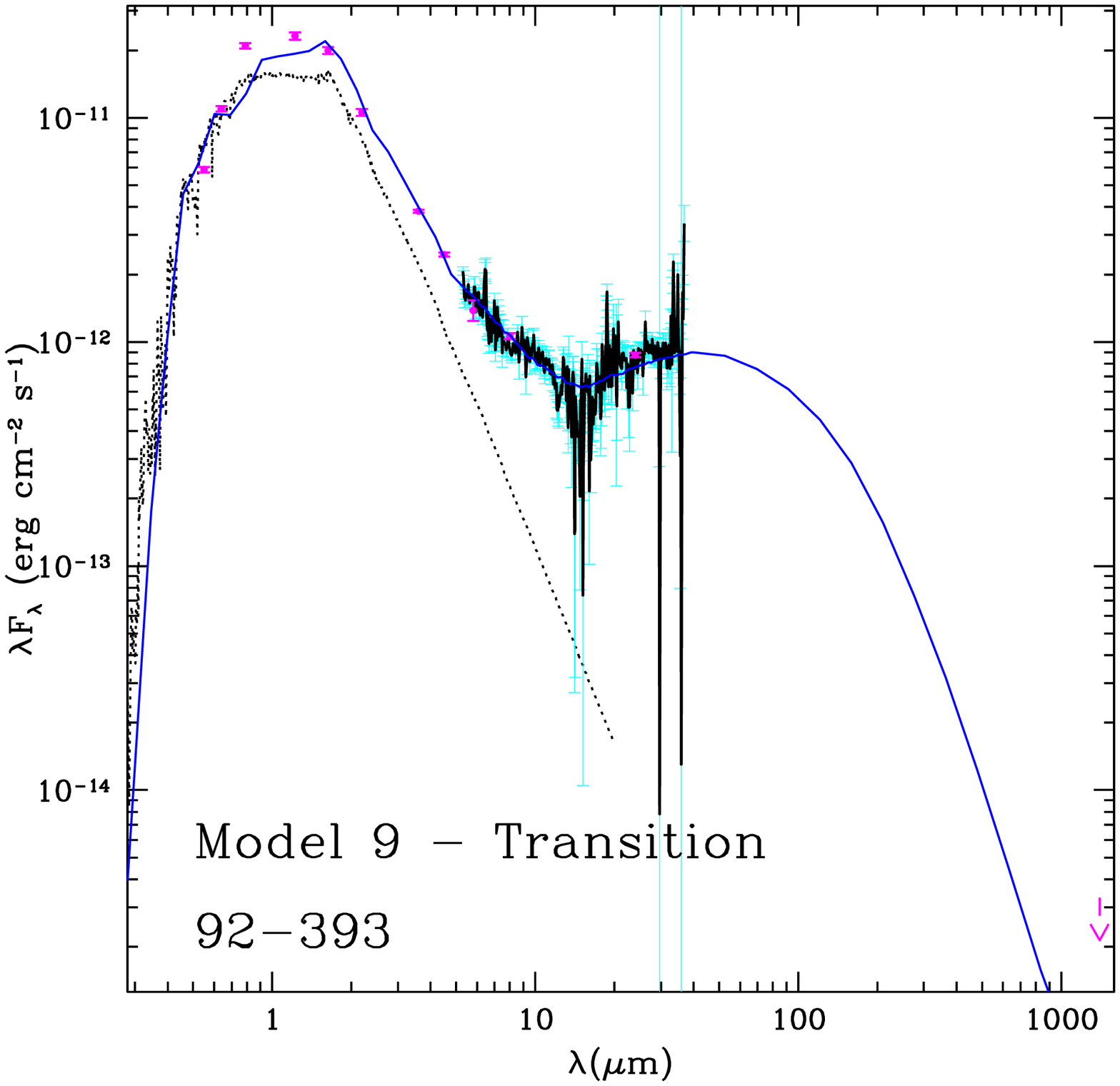}{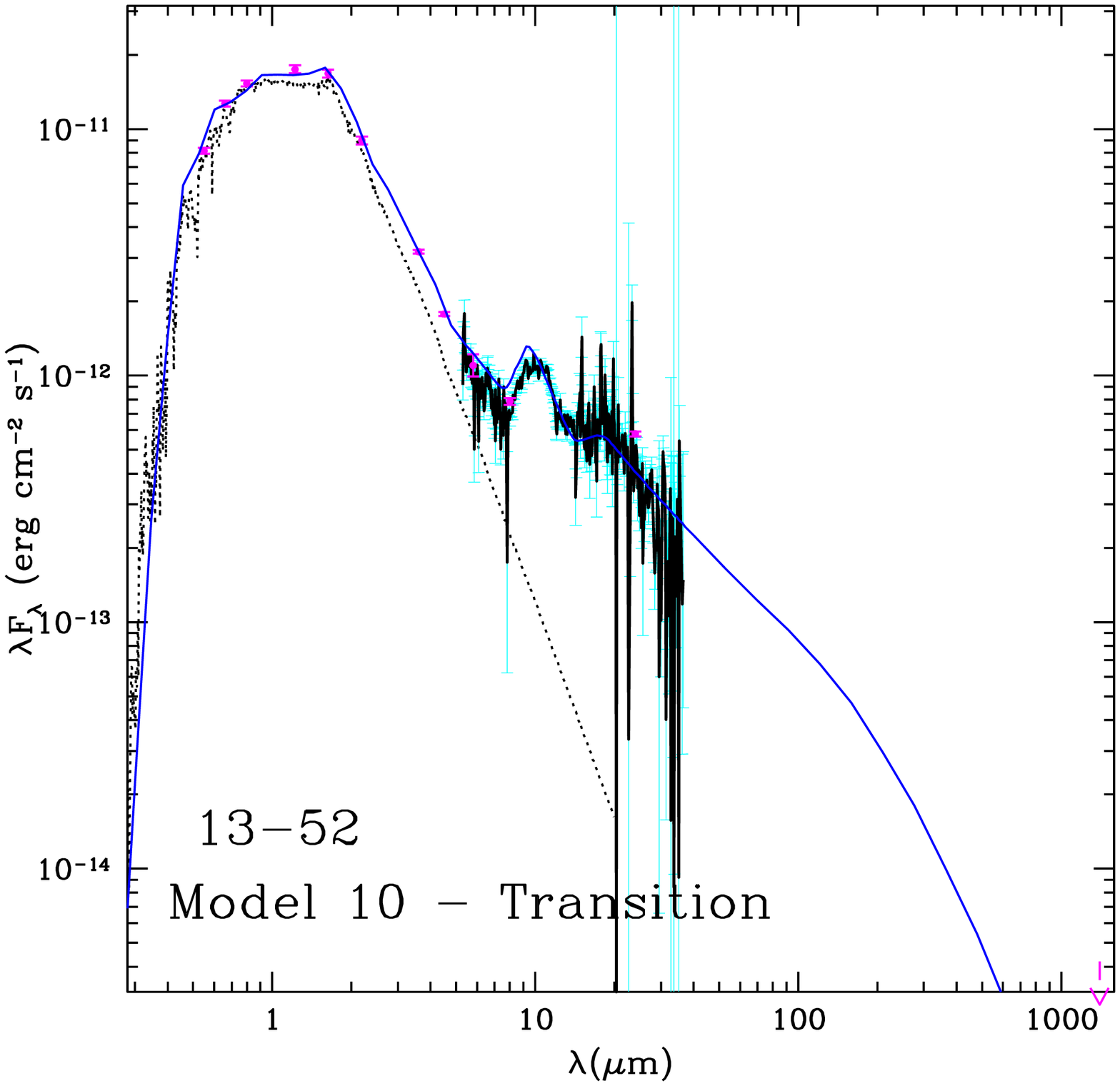}
\caption{Models for settled transition disks with no turn up and no 
silicate feature (left, \#9) versus
silicate feature (right, \#10), compared to the SEDs and IRS spectra of their 
prototypes 92-393 and 13-52, respectively.An upper limit at 1.3mm is shown as a 
down-pointing arrow. \label{modelsE-fig}}
\end{figure}

We regulated the amount of crystalline
silicates in the disk by assuming that the innermost disk is fully
crystalline up to a given (variable) radius (r$_{crys}$), after which the amount of
crystals decays as predicted by simple mixing models  (see Pavlyuchenkov \& Dullemond 2007)
until the grains are totally amorphous.
Given that our models are to be understood as a representation
of a large collection of objects that have different silicate
features, we did not try to reproduce exactly the silicate profiles of the prototypes nor
to derive their mineralogy, but to obtain roughly similarly strong 
or weak features. Therefore, we gave priority to
studying the amorphous grain population, excluding crystalline
grains in the cases where their presence is not important/evident. 
For simplicity, we also do not assume any radial dependency of
the chemical composition of the dust, since our analysis of the short- and
long-wavelength regimes of the IRS spectra does not provide conclusive evidence
for it. In most models, dust populations with different grain sizes for the
innermost and outer disk are required in order to fit the SED.
Here we followed a similar approach as for the crystalline grains, assuming 
a given uniform composition up to a radius r$_{comp}$, after which it is
progressively replaced by the outer disk composition (Pavlyuchenkov \& Dullemond 2007).
Table \ref{models-table} lists the properties of the best-fit models, which are displayed
in Figures \ref{modelsA-fig}-\ref{modelsE-fig}.

\subsubsection{Model 1: Normal CTTS \label{11-2037}}

A normal CTTS disk can be easily reproduced with a relatively massive disk
(M$_{disk}\sim$0.01 M$_*$, assuming a gas to dust ratio 100)
with grains varying between 0.1 and 10000 $\mu$m. Large grains are required
in order to fit the millimeter data within a reasonable disk mass. Nevertheless,
we find that even such a disk presents some degree of settling, as the
assumption of hydrostatic equilibrium between gas and dust invariably produces
a too large near-IR excess to fit our prototype \#1. This group is the most
numerous among the objects observed here, as well as those with IRAC and MIPS data 
from SA06. This suggests that even the most pristine-like
disks in the Cep OB2 region have already experienced substantial grain growth
and settling, which is consistent with the findings of Manoj et al. (2011)
in the Chamaeleon I region.

\subsubsection{Model 2: Massive disk   \label{13-277}}

Reproducing a very massive disk like that of GM Cep does not lack difficulties.
The relatively low far-IR flux, compared to the millimeter data, is hard
to reconcile unless other elements are added to the model. In Sicilia-Aguilar et al.
(2008a) we tried an extra black body-like component, that could be interpreted
as an increase of flaring at some 30-50 AU. With RADMC, we manage to reproduce the
strong near- and mid-IR flux with a very massive (0.052 M$_{\odot}$, assuming a gas to dust ratio 100) and
strongly flaring or relatively thick disk (H$_{rdisk}$/R$_{disk}$=0.2 at the outer disk radius) with a small
grain population in the innermost disk (only 0.1 $\mu$m grains) that turns into a
mixed population (0.1-100 $\mu$m) after 3 AU. Nevertheless, this model still fails to
reproduce the very strong silicate feature, and its flux in the mid-IR is higher than
observed. Assuming a smaller disk mass lowers the mid-IR flux, but this fails to
produce the high millimeter flux. Given that its high-resolution IRS
spectrum suggest that the object is extended (see Section \ref{hires}), the most
plausible solution is the presence of an envelope or surrounding nebulosity 
heated by the star, in which case the disk mass could be lower. Two more disks in
Tr 37 appear very flared as well, with SED slopes similar to GM Cep, 21374275 and 93-720.
Nevertheless, the lack of millimeter detection for 93-720 suggests a much lower disk mass,
and we have no evidence of the presence of an envelope for any of them, although
21374275 is embedded in a small patch of nebulosity near CCDM+5734.

\subsubsection{Model 3: Depleted CTTS \label{13-236}}

Several objects in Tr 37 show very strong near-IR excesses, comparable to
typical CTTS, but comparatively low mid-IR fluxes. Their silicate
features are usually small (but noticeable). Such SEDs require a slightly flattened/settled
disk, with vertical scale height H$_{rdisk}$/R$_{disk}$=0.13  and a normal dust 
population including submicron to millimeter grains, but the disk mass must
be always extremely low ($<$2 10$^{-4}$ M$_*$, assuming a gas to dust ratio 100). 
Such a low disk mass would be consistent with a classification within the
homologously depleted TD (Currie et al. 2009; Currie \& Sicilia-Aguilar 2011),
although their near-IR excesses are strong and their accretion rates are in general  
high ($\sim$5$\times$10$^{-9}$M$_\odot$/yr for the prototype 13-236). This last point suggests that the total gas mass is probably much
higher than what we estimate from dust observations.
Some extra small grains in the innermost disk, not following
the collisional distribution (see option B in Table \ref{models-table}), 
a puffed inner rim, or differential settling, may account for the higher near-IR excesses observed,
compared to a pure settled model, although too much differential settling would result in a
silicate feature much stronger than observed (Dullemond \& Dominik 2008; see also our
model \#6). A disk composed of only submicron grains,
where grain coagulation has not been able to proceed, would
also produce a similar SED.
The very low mass of the disk and strong small dust depletion could also be indicative of
planetesimal formation. In this case, the total gas mass
of the disk could be higher, thus explaining the relatively large 
accretion rates observed. Note that we include within this group the
star 21364762, given its steep IRS spectrum with very weak silicate features,
although its flux at $>$14$\mu$m is uncertain due to nebular emission,
and from the high accretion rate suggested by its veiled spectrum
(Sicilia-Aguilar et al. 2005), it could be that its disk is more massive
than the rest of type \#3 objects.

\subsubsection{Model 4: Globally settled/depleted disk   \label{11-1209}}

In order to reproduce this type of SED, with nearly zero near-IR excess and strongly 
settled at all wavelengths, a low disk mass ($\sim$0.005M$_*$, assuming a gas to dust ratio 100), very small vertical pressure
scale height (H$_{rdisk}$/R$_{disk}$=0.02), and relatively large 
minimum grain size ($>$6 $\mu$m) are required. As in model \#3, the low
disk masses and settled SEDs make them candidates for homologous dust depletion
(Currie et al. 2009), despite their full gas masses are hard to quantify.
Nevertheless, the models that reproduce the steep slope and small excess in the
near- and mid-IR usually generate too little flux at around $\sim$5-6 $\mu$m. Adding an 
extra population with small grains in the inner disk (as we do in models \#5 and \#10, see
below) does not increase the flux in this region due to self-shadowing, and increasing the
disk thickness or flaring is not possible if we want to reproduce the mid-IR SED. Therefore,
we speculate that these SEDs, strongly affected by settling, may still have some
variations in the vertical scale height with radius, a puffed-up inner rim or
gap edge, and/or some degree of differential
settling resulting in different vertical scales depending on particle size.

\subsubsection{Model 5: ``Kink" disk   \label{13-1250}}

The presence of disks with low near-IR excess and a sudden 
change of slope around 6--12 $\mu$m is 
characteristic of Tr 37, as appears reflected in the median disk SED
(SA06). The global structure of the disk can be
well reproduced by a model with two different grain distributions, one
for the innermost disk ($\lesssim$5 AU) with a large minimum grain size
($\sim$10 $\mu$m), and another one for the outer disk with smaller grains.
A small vertical scale height is also necessary. Nevertheless, such a distribution
cannot explain the strong silicate features generally observed. An extra small dust
component in the innermost disk is needed. Instead, assuming that the innermost disk 
($\lesssim$0.5-2 AU) has a small grain population (e.g. 0.1 $\mu$m grains) and the rest is populated by a
collisional distribution with larger grains (minimum grain size 1-10 $\mu$m)
can reproduce the near-IR and silicate feature quite well (with a tendency to
overestimating the near-IR flux), but fails to reproduce the high mid-IR 
excess and the ``kink" shape. We therefore believe that these objects suffer from
stronger grain growth in the inner disk than in the outer disk, but the (inner) dust
distribution may be anomalous. They can be considered then as a
special case of Model \#9 (see below) with an extra small dust population in the innermost
disk. An extra population of small grains has been also included to explain
the SEDs of pre-transitional disks with inner gaps (Espaillat et al. 2010), which
have similarly low near-IR excesses, strong silicate features, and strong excess
at $>$6$\mu$m.

\subsubsection{Model 6: Old surviving, dust-depleted disk with strong silicate \label{01-580}}

This model is based on the strongly accreting (\.{M}$\sim$4$\times$10$^{-8}$M$_\odot$/yr)
but dust-depleted disk present around 01-580, the only accreting star in the 12 Myr-old 
cluster NGC 7160. The star 91-506 in Tr 37 has similar IR colors, although it is younger and
its higher 24$\mu$m flux and millimeter data indicate that it has a more massive disk. 
The SED is characterized
by a low mid-IR excess as in the models of depleted disks, together with a strong near-IR
and very strong silicate feature. A very low disk mass ($\sim$5$\times$10$^{-4}$M$_*$,
assuming a gas to dust ratio 100),
together with strong settling (vertical pressure scale height 
H$_{rdisk}$/R$_{disk}$=0.09 ) are needed
in order to produce the flattened SED, as in models \#3 and \#4. But, in contrast
to the mentioned models, this prototype has a very strong silicate emission, which
requires the presence of small grains ($\sim$0.1$\mu$m) in a non-negligible portion of the
inner disk ($\sim$2-3 AU). Alternatively, differential settling with a disk atmosphere
populated only by small grains could also explain the observed SED. Indeed, 
Dullemond \& Dominik (2008) suggested that strong silicate emission in disks with
weak mid-IR excesses would be a sign of strong size-dependent dust sedimentation. As in cases
\#3 and \#4, our model does not provide enough near-IR excess, which also calls for
flaring or vertical scale changes between the innermost and outer disk,
differential (size-dependent) settling, and/or presence of extra puffed-up walls around
gaps. The lack of dust mass compared to the accretion rate may also hint
the formation of planetesimals or planets within the disk.

\subsubsection{Model 7: Classical TD with weak silicate emission  \label{24-1796}}

By classical TD we understand objects with reduced near-IR excess and
24$\mu$m fluxes comparable or higher than those of normal CTTS (Hern\'{a}ndez
et al. 2007).
Simple models with two populations of grains for the inner and outer
disk cannot reproduce both the shape and silicate emission observed
for our model \#7 (prototype 24-1796). In order to reproduce the general shape of the
SED, we need a low disk mass and a low vertical scale height, in addition to a
change in grain population around $\sim$5 AU. Assuming a collisional
distribution of grains 1-10000$\mu$m in the innermost 5 AU and a
submicron (0.1 $\mu$m) dust population at longer distances, 
we produce a relatively good fit to the data, but
the excess at 5-12 $\mu$m is too high. A model with a physical dust hole
(see model \#8) also produces a good fit, although in this
case the excess at 5-12 $\mu$m is too low. We thus speculate that this
SED type may result from a disk with either an optically thin dust population 
in its innermost part,
or a change in flaring/thickness between the innermost and outermost disk. A secondary
wall at few AU may also help to reproduce 
the large flux in the mid-IR, which can otherwise only be attained with
a submicron dust grain population in the outer disk. 

\subsubsection{Model 8: Classical TD with strong silicate emission  \label{14-11}}

In order to reproduce this type of disk, with prototype object 14-11, a physical hole 
with a minimum size 2 AU (t$_{in}<$250-150 K) is required. Settled dust distributions with
small grains are unable to reproduce both the negligible near-IR excess
and the high mid-IR flux with a strong turn-up around $\sim$8 $\mu$m. 
The best results to reproduce the global shape of the spectrum are attained
with holes $\sim$3.5-6 AU in size, which are roughly in agreement with
our toy model presented in SA07 (consisting of a single black body with T$\sim$70 K
to model an inner disk rim). In order to produce the observed silicate emission, 
an optically thin population of small grains needs to be included in the inner rim 
(or inside the hole)
of the disk, as it has been done by Espaillat et al. (2010) for pre-transitional disks. 
A change in composition (from submicron grains to large grains) around 6 AU produces good 
results in a model with a $\sim$5.5 AU hole.
The vertical scale height of the disk is 
harder to determine with the available data: the large mid-IR excess can
be obtained using a relatively large vertical scale height (H$_{rdisk}$/R$_{disk}$$\sim$0.3)
with a large ($\sim$10 $\mu$m) minimum grain size, or a smaller H$_{rdisk}$/R$_{disk}$ with a dust 
distribution containing small grains ($\sim$0.1 $\mu$m minimum grain size).

\subsubsection{Model 9: Settled TD with weak silicate emission  \label{92-393}}

The small near-IR excess and relatively flat SED of our prototype \#9
(star 92-393) can
be reproduced with a two-population grain model, including large grains
in the innermost disk (20-10000 $\mu$m below 15 AU) and small grains only
in the outermost part (0.1$\mu$m). Including a collisional distribution with
small grains in this part fails to reproduce the high mid-IR flux, and a single,
large-grain distribution also underestimates the mid-IR excess. No
clean hole is required in this case, which is also in agreement with our
observations of accretion in the SED prototype 92-393. The mid-IR bump could also
be explained with a change in flaring/extra wall at 10-15 AU. This type
of SED is thus a good example of inside-out evolution. The relatively low
disk mass (5$\times$10$^{-3}$M$_*$, assuming a gas to dust ratio of 100) 
could also be explained with planetesimal
formation. The formation of planetesimals or planets in the innermost 10-15 AU
would reduce the dust contain we see in this region, and
could also cause the presence of a secondary wall or change in disk flaring/thickness at
this distance.

\subsubsection{Model 10: Settled TD with strong silicate emission  \label{13-52}}
 
In order to reproduce the very small near-IR excess of a disk similar to
our prototype \#10 (object 13-52) we need a very small vertical 
pressure scale height (H$_{rdisk}$/R$_{disk}$$<$0.03)
 in combination with large grains (minimum grain size $>$1-10 $\mu$m). 
Nevertheless, to produce the 
strong silicate feature we need to include some optically thin small dust (0.1$\mu$m). This
small amount of dust is also enough to produce the tiny excess over the
photosphere observed. Nevertheless, assuming a collisional distribution
with grains as small as 0.1$\mu$m would produce too much near-IR excess and
dilute the silicate feature, so the small grain component needs to be decoupled
from the general dust distribution.
A physical inner hole is not required in this case, as long as the main dust distribution
has a large enough minimum size and the disk is very settled, although the lack of
accretion in our prototype requires either a very low gas content or a 
physical hole or gap within the inner large-grain disk. This, together with the mass depletion
inferred from the very low mid-IR fluxes, makes these type of disks good candidates
to look for giant planet formation, where accretion onto a planet could explain the
lack of significant \.{M}.

\section{Discussion: Structure and properties of disks in evolved clusters \label{discussion}}

\subsection{Dust evolution and stellar properties\label{mineral}}

In this section, we contrast the results of our mineralogy analysis
(Section \ref{fit}) with the known properties of the stars in our sample.
Considering the limitations due to S/N and unresolved disk structure, we
examine the correlation between different stellar and disk properties
and the disk mineralogy inferred from the silicate feature. Several 
trends have been observed in the past. In our previous study (SA07),
we found a correlation between the strength of the silicate 
peak (related to the grain size) and the age, in the sense that younger
stars appeared to have
weaker silicate peaks and thus larger grains in their disks
atmospheres. This could be interpreted as an effect of turbulence
stirring larger grains into the disk atmospheres of the younger
and more turbulent stars, with older and more quiescent disks being more
settled (such an effect has also been observed by Fang et al. 2011). 
We also found that the presence of flattened
SEDs without silicate emission appeared more frequently among
M-type stars than for K-type objects. 

\begin{figure}
\plotone{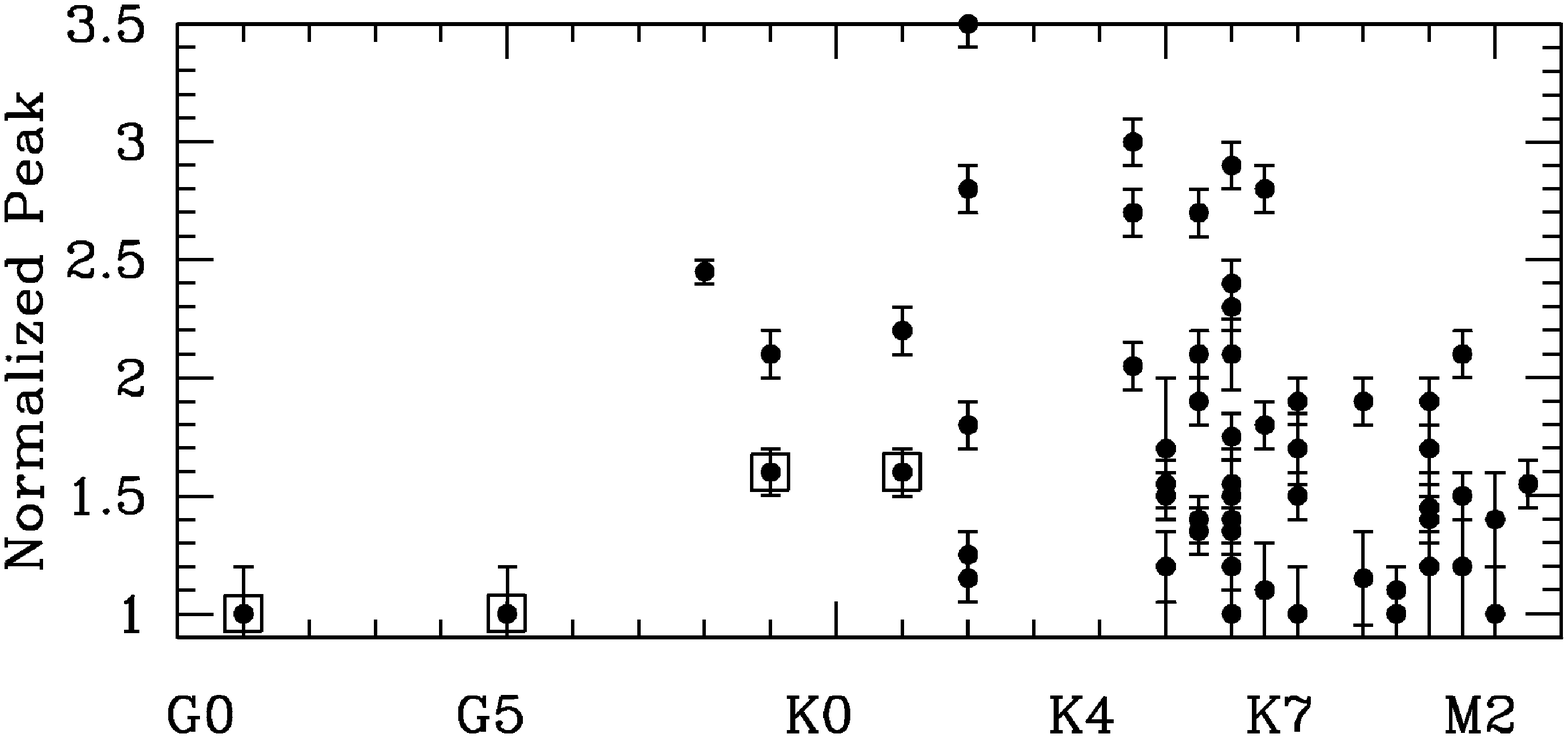}
\caption{Normalized silicate peak versus spectral type. The intermediate mass stars
(KUN-196, CCDM+5734, DG-481, and also the F9 star 21374275) are marked with boxes and shifted by 20
subtypes for a better display in the figure, with their spectral types
being  B9, A3+A5, A7, and F9, respectively. Stars with spectral types M0 or later
have normalized peak fluxes lower than earlier objects, as it had been suggested in SA07.
The errors in the spectral types (not shown) are 0.5 to 1 subtype. \label{stypeak-fig}}
\end{figure}

Examining the new data, we confirm again the trend
of the sources with later spectral types (M) to have weaker 
silicate emission among the solar-type stars (Figure \ref{stypeak-fig}).
Excluding the intermediate-mass (spectral types A and B) objects, which strongly
differ in properties from the rest of the sample, we obtain a
Spearman rank correlation coefficient of -0.37 (range -0.61 --- -0.06), 
with a probability p=0.006 that the two quantities are uncorrelated.
Given the correlation between silicate peak and grain size, this
relation could be related to differences in disk structure, transport
and/or grain processing
depending on stellar mass, even though our sample contains objects
down to spectral type M2.5 only. Weak silicate features have also
been observed for later-type stars in different regions
(Sicilia-Aguilar et al. 2008b, 2009; Furlan et al. 2011).
The fact that in disks around colder, less luminous
stars, a given wavelength traces a closer-in part of the disk than in more
luminous objects could be partially responsible for these differences, but
we emphasize that our sample is dominated by late-K and early-M stars, which
are not so different in temperature and luminosity. 
In addition, we observe a tendency of M-type stars to have
steeper SEDs, which has also been observed even in very young
regions (e.g. the Coronet cluster; Sicilia-Aguilar et al. 2008b, see also
Muzerolle et al. 2010; Sz\"{u}cs et al. 2010). This could be caused by a larger 
degree of dust settling or grain growth in the innermost disk,
although further evidence is needed from detailed modeling of a larger
number of low-mass objects. In addition, the differences between solar-type 
and M-type stars are not evident in all regions (Sicilia-Aguilar et al. 2009; 
Furlan et al. 2011), and other parameters (like the accretion rates) may 
also affect the SED appearance (Manoj et al. 2011).

Although there is still a weak trend for weaker silicate features being more
frequent among the youngest disks, its significance is weaker after adding
the new sample (Spearman correlation coefficient 0.21, in the range
-0.11 --- 0.49, with probability p=0.14 that there is no correlation
between both quantities; note that we exclude for this analysis the
stars with spectral types G and earlier, which have uncertain ages). 
Our extended sample also lacks strong correlations
between the normalized silicate peak or grain size and the accretion
rate, which is related to the turbulence level (using the most recent
accretion rate estimates from Sicilia-Aguilar et al. 2010). 
Performing a Spearman rank correlation 
test, the probability that the accretion rate and silicate peak/grain 
size are uncorrelated are 0.46 and 0.67, respectively. 

We had previously attributed these trends of silicate strength
with age and accretion rate to the effect of turbulence, which is thought
to be stronger at younger ages. (SA07). The main difference 
is that, instead of being focused on normal CTTS disks
as that of SA07, our new dataset includes a large number of 
objects with settled and evolved
disks, including many disks with signs of inside-out evolution.
Exploring the different RADMC disk models, we find that settling combined with 
grain growth may also increase the large grain population in the
disk atmosphere, as it happens in cases with strong turbulence,
so the net effect of turbulence may not 
be evident nor easy to quantify. Recent theoretical (Zsom et al. 2011) and 
observational (Fang et al. 2011) studies suggest that the connection
between disk turbulence and the grain size in the disk atmosphere is real.

When studying the age effects, we must also keep in mind that old disks
($>$8 Myr) are typically rare, so studying the disk mineralogy
over the whole disk life span is always limited by the paucity
of old objects. Moreover, given that half of the disks have
dispersed by the age of $\sim$4 Myr (SA06;
Hern\'{a}ndez et al. 2007), the long-surviving disks are probably
those with initially large masses (that can stand long periods
of viscous evolution and photoevaporation) and/or those for which
coagulation and/or settling are far less efficient than in the
typical objects. In this last case, 
very old objects may not be representative of the typical processes
occurring in most aging disks.

Finally, as it has been observed before, we do not find any evident
correlation between the crystalline fraction and the stellar properties.
The fact that crystallinization
and amorphization may occur at variable rates over the 
lifetime of the star (\'{A}brah\'{a}m et al. 2009; Glauser
et al. 2009) is probably the main cause of this lack of correlation.
Since the example of EX-Lupi (\'{A}brah\'{a}m et al. 2009) is the most remarkable one regarding
strong and fast variation of grain properties, we have examined
the mineralogy of the sources with known accretion variations (of
1 order of magnitude or more; Sicilia-Aguilar et al. 2010). 
Looking at these sources (72-875, 11-2146, 14-141, 11-2031, 13-1238,
82-272, GM Cep, 13-236, and 13-1250), we observe a tendency 
to having large (few microns) amorphous grains, but no 
evidence of any trend to higher or lower crystallinity fractions.
We also do not observe any trend relating grain size and crystallinity fraction and,
as had been previously found, crystalline grains are always
small.

\subsection{Dust evolution and disk structure \label{transition}}

Here, we study the results of our mineralogy analysis
(Section \ref{fit}), contrasting them to global disk properties
as derived from the SED slope at different wavelengths.
Several previous studies have suggested that the silicate emission
can be used as a tracer of the global disk structure. Bouwman et al. (2008) 
found a correlation between the amorphous silicate feature and
the SED shape, suggestive of dust settling. Lommen et al. (2010)
compared the millimeter emission with the strength and characteristics of the
silicate feature, finding a weak correlation that could indicate
global dust evolution throughout the disk, as has been suggested by 
Currie et al. (2009). Our IRS sample was especially selected to contain
disks with very different structures, in order to study whether
the grains in the disk atmosphere are tracers of more global processes
affecting the whole structure of the disk. Many of the SEDs of low-mass
disks in Cep OB2 are substantially different from those seen in younger
regions, showing evidence of dust settling (at a global scale or with
a radial dependence, as in ``kink'' disks), global grain growth and dust depletion, 
and inside-out evolution (from differential
growth/settling to inner holes in the TD).
Transition and ``kink'' disks (similar to the so-called pre-transitional; Espaillat et al.
2010), having a very low and/or negligible near-IR excess and
an otherwise normal mid-IR excess, are prime candidates to
look for inside-out disk evolution (as predicted by Hayashi et al. 1985)
and to search for evidence of the different mechanisms that can
evacuate the innermost disk of small dust (coagulation, planet formation,
photoevaporation, binarity). On the other hand, fully settled disks
and those with very low dust masses are suggestive of global or
homologous disk evolution 
(Currie et al. 2009) happening in both the innermost and the outer disk.

\begin{figure}
\plotone{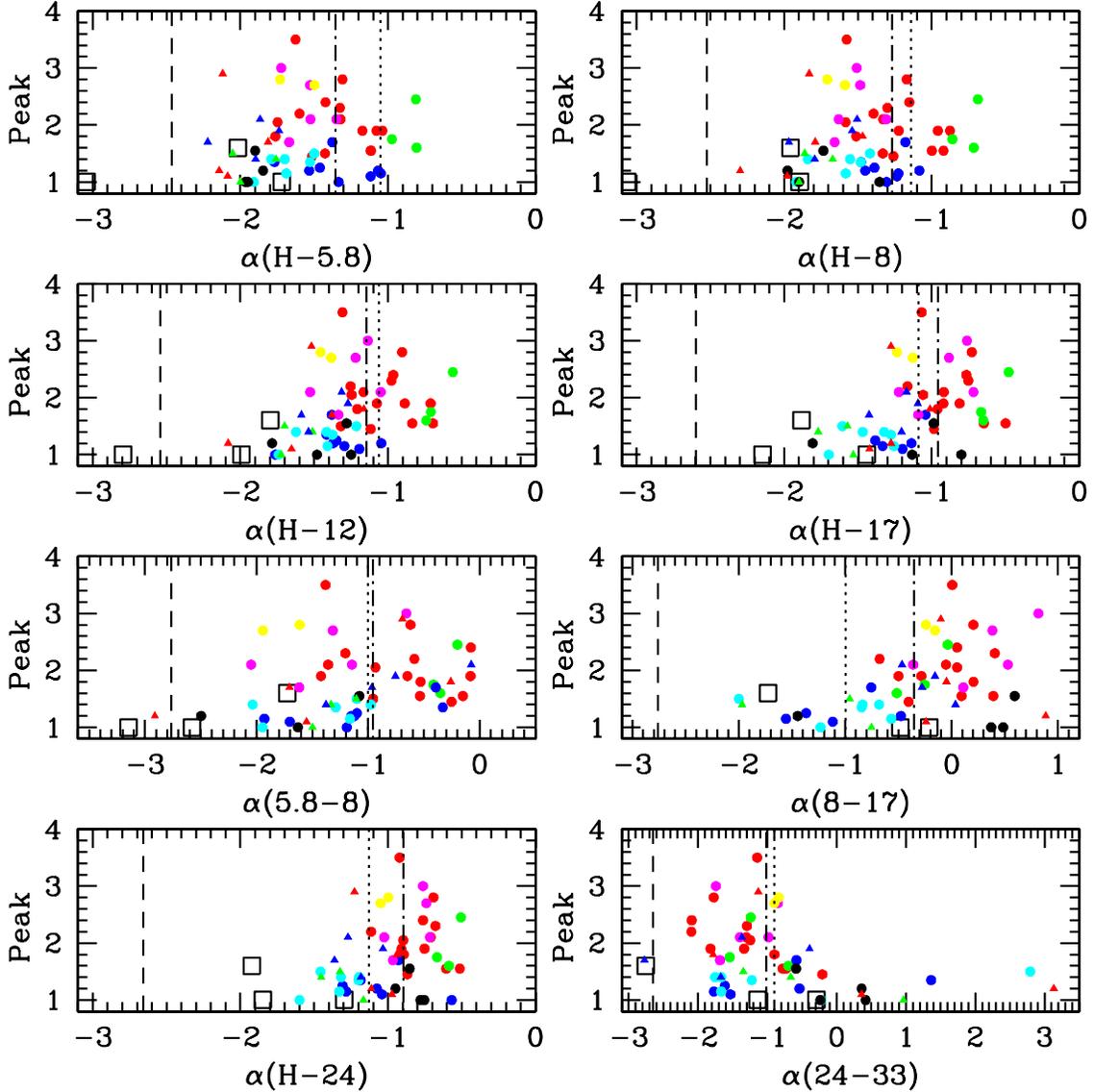}
\caption{Disk slope ($\alpha$) at different wavelengths versus normalized
silicate peak, which is a tracer of grain size.
The different $\alpha$ trace the disk structure at different disk radii. There is a tendency of more settled
and transitional disks to have lower silicate peaks (or absence of silicate emission) 
and thus larger grain sizes. This suggest the importance of
grain growth for the innermost disk structure. 
The correspondence of SED class and symbol is as follows: red circles = \# 1, green
circles = \# 2, blue circles = \#3, cyan circles = \#4, magenta circles = \#5,
yellow circles = \#6, black circles = \#7, red triangles = \#8, green triangles = \#9,
blue triangles = \#10.
Intermediate-mass (spectral types B and A) objects are marked with
a box. Among the vertical lines, the dashed line represents the 
slope of a bare photosphere (spectral type K7), 
the dotted line is the slope of a geometrically thin, optically thick disk (Kenyon \& Hartmann 1987),
and the dotted-dashed line is the slope for the median of the Taurus disk (Furlan et al. 2006).
\label{alphasize-fig}}
\end{figure}

\begin{figure}
\plotone{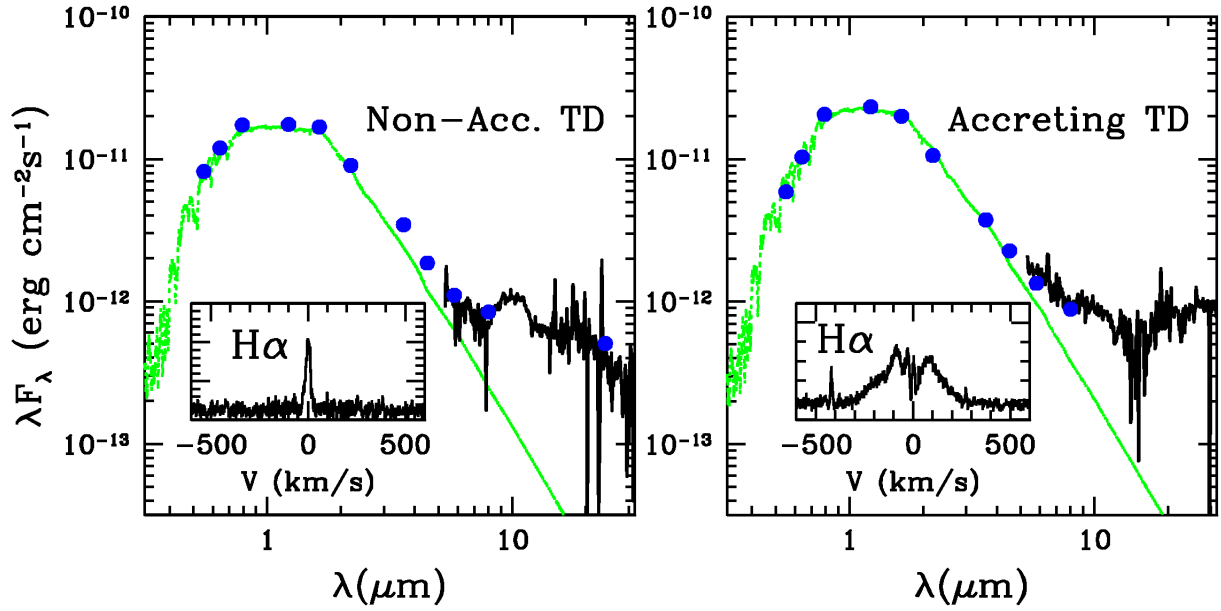}
\caption{Comparison of the SEDs (including silicate features) of an accreting and a non-accreting
transition disk with similar spectral types. \label{TO-fig}}
\end{figure}

We have explored the disk mineralogy derived from the
silicate feature (in particular, the flux at the normalized
silicate peak and the grain size) as a function of disk structure.
Figure \ref{alphasize-fig} displays the SED slopes ($\alpha$, see Section \ref{radmc}) at 
different wavelengths
together with the  peak height of the normalized silicate feature.
Note that, for most transitional and ``kink" disks, the slope changes sign between
8 and 24 $\mu$m, due to the turn-up in the SED. We find that objects with the
most settled inner disks (including TD) have low silicate
peaks, which are typical of large grain sizes. A Spearman rank correlation test
confirms a strong correlation between the peak height and the SED slope
at any wavelength (see Table \ref{correlation-table}), with probabilities
that the two quantities are uncorrelated varying between 6$\times$10$^{-5}$
and 0.04. The correlation is stronger for slopes including short and long
wavelengths, which trace the disk structure over several AU. 
This points out that strong grain growth in the innermost
disk is probably the main cause of the SED geometry of transitional and ``kink"
disks, and it also suggests that grain growth strongly affects the disk structure
(for instance, due to increased settling). 
In addition, the accreting TD tend to have have comparatively weaker
silicate peaks than the non-accreting ones.
If their reduced IR excess is caused by grain growth in the innermost disk 
it could explain why those objects are still accreting, as grain growth does
not need to affect the gas content (see disk model \#9 and Figure \ref{TO-fig}). 
The presence of large ($>$1 $\mu$m) and processed grains
(with normal crystalline fractions) in all the TD, including
those with SEDs consistent with inner holes, offers a strong
contrast to the findings of Watson et al. (2009) in Taurus, where the silicate
features of TD with inner holes are nearly pristine. This is again a 
confirmation that the TD that are found in clusters of different
ages do not necessarily correspond to the same type of physical objects
(Sicilia-Aguilar et al. 2010). If the presence of pristine grains in the
Taurus TD is due to dust replenishment from the outer disk, it could also
indicate changes in the transport mechanisms or in the global dust processing.
Differences in the structure of TD with various ages is also
to be expected taking into account the different timescales of
the various disk clearing mechanisms (planet formation, grain growth, photoevaporation;
Alexander \& Armitage 2009).

\begin{figure}
\plotone{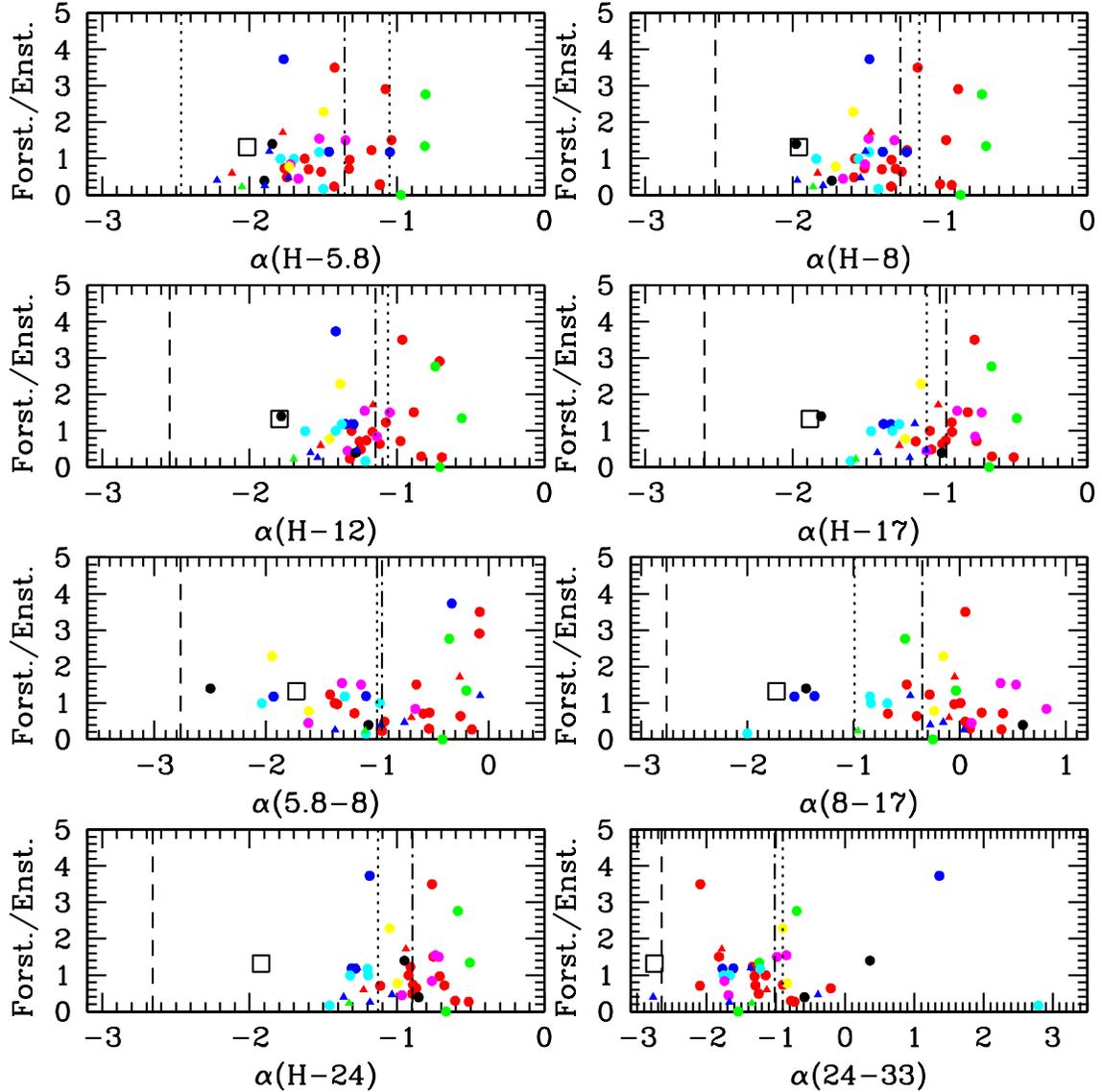}
\caption{Disk slope ($\alpha$) at different wavelengths versus forsterite to enstatite ratio. 
The different $\alpha$ trace the disk structure at different disk radii. There is a 
weak tendency of very flared disks to have higher forsterite to enstatite ratios. 
The correspondence of SED class and symbol is as follows: red circles = \# 1, green
circles = \# 2, blue circles = \#3, cyan circles = \#4, magenta circles = \#5,
yellow circles = \#6, black circles = \#7, red triangles = \#8, green triangles = \#9,
blue triangles = \#10.
Intermediate-mass (spectral types B and A) objects are marked with a box.
Among the vertical lines, the dashed line represents the slope of a bare photosphere (spectral type K7), 
the dotted line is the slope of a geometrically thin, optically thick disk (Kenyon \& Hartmann 1987),
and the dotted-dashed line is the slope for the median of the Taurus disk (Furlan et al. 2006).
\label{alphacryst-fig}}
\end{figure}

We have also explored the potential correlation between crystallinity and
disk structure, finding no significant results
(Table \ref{correlation-table}).
There is a marginal correlation (p=0.04-0.09 in the
Spearman rank test) between the forsterite to enstatite ratio and
the innermost disk structure (Figure \ref{alphacryst-fig}). This is consistent with 
Bouwman et al. (2008), who suggested that
young disks (which have typically large near-IR excesses) 
would tend to form only forsterite in their
inner disks due to their non-equilibrium conditions. 
The slopes at longer wavelengths do not show any correlation with
the forsterite to enstatite ratio, so the effect seems to be
exclusive of the innermost disk, although the poor S/N of most spectra
at long wavelengths does not let us test this hypothesis.
Nevertheless, the forsterite to enstatite ratio is never
too large in Cep OB2, so the trend should be tested on a larger sample. 
We do not find the weak trend of Watson et al. (2009) that there are
more crystals at shorter wavelengths (innermost disk regions): For the
6 objects that have enough S/N at longer wavelengths, we find a similar
crystallinity fraction, with small variations, although our sample
is too small for statistically robust results. In addition, detecting crystalline grains
at longer wavelengths is easier than tracing the amorphous ones, so our
long-wavelength crystallinity fraction may be overestimated.

\subsection{Disk dispersal and survival and the nature of transition disks in Cep OB2\label{dispersal}}

From the analysis of the previous sections, in particular, the comparisons
with the model SEDs obtained with RADMC (Section \ref{radmc}) and the trends 
observed with mineralogy (Sections \ref{mineral} and \ref{transition}), we can draw 
general results about disk evolution in intermediate-aged clusters.
Although we do not have millimeter observations (or detections) for all disks,
our models revealed that an important deal of information about disk mass
(low or high) as well as dust distribution (presence of small or large grains) 
in the planet-formation area can be already inferred from the $\sim$20-30$\mu$m 
region in the IRS spectra, as it had been previously found by Currie \& Sicilia-Aguilar
(2011). Of course, a better constrain on grain growth
and grain distributions will be only provided by deeper millimeter/submillimeter 
observations, or by far-IR data from Herschel. 

We observe important evidence of dust and disk structure evolution.
On one hand, the low millimeter fluxes and the
presence of objects with low mid-IR excesses suggest 
that dust evolution and grain growth up to millimeter sizes (and probably larger) is
generalized throughout the disk. 
Not only some disks require relatively large grains in the
innermost part, but a substantial portion of the dust mass must be
forming large grains, as expected from global dust evolution (Currie et al. 2009). 
The inferred disk masses are sometimes very small, assuming a normal gas to dust
ratio ($\sim$100), even in cases
where we still detect strong accretion (e.g. 11-2146, 13-236, 13-669, 82-272, 12-1091,
among others). We thus expect these disks to have far more gas than
what their dust emission suggests, which would also explain the lack of correlation
between the accretion rate and the dust mass derived from millimeter observations
already mentioned in Section \ref{iram}. This could indicate an anomalous gas
to dust ratio, together with a probable large maximum grain size.
In case of very dust depleted disks like 13-236,
the very low dust masses required to fit the SEDs could even suggest 
the formation of planets or planetesimals within the disk.
We also find that even the most pristine-like disks in Cep OB2 require
some settling in order to properly fit their SEDs (in fact, GM Cep is
the only one that does not require settling but rather extra flaring).
Such a generalized dust settling has also been observed even in younger
regions, like Chamaeleon I (Manoj et al. 2011).

In addition to this global grain growth and settling,
we also observe significant evidence of inside-out dust evolution.
More than half of the disks in Tr 37 show strong variations in the SED
slope between the innermost part ($<$6-10$\mu$m) and the rest,
consistent with inside-out evolution (SA06). In some cases, the near-IR
excess is low enough to empirically classify the object as transitional 
([3.6]-[4.5]$<$0.2). Our modeling of selected objects
proves that non-standard disk structures
can reproduce the SEDs of ``kink" and transitional disks. 
Objects with a strong change of slope (``kink" SED) invariably
require different grain populations in the innermost and
outer disk, as long as the vertical scale height is assumed to
run smoothly as a single power-law throughout the disk. Surprisingly, a few of these
disks do not require the grains in the innermost disk to be
larger: a non-collisional distribution populated by submicron
grains alone in a settled innermost disk can reproduce the near-IR
excess and silicate feature quite well. Nevertheless, the large
``kinks" typically require the inclusion of some several-micron grains. 
We can interpret this result if small and large grains are not
well mixed, as could result if large grains are settled, leaving
a disk atmosphere populated mostly by small dust.
Other authors have also proposed adding a small dust population within 
a disk gap for such objects with very 
strong silicate features (Espaillat et al. 2010). Including such
a small dust population could improve our fits in the cases where the
general structure of the disk that reproduces the global SED shape fails to
produce strong enough silicate features. Size-dependent dust settling
(with most large grains being confined to the optically thick midplane of
the disk) could also be invoked to increase the strength of the silicate emission
(Dullemond \& Dominik 2008). 
Conversely, inside-out grain growth and/or grain filtering at gap
edges due to the presence of planets (Rice et al. 2006) could 
explain different grain populations between the inner and the outer disk.
Grain filtering due to planet formation, resulting in small grains in the innermost disk, also
has the interesting side effect of increasing the gas to dust ratio (Rice et al. 2006), which 
could explain the relatively large accretion rates observed in Tr 37.

Clean inner holes are required to reproduce the objects with smaller near-IR
excesses and large turn-ups at longer wavelengths. Since many of these objects do not
seem to be actively accreting (in contrast with the objects with near-IR
excesses, all of which display indicators of ongoing accretion), the presence of a physical hole
is indeed the best explanation. In fact, the strong silicate
features observed in many of these non-accreting transition 
disks suggest that the disks still contain a large population of
small grains, albeit at a distance larger than the dust 
destruction radius. 
The disks with inner holes and no accretion are of great relevance, since they 
could be produced by the recent formation
of a (giant) planet at few AU (Lin \& Papaloizou 1986, Bryden et al. 1999,
Quillen et al 2004), which would disrupt the disk and prevent accretion
by opening a gap. An alternative explanation for TD without accretion is
photoevaporation by the central star (Alexander et al. 2006), although the
number of non-accreting TD in Tr 37 seems in conflict with the timescales
predicted for photoevaporation (Gorti et al. 2009).
On the other hand, strong grain growth in the innermost disk is
required to explain the transitional disks with weak or no silicate 
features. This result agrees well with the fact that all of
these objects still show active accretion (see Figure \ref{TO-fig}), 
so the innermost disk must still contain a large amount of gas. 
In both cases (with and without hole), the flux mid-IR region 
reveals either inside-out evolution (objects with high mid-IR flux 
require the presence of small grains at a few AU
distance) or global grain growth (in objects with very steep SEDs in the mid-IR).

Finally, we find two puzzling objects in the region. One of them is
the massive disk around GM Cep (see Section \ref{hires} for a detailed
discussion). Although the high-resolution IRS observations
suggest that there could be some remnant of an envelope, the high disk
mass and the strong (and variable) accretion rate are anomalous for a
star in a 4 Myr-old cluster. The other one is the star 01-580 in NGC 7160,
which despite being located in a cluster with median age 12 Myr, has a very
strong silicate feature, strong near-IR excess, and an accretion rate 
of several times 10$^{-8}$ M$_\odot$/yr (Sicilia-Aguilar et al. 2005, 2010). 
Considering its accretion rate, the disk around 01-580 must have been
more massive than the one around GM Cep and similar to that of GW Ori
(Mathieu et al. 1995). Such objects pose strong constraints to the possibility
of planet formation via gravitational instability (Boss 1997, 2003),
and also suggest that strong and global grain growth may not always be efficient.
Therefore, also these anomalous objects may provide important information
to understand the physical processes involved in planet formation.

\subsection{GM Cep: Jets, photoevaporation, and extended emission \label{hires}}

\begin{figure}
\plotone{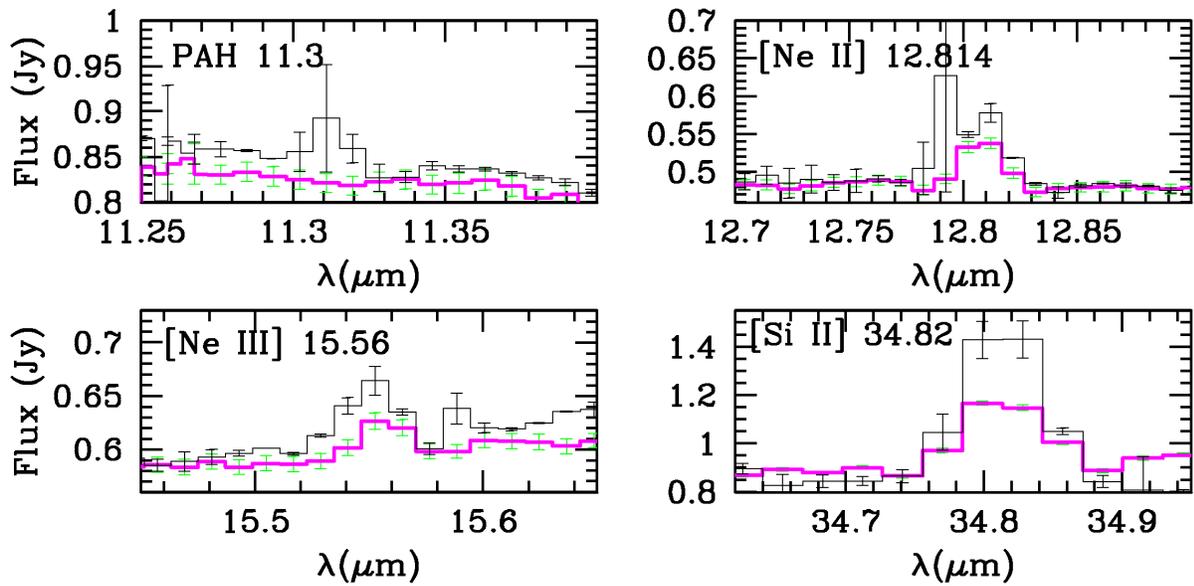}
\caption{Lines observed in the high-resolution IRS spectrum of 13-277 (GM Cep).
The PSF-extracted spectrum is shown as a thick magenta histogram; the full aperture
spectrum is displayed as a thin black histogram. The differences in
both extractions could be indicative of extended emission. \label{hires-fig}}
\end{figure}

Of the two objects observed with the high-resolution IRS modules, GM Cep and 21392541,
only the former was detected. This object has the brightest
mid-IR emission among all the TTS in Tr 37, being twice as bright as the Herbig Be
star MVA-426 at 70 $\mu$m (SA06). We examined the spectrum
looking for gas lines and found evidence of [Ne II] 12.81$\mu$m,
[Ne III] 15.56$\mu$m, and [Si II] 34.82$\mu$m in emission about 5-7$\sigma$ over
the continuum (see Figure \ref{hires-fig}) in the PSF-extracted spectrum.
Examining the full aperture spectrum, we also find marginal evidence of PAH
emission at 11.3$\mu$m. We do not find evidence of molecular emission in
the spectrum, but it must be noted that the strong mid-IR continuum emission
of the source does not allow to exclude weak lines due to a contrast
problem. We find significant differences between the PSF-extracted
and full aperture spectra in all the [Ne II], [Ne III], and [Si II] lines
(see  Figure \ref{hires-fig}), suggestive of extended emission. In addition to the
differences in the extraction, [Ne III] and [Si II] are more typical of
H II and photodissociation regions (Osterbrock 1989), which also points
in the direction of nebular contamination near the source. The relatively
low contrast of the lines against the strong continuum emission 
does not allow for further interpretation in terms of
dynamics. So far, no extended material (cloud nor envelope) has been
detected around GM Cep at millimeter (Patel et al. 1998) nor $Spitzer$ wavelengths, 
but the presence of a remnant envelope or cloud material could explain the high millimeter fluxes
observed for this source, as well as the difficulties in fitting its SED 
with a simple disk model (Sicilia-Aguilar et al. 2008a). Moreover,
if a remnant envelope or cloud material are present, as it is seen around other relatively isolated
objects in the cluster (e.g., around the group formed by the binary 
CCDM+5734 and the F9 star 21374275; SA06), it 
could indicate that GM Cep is younger than the main Tr 37 cluster,
which would be more in agreement with its high and variable accretion rate. 

[Ne II] emission requires X-ray/EUV ionization (Glassgold et al. 2007), which
could be due to the normal stellar activity, to the presence of jets/winds, and/or
to active photoevaporation (Pascucci et al. 2007; Lahuis et al. 2007;
Hollenbach \& Gorti 2009; G\"{u}del et al. 2010; Pascucci \& Sterzik 2009). 
GM Cep was one of the Tr~37 stars detected in X-ray with Chandra (Mercer et al. 2009),
with observed luminosity L$_X$=1.1$\times$10$^{30}$ erg s$^{-1}$ (3.25$\times$10$^{30}$ 
erg s$^{-1}$ after correcting for absorption). The IRS spectrum reveals an 
integrated [Ne II] luminosity L$_{[Ne II]}$=2$\times$10$^{30}$ erg s$^{-1}$.
Both L$_X$ and L$_{[Ne II]}$ appear higher than the typical values found in CTTS
(G\"{u}del et al. 2010), except when sources with strong jets are considered.
While no definite confirmation of an extended jet has been found for GM Cep, 
its optical spectra reveal the presence of a strong stellar wind (Sicilia-Aguilar et al. 2008a). 
The double-peaked O I emission at 8446\AA\ is also suggestive
of a bipolar wind/outflow, so the strong [Ne II] emission observed in
GM Cep could be related to jet emission, as it has been detected toward T Tau
(van Boekel et al. 2009). Active photoevaporation and puffed-up disks have also been 
suggested for increased [Ne II] emission. The high mid-IR fluxes observed in
GM Cep suggest a very flared disk, but due to the strength of the [Ne II] emission
(more than one order of magnitude of the value to be considered within the
normal CTTS range for its L$_X$;  G\"{u}del et al. 2010), the presence of
a jet is the most plausible candidate.

\section{Summary and conclusions \label{conclu}}

We present new $Spitzer$/IRS spectra for 31 TTS
and IRAM 1.3mm observations for 34 low- and intermediate-mass
stars, all within the clusters Tr 37 and NGC 7160 in the Cep OB2
region. The observations were selected to cover a representative
sample of SEDs and disk types observed in Tr 37,
together with the only 2 low-mass disks detected in NGC 7160 (SA06).
Including our previously published IRS spectra (SA07), we analyze
a total of 56 IRS spectra of solar-type stars plus 3 of
intermediate-mass objects with silicate features.
We study the new data, together with the previously published
optical, near- and mid-IR observations, in order to examine the
correlations between silicate dust features and the 
stellar properties, accretion, and disk structure.

We analyze the silicate emission features with a TLTD method, finding
that they are consistent with a mixture of amorphous silicates (with
olivine and forsterite stoichiometry), amorphous silica, and crystalline
forsterite and enstatite grains with sizes 0.1-6$\mu$m. No other components
are required in order to reproduce the observed features, and in general,
the amount of crystals in the disk does not exceed 20\%, with a typical
value being 5-10\%. The lack of S/N at long wavelengths in most objects does not
allow to explore the radial dependence of the crystallinity fraction nor
chemical composition. 

A large number of disks present strong signs of dust evolution: settled SEDs,
evidence of optically thin inner disks, presence of ``kinks'' or slope
changes in the SEDs. We thus characterize the global disk structure
by measuring the SED slope at different wavelengths, and create a series
of prototype disks by means of the RADMC radiative transfer code. This allows us to
determine what types of disk geometry and dust distributions reproduce
the behaviors observed in our sample. 

The comparison with the models reveal that even the most typical CTTS
disks in Cep OB2 have already suffered strong grain growth and settling.
This general requirement of strong settling has also been observed in
younger regions (e.g. Chamaeleon I; Manoj et al. 2011), and thus seems
to occur very early in the disk lifetime. In addition, we find that about
half of the disks present signs of inside-out evolution, seen as a variable
SED slope at different wavelengths (with a large negative slope at short
wavelengths, changing to a near-zero or positive slope at longer $\lambda$). 
Such a SED geometry can be reproduced by radially dependent grain populations
or by changes in the disk flaring/thickness with radius.

We find that TD (especially, accreting ones) and 
``kink" disks appear strongly correlated to grain growth in the 
innermost disk and typically show flattened and small silicate features. Even the
objects whose SEDS can be explained by inner holes show processed grains, 
in contrast with what is seen in the TD with holes in Taurus (Watson
et al. 2009), which tend to have nearly pristine silicate features.
This may indicate that TD are physically different objects
at different ages, as has been previously suggested (Alexander \& Armitage 2009;
Sicilia-Aguilar et al. 2010). While TD in Taurus could be mostly related to
giant planet formation (still subject to migration, due to the high mass of the
disks), substantial grain growth and settling (and eventually, photoevaporation)
can explain the older TD in Tr 37. It could also be a sign that, at older
ages, the mechanisms replenishing the inner disk with pristine grains are not
effective, or the outer disk has also suffered grain processing. 

In order to reproduce the near- and mid-IR SEDs of transitional and ``kink" disks, 
very settled disk structures and, in some cases, radially-dependent dust distributions are
required, in addition to typically very low disk masses (M$_{disk}<$5$\times$10$^{-4}$ M$_*$,
considering a standard gas to dust ratio of 100). This
suggest that, although evolution seems to operate faster in the innermost disk,
there is also substantial evolution and dust depletion at a more
global scale, as it has been suggested by Currie
et al. (2009) and Lommen et al. (2010). We also find that many of our SEDs require 
the presence of submicron-only ($\sim$0.1$\mu$m) grains in the inner disk. This
could be caused either by differential settling, with larger grains being preferentially
in the midplane, or by a radially dependent grain population. Populating the inner
disk with small dust grains could be seen as inconsistent with inside-out
evolution, but may be explained by dust filtering at the edges of the gaps
formed by planets (Rice et al. 2006).

The differences in disk structure in our varied sample dilute the observed
correlation between accretion rate (or turbulence) and the silicate feature,
although a correlation is still seen in samples of normal CTTS disks (SA07; Fang
et al. 2011). The relatively strong accretion rates observed in the stars in the Cep OB2 
region (of the order of few times 10$^{-9}$ M$_\odot$/yr) are in contrast with the
small dust masses inferred from millimeter data and radiative transfer
models, which could be an indication of a larger-than-normal gas to (small) dust
ratio. This also points out the importance of taking into account the gas content 
to determine the evolutionary status and the fate of disks, including the
possibility of planet formation.

Finally, the IRS high-resolution spectrum obtained for the variable star
GM Cep reveals that it is probably extended and thus surrounded by 
cloud material or a
remnant envelope, which could be responsible for its strong and variable
accretion rate. Its IR [Ne II] emission is also consistent with the 
presence of jets, which is in agreement with the strong wind signatures
observed in its optical spectrum.

We would like to thank  R. Zylka for his help with the IRAM observations and
data reduction, in particular, his MOPSIC scripts. We are also indebted to
F. Lahuis for his help with the reduction of
high-resolution IRS spectra, and the Spitzer Science Center Helpdesk for their
help with the MOPEX software. 
We also thank A. Mart\'{\i}nez-Sansigre
and V. Roccatagliata for their participation in the IRAM observations, and 
the referee for the careful report that helped to
clarify our paper.
A.S.-A. acknowledges support from the
Deutsche Forschungsgemeinschaft (DFG) grant SI-1486/1-1.
This work is based on observations made with the Spitzer Space Telescope, which is operated 
by the Jet Propulsion Laboratory, California Institute of Technology under a contract with NASA. It also makes use of
data products from the Two Micron All Sky Survey, which is a joint project of the University
of Massachusetts and the Infrared Processing and Analysis Center/California Institute of Technology,
funded by the National Aeronautics and Space Administration and the National Science
Foundation.

\clearpage

\clearpage

\clearpage

\clearpage

\clearpage

\clearpage

\clearpage

\clearpage

\clearpage

\clearpage

\clearpage

\clearpage

\clearpage

\clearpage

\clearpage

\begin{deluxetable}{lcccccccccl}
\tabletypesize{\scriptsize}
\rotate
\tablenum{1}
\tablecolumns{11} 
\tablewidth{0pc} 
\tablecaption{Summary of IRS observations and stellar properties. \label{obs-table}} 
\tablehead{
 \colhead{Target} & \colhead{Coordinates(J2000)} & \colhead{Mode} & \colhead{Int.Time(s)} & \colhead{Sp.T.} & \colhead{A$_V$ (mag)} & \colhead{Age (Myr)} &\colhead{SED type} & \colhead{\.{M} (10$^{-8}$M$_\odot$/yr)} & \colhead{Sil. Peak} & \colhead{Comments}}
\startdata
{\bf Tr 37}\\
73-758  & 21:35:08.39 +57:36:03.2 & Low & 59-59-19-19 & 	K6.5 & 1.9    & 1.8 & 8 & 0 &		1.8$\pm$0.1					& \\
72-1427 & 21:35:16.32 +57:28:23.0 & Low & 44-183 & 		M1.0 & 1.4    & 3.3 & 8 &  0.15$^{+0.28}_{-0.07}$ &		 1.7$\pm$0.2			& \\
73-472 & 21:35:18.61 +57:34:09.2 & Low & 1451-1451-975-975 & 	K5.0 & 1.7    & 3.3 & 7 & A &			 1.2$\pm$0.1					&  \\
73-311 & 21:35:24.51 +57:33:01.1 & Low & 366-366-315-315 & 	M1.5 & 1.1    & 2.8 & 4 & A &	 1.5$\pm$0.1						&  \\
73-71 & 21:35:30.21 +57:31:16.4 & Low & 1934-1934-975-975 & 	K6.0 & 2.1    & 1.8 & 4 & 2.1$^b$ &	 1.4$\pm$0.1					&     \\
72-875 & 21:35:49.80 +57:24:04.6 & Low & 44-183-94-44 & 	M0.5 & 2.0    & 2.6 & 7 & 0.11$^{+0.23}_{-0.06}$/0 &	 1.0$\pm$0.2			& SB1:, variable \.{M} \\
21362507 & 21:36:25.08 +57:27:50.3 & Low & 44-183-94-44 & 	M0.0 &  3.5   & 26 & 1 & 0.09$^{+0.16}_{-0.05}$ &	 1.9$\pm$0.1			& SB1\\
\enddata
\tablecomments{Disked stars observed with IRS and their characteristics.
Spectral types and extinctions from Contreras et al.(2002; marked with $^1$)
and Sicilia-Aguilar et al. (2004, 2005); the data for 13-277 from Sicilia-Aguilar
et al. (2008). 
Spectral types and extinction obtained from SED fitting (see text) 
are marked with $^f$. The ages are derived from the V vs V-I diagram and the
Siess et al. (2000) isochrones, except for the intermediate-mass stars (marked with *), 
for which we take it to be the cluster median. Early K-G stars and/or those affected
by scattering in the V-I colors tend to have anomalously large ages, not
taken into account in the analysis. 
The SED types refer to the model prototypes in Section \ref{radmc}, while intermediate-mass
stars are marked as ``D", given they are most likely debris disks or dust-depleted
TD with large inner holes.
Accretion rates and presence/absence of accretion from Sicilia-Aguilar et al.(2005,2006b, 2010). 
Accretion 
rates marked with $^b$ come from non-simultaneous data and may be more uncertain due 
to the intrinsic stellar variability. Objects marked with ``A" have broad H$\alpha$ and are
thus accreting, although the accretion rate has not been measured. 
Objects marked with ``PA" have strong H$\alpha$ EW and are most likely accreting; conversely, objects
marked with ``0*'' have small and narrow H$\alpha$ emission and are probably not accreting. 
Uncertain values are marked with ``:". The SED type number corresponds to the models in Table \ref{diskproto-table},
see text. A few cases where the disks around intermediate-mass stars may correspond to 
debris disks are marked with "D".}
\end{deluxetable}

\clearpage

\begin{deluxetable}{lcccccl}
\tabletypesize{\scriptsize}
\tablenum{2}
\tablecolumns{7} 
\tablewidth{0pc} 
\tablecaption{IRAM 1.3mm Observations \label{iram-table}} 
\tablehead{
 \colhead{Name}  &  \colhead{Program(s)}  & \colhead{Nr. of scans}& \colhead{Tau} & \colhead{Flux (mJy)} & \colhead{M$_{disk}$ (M$_\odot$)} & \colhead{Comments} } 
\startdata
{\bf Tr 37}\\
73-758	& 132-07,145-08	& 6	& 0.2-0.4	& $<$1.8	& $<$0.009	&	\\
73-311	& 132-07,145-08	& 5	& 0.2-0.3	& $<$1.8	& $<$0.009	&	\\
73-71	& 132-07,145-08	& 6	& 0.2-0.3	& $<$1.5	& $<$0.008	&	\\
MVA-1312& 146-06	& 3	& 0.2-0.3	& $<$4.8	& $<$0.024	&	\\
KUN-314s& 146-06	& 3	& 0.2-0.7	& $<$6.3  & $<$0.032	  & Nebulosity contamination	  \\
11-2146	& 146-06	& 3	& 0.2-0.7	& 2.9$\pm$0.6	& 0.015	&	\\
11-1209	& 146-06	& 3	& 0.2-0.4	& $<$2.4	& $<$0.012	&	\\
\enddata
\tablecomments{Summary of IRAM 30m observations. To derive the total disk mass, the
opacity of Beckwith et al. (1990; $\kappa_\nu$=2 cm$^2$ g$^{-1}$) and a gas to dust
ratio of 100 have been applied. Upper limits correspond to 3$\sigma$ for non-detections. }
\end{deluxetable}

\clearpage

\begin{deluxetable}{lccccc}
\tabletypesize{\scriptsize}
\tablenum{3}
\tablecolumns{6} 
\tablewidth{0pc} 
\tablecaption{$Spitzer$ IRAC and MIPS fluxes \label{spitzer-table}} 
\tablehead{
 \colhead{Target} & \colhead{IRAC 3.6$\mu$m ($\mu$Jy)} & \colhead{IRAC 4.5$\mu$m ($\mu$Jy)} & \colhead{IRAC 5.8$\mu$m ($\mu$Jy)} & \colhead{IRAC 8.0$\mu$m ($\mu$Jy)} & \colhead{MIPS 24$\mu$m ($\mu$Jy)}}
\startdata
{\bf Tr 37}\\
73-758 & --- & 4264$\pm$63 & 4308$\pm$206 & 6811$\pm$91 & 15638$\pm$101 \\
72-1427$^I$ & 4663$\pm$50 & 3721$\pm$54 & 2932$\pm$143 & 3013$\pm$93 & 13831$\pm$484 \\
73-472$^I$ & 9985$\pm$68 & 8254$\pm$77 & 5865$\pm$155 & 3934$\pm$82 & 24402$\pm$100 \\
73-311 & 4373$\pm$48 & 4451$\pm$66 & 4205$\pm$156 & 4388$\pm$88 & 4418$\pm$101 \\
73-71 & --- & 8258$\pm$110 & 6028$\pm$157 & 5412$\pm$195 & 9652$\pm$102 \\
72-875 & --- & --- & --- & --- & 19814$\pm$1204 \\
21362507 & 3060$\pm$58 & 3013$\pm$61 & 3021$\pm$224 & 4872$\pm$26 & 9196$\pm$103 \\
\enddata
\tablecomments{IRAC and MIPS fluxes for the targets observed with IRS.
Most of the objects had already been presented in SA06, and their newly reduced
values are in good agreement with those. New IRAC and MIPS detections (due to the
inclusion of archive data from the region, better data from the new pipelines,
and/or new members found in Sicilia-Aguilar et al. 2006b) 
are marked with $^I$ and $^M$, respectively. IRAC and MIPS photometry for 21364762 is
contaminated by the strong background, and 93-720 is located out of the IRAC and
MIPS fields.}
\end{deluxetable}

\clearpage

\begin{deluxetable}{lccccl}
\tabletypesize{\scriptsize}
\tablenum{4}
\tablecolumns{6} 
\tablewidth{0pc} 
\tablecaption{Optical data\label{opt-table}} 
\tablehead{
 \colhead{Target} & \colhead{U} & \colhead{V} & \colhead{R} & \colhead{I} & \colhead{References}}
\startdata
{\bf Cousins System photometry}\\
{\bf Tr 37}\\
73-758 & --- & 17.062 & 15.845 & 14.761 &  1   \\
72-1427 & --- & 18.080 & 16.822 & 15.580 &  1   \\
73-472 & --- & 16.851 & 15.755 & 14.805 &  1   \\
73-311 & --- & 18.378 & 17.126 & 15.887 &  1   \\
73-71 & --- & 16.989 & 15.813 & 14.694 &  1   \\
72-875 & --- & 18.367 & 17.382 & 16.178 &  1   \\
\enddata
\tablecomments{Summary of optical magnitudes for the analyzed sources. References: (1) Sicilia-Aguilar et al. (2005); 
(2) De Graeve (1983); (3) Marschall \& van Altena(1987); (4) Sicilia-Aguilar et al. (2010).}
\end{deluxetable}

\clearpage

\begin{deluxetable}{lcccc}
\tabletypesize{\scriptsize}
\tablenum{5}
\tablecolumns{5} 
\tablewidth{0pc} 
\tablecaption{2MASS data\label{2mass-table}} 
\tablehead{
 \colhead{Target} & \colhead{2MASS ID} & \colhead{J} & \colhead{H} & \colhead{K} }
\startdata
{\bf Tr 37} \\
73-758 & 21350835+5736028 & 13.330$\pm$0.028 & 12.533$\pm$0.032 & 12.163$\pm$0.028      \\
72-1427 & 21351627+5728222 & 14.005$\pm$0.031 & 13.067$\pm$0.029 & 12.673$\pm$0.028      \\
73-472 & 21351861+5734092 & 13.219$\pm$0.024 & 12.299$\pm$0.031 & 11.829$\pm$0.022      \\
73-311 & 21352451+5733011 & 14.014$\pm$0.034 & 13.115$\pm$0.037 & 12.674$\pm$0.030      \\
73-71 & 21353021+5731164 & 12.979$\pm$0.034 & 12.105$\pm$0.037 & 11.715$\pm$0.029      \\
72-875 & 21354975+5724041 & 14.092$\pm$0.031 & 13.006$\pm$0.027 & 12.500$\pm$0.026      \\
21362507 & 21362507+5727502 & 14.952$\pm$0.044 & 14.031$\pm$0.048 & 13.518$\pm$0.040      \\
\enddata
\tablecomments{Summary of 2MASS magnitudes for the analyzed sources. Most of the data are
already published in Sicilia-Aguilar et al. (2004, 2005).}
\end{deluxetable}

\clearpage

\begin{deluxetable}{lccccc}
\tabletypesize{\scriptsize}
\tablenum{6}
\tablecolumns{6} 
\tablewidth{0pc} 
\tablecaption{Silicate Composition in the 7-14$\mu$m region \label{silshort-table}} 
\tablehead{
 \colhead{Name} & \colhead{0.1$\mu$m A.Ol.}  & \colhead{0.1$\mu$m A.Py.}  & \colhead{0.1$\mu$m Fors.}  & \colhead{0.1$\mu$m Enst.}  & \colhead{0.1$\mu$m Sil.}   \\
 	        & \colhead{1.5$\mu$m A.Ol.}  & \colhead{1.5$\mu$m A.Py.}  & \colhead{1.5$\mu$m Fors.}  & \colhead{1.5$\mu$m Enst.}  & \colhead{1.5$\mu$m Sil.}   \\
  	        & \colhead{6.0$\mu$m A.Ol.}  & \colhead{6.0$\mu$m A.Py.}  & \colhead{6.0$\mu$m Fors.}  & ---  & \colhead{6.0$\mu$m Sil.}   \\ } 
\startdata
{\bf Tr 37}\\
73-758 & 0.7$^{+2.6}_{-0.7}$ & --- & 2.0$^{+0.4}_{-0.3}$ & 0.1$^{+0.4}_{-0.1}$ & ---  \\
 & 12$^{+5}_{-6}$ & 17$^{+11}_{-8}$ & 0.7$^{+1.2}_{-0.6}$ & 1.6$^{+1.3}_{-1.1}$ & 0.2$^{+0.7}_{-0.2}$  \\
 & 7$^{+18}_{-7}$ & 55$^{+10}_{-15}$ & --- & --- & 4$^{+4}_{-3}$  \\
73-472 & 30$^{+27}_{-20}$ & 9$^{+12}_{-6}$ & 2.5$^{+1.3}_{-1.2}$ & 0.5$^{+1.2}_{-0.4}$ & 0.1$^{+0.9}_{-0.1}$  \\
 & 21$^{+55}_{-21}$ & 8$^{+9}_{-7}$ & 0.6$^{+1.6}_{-0.6}$ & 2.7$^{+3.4}_{-2.3}$ & 0.4$^{+1.2}_{-0.4}$  \\
 & 25$^{+38}_{-22}$ & --- & --- & --- & 0.1$^{+0.9}_{-0.1}$  \\
73-311 & 2$^{+3}_{-2}$ & 3$^{+5}_{-2}$ & --- & 1.5$^{+0.8}_{-0.8}$ & 0.5$^{+0.6}_{-0.3}$  \\
\enddata
\tablecomments{Silicate fit for the objects in Cep OB2 using
the Two Layer Temperature Distribution method (J09) in the short wavelength range (7-14$\mu$m). 
Only those objects with enough S/N to significantly constrain the dust mineralogy
are listed here. The values are percentages in mass fraction. If no value is given
for a certain species/size, it means that it is present at 2\% level
or less.}
\end{deluxetable}

\begin{deluxetable}{lccccc}
\tablenum{7}
\tablecolumns{6} 
\tablewidth{0pc} 
\tablecaption{Silicate Composition in the 17-35$\mu$m region \label{sillong-table}} 
\tablehead{
 \colhead{Name} & \colhead{0.1$\mu$m A.Ol.}  & \colhead{0.1$\mu$m A.Py.}  & \colhead{0.1$\mu$m Fors.}  & \colhead{0.1$\mu$m Enst.}  & \colhead{0.1$\mu$m Sil.}   \\
 	        & \colhead{1.5$\mu$m A.Ol.}  & \colhead{1.5$\mu$m A.Py.}  & \colhead{1.5$\mu$m Fors.}  & \colhead{1.5$\mu$m Enst.}  & \colhead{1.5$\mu$m Sil.}   \\
  	        & \colhead{6.0$\mu$m A.Ol.}  & \colhead{6.0$\mu$m A.Py.}  & \colhead{6.0$\mu$m Fors.}  & ---  & \colhead{6.0$\mu$m Sil.}   \\ } 
\startdata
{\bf Tr 37}\\
11-2146 & 6$^{+22}_{-6}$ & 10$^{+10}_{-8}$ & 3$^{+3}_{-1}$ & --- & ---  \\
 & 49$^{+9}_{-18}$ & $<$7 & 2$^{+6}_{-2}$ & 2$^{+1}_{-1}$ & ---  \\
 & --- & 25$^{+11}_{-10}$ & $<$7 & --- & 2$^{+2}_{-1}$  \\
CCDM+5734 & --- & 19$^{+14}_{-18}$ & 5$^{+1}_{-1}$ & --- & ---  \\
 & 61$^{+7}_{-7}$ & 11$^{+12}_{-11}$ & --- & --- & ---  \\
 & --- & 2$^{+6}_{-2}$ & --- & --- & ---  \\
82-272 & --- & 54$^{+16}_{-22}$ & 8$^{+8}_{-1}$ & 2$^{+2}_{-1}$ & ---  \\
\enddata
\tablecomments{Silicate fit for the objects in CepOB2 using
the Two Layer Temperature Distribution method (J09) in the long wavelength range (17-35$\mu$m). 
Only objects with good S/N in this region are presented.
The values are percentages in mass fraction. If no value is given
for a certain species/size, it means that it is present at 2\% level
or less. }
\end{deluxetable}

\begin{deluxetable}{lccccl}
\tabletypesize{\scriptsize}
\tablenum{8}
\tablecolumns{6} 
\tablewidth{0pc} 
\tablecaption{Grain sizes and crystalline fraction in the 7-14$\mu$m region \label{sizecrystshort-table}} 
\tablehead{
 \colhead{Name}  & \colhead{Size (am.) /~$\mu$m} & \colhead{Size (cryst.) /~$\mu$m} & \colhead{Cryst.Fract.} & \colhead{Forst/Enst} & \colhead{Comments}} 
\startdata
{\bf Tr 37}\\
73-758 & 4.6$^{+0.2}_{-0.5}$ & 0.8$^{+0.1}_{-0.4}$ & 4$^{+2}_{-1}$ & 2$^{+2}_{-2}$ & TDn   \\
73-472 & 1.7$^{+1.9}_{-1.2}$ & 0.7$^{+0.2}_{-0.4}$ & 5$^{+5}_{-3}$ & 1$^{+4}_{-2}$ &  TDa, Low S/N  \\
73-311 & 4.7$^{+0.3}_{-0.8}$ & 0.6$^{+3.3}_{-0.5}$ & 3$^{+5}_{-1}$ & 0.2$^{+4.3}_{-0.2}$ &    \\
73-71 & 4.9$^{+0.2}_{-0.2}$ & 3.2$^{+0.6}_{-1.3}$ & 8$^{+3}_{-2}$ & 1.0$^{+0.9}_{-0.8}$ &    \\
21362507 & 4.3$^{+0.3}_{-0.7}$ & 1.5$^{+1.0}_{-0.2}$ & 7$^{+6}_{-2}$ & 3$^{+5}_{-3}$ &    \\
11-2146 & 3.6$^{+0.3}_{-0.7}$ & 1.1$^{+0.2}_{-0.4}$ & 4$^{+3}_{-2}$ & 0.7$^{+1.3}_{-0.7}$ &  Variable \.{M}, Continuum uncertain  \\
11-2037 & 3$^{+1}_{-2}$ & 1.0$^{+0.1}_{-0.1}$ & 12$^{+6}_{-3}$ & 0.5$^{+0.4}_{-0.3}$ &    \\
\enddata
\tablecomments{Average grain sizes for amorphous and crystalline silicates 
and crystalline fraction (mass fraction of crystalline silicates) from the TLTD fit
in the 7-14$\mu$m region. Only those objects with enough S/N to significantly constrain the dust mineralogy
are listed here. The colon (:) indicates uncertain
values in spectra with poor S/N. Grain sizes of 0.1 or 6$\mu$m indicate the presence of very small
or large grains, but given that our model does not include smaller nor larger grains, the uncertainties cannot
be well determined, we thus mark these objects with ``*". The comments include whether the star presents a variable
accretion rate ("Variable \.{M}), whether it is an accreting or non-accreting transition disk (TDa, TDn, respectively),
and whether the fitted continuum level may be uncertain as described in Section \ref{radmc}.}
\end{deluxetable}

\begin{deluxetable}{lccccl}
\tablenum{9}
\tablecolumns{6} 
\tablewidth{0pc} 
\tablecaption{Grain sizes and crystalline fraction in the 17-35$\mu$m region  \label{sizecrystlong-table}} 
\tablehead{
 \colhead{Name}  & \colhead{Size (am.) /~$\mu$m} & \colhead{Size (cryst.) /~$\mu$m} & \colhead{Cryst.} & \colhead(Forst/Enst) & \colhead{Comments}} 
\startdata
{\bf Tr 37}\\
11-2146 & 2.5$^{+0.6}_{-0.8}$ & 0.8$^{+0.2}_{-0.1}$ & 5$^{+6}_{-1}$ & 1.3$^{+2.8}_{-0.6}$ &    \\
CCDM+5734 & 1.3$^{+0.2}_{-0.3}$ & 0.1* & 5.1$^{+0.4}_{-0.5}$ & --- &   TD/D \\
82-272 & 0.6$^{+0.9}_{-0.2}$ & 0.1$^{+0.4}$* & 8.9$^{+9.1}_{-0.9}$ & 3$^{+6}_{-1}$ &    \\
13-277 & 2.3$^{+0.3}_{-0.4}$ & 0.4$^{+0.2}_{-0.2}$ & 1.0$^{+0.1}_{-0.1}$ & 2$^{+2}_{-1}$ &    \\
21393104 & 5$^{+1}_{-3}$ & 0.6$^{+0.6}_{-0.3}$ & 5$^{+3}_{-1}$ & 2$^{+5}_{-2}$ &    \\
93-720 & 0.7$^{+0.4}_{-0.3}$ & 0.4$^{+0.2}_{-0.2}$ & 4.8$^{+0.9}_{-1.1}$ & 1.3$^{+0.6}_{-0.6}$ &    \\
\enddata
\tablecomments{Average grain sizes for amorphous and crystalline silicates 
and crystalline fraction (mass fraction of crystalline silicates) from the TLTD fit
in the 17-35$\mu$m range. Only objects with enough S/N in this region are presented.
The intermediate-mass star CCDM+5734 is marked with TD/D since it is probably a transitional disk with
a very large inner hole or a debris disk.
The colon (:) indicates uncertain
values in spectra with poor S/N. Grain sizes of 0.1$\mu$m indicate the presence of very small
grains, but given that our model does not include smaller grains, the uncertainties cannot
be well determined, we thus mark these objects with ``*".}
\end{deluxetable}

\clearpage

\begin{deluxetable}{lcccccl}
\tabletypesize{\scriptsize}
\rotate
\tablenum{10}
\tablecolumns{6} 
\tablewidth{0pc} 
\tablecaption{Summary of stellar and SED properties of the disk prototypes  \label{diskproto-table}} 
\tablehead{
 \colhead{Model}  & \colhead{Prototype} & \colhead{T$_{eff}$ (K)} & \colhead{R$_*$ (R$_\odot$)} & \colhead{M$_*$ (M$_\odot$)} & \colhead{M$_{disk}$ (M$_\odot$)} & \colhead{Comments}} 
\startdata
Normal (1) 		& 11-2037 & 4770. & 2.3 & 1.3 & 0.015 		&  Normal disk with silicates  \\
Massive (2) 		& 13-277  & 5520. & 5.9 & 2.6 & 0.07 		&  Very massive, strongly accreting disk \\
Normal/Depleted (3) 	& 13-236  & 4900. & 2.3 & 1.6 & $<$0.020 	&  Settled disk with silicates  \\
Settled/Depleted (4) 	& 11-1209 & 4205. & 2.9 & 1.0 & $<$0.012 	&  Settled disk with no/marginal silicates\\
Kink (5) 		& 13-1250 & 4470. & 2.2 & 1.3 & 0.012 		&  Disk with strong slope change or kink\\
Old CTTS disk(6)	& 01-580  & 4470. & 1.5 & 1.2 & $<$0.012	&  Depleted disk with strong silicate feature\\
Classical TD (7)	& 24-1796 & 4060. & 1.4 & 0.9 & --- 		&  Accreting TD without silicate, strong turn-up \\
Classical TD, sil (8)	& 14-11   & 3650. & 2.1 & 0.4 &  $<$0.001 	&  Non-accreting TD with silicate, strong turn-up \\
Settled TD, no sil (9)	& 92-393  & 3580. & 2.1 & 0.6 &  $<$0.009 	&  Accreting TD without silicate, settled \\
Settled TD, sil (10)	& 13-52   & 4060. & 1.5 & 0.8 & --- 		&  Non-accreting TD with silicate, settled  \\
\enddata
\tablecomments{Modeled disk prototypes.Temperatures use the correspondence between spectral types 
and T$_{eff}$ from Kenyon \& Hartmann (1995). Spectral types, stellar radii and masses from
Sicilia-Aguilar et al. (2006b, 2010).  }
\end{deluxetable}

\begin{deluxetable}{lccccccccc}
\tabletypesize{\scriptsize}
\tablenum{11}
\tablecolumns{10} 
\tablewidth{0pc} 
\tablecaption{SED slope at different wavelengths  \label{alpha-table}} 
\tablehead{
 \colhead{Name}  & \colhead{$\alpha$(H-5.8)} & \colhead{$\alpha$(H-8)} & \colhead{$\alpha$(H-12)} &
 \colhead{$\alpha$(H-17)} & \colhead{$\alpha$(3.6-5.8)} & \colhead{$\alpha$(H-24)} & \colhead{$\alpha$(5.8-8)} & \colhead{$\alpha$(8-17)} & \colhead{$\alpha$(24-33)} } 
\startdata
73-758 & -1.77 &  -1.47 & -1.16 & -1.01 & --- & -0.94 & -0.26 & -0.05 & -1.78  \\
72-1427 & -1.81 &  -1.79 & -1.38 & --- & -1.93 & --- & -1.71 & --- & ---  \\
73-472 & -1.85 &  -1.98 & -1.78 & -1.81 & -2.08 & -0.95 & -2.50 & -1.45 & 0.36  \\
73-311 & -1.50 &  -1.42 & -1.21 & -1.60 & -1.05 & -1.46 & -1.10 & -2.00 & 2.79  \\
73-71 & -1.79 &  -1.84 & -1.62 & -1.47 & --- & -1.32 & -2.04 & -0.68 & -1.75  \\
72-875 & -1.95 &  -1.35 & -1.25 & -0.80 & --- & -0.78 & 0.99 & 0.37 & 0.42  \\
21362507 & -1.08 &  -0.88 & -0.71 & --- & -1.06 & --- & -0.08 & --- & ---  \\
21364762 & -1.33 &  -1.30 & -1.76 & --- & --- & -0.57 &-1.19 & --- & ---  \\
\enddata
\tablecomments{SED slopes (estimated as 
$[log(\lambda_1 F_\lambda1)-log(\lambda_2 F_\lambda2)]/(log\lambda_1-log\lambda_2)$)
used to analyze the disk structure. The slopes are calculated using preferentially the
IRS data (see text), and the IRAC and/or MIPS data for wavelengths not included in
the IRS range and/or in case the IRS data presented a problem for particular objects.
H photometry is taken from 2MASS.}
\end{deluxetable}

\begin{deluxetable}{lcccccl}
\tabletypesize{\scriptsize}
\rotate
\tablenum{12}
\tablecolumns{7} 
\tablewidth{0pc} 
\tablecaption{Properties of the prototype disk models  \label{models-table}} 
\tablehead{
 \colhead{Model}  & \colhead{M$_{disk}$ (M$_\odot$)} & \colhead{R$_{disk}$ (AU)} & \colhead{Grain min-max, in/out ($\mu$m)}  & \colhead{H$_{rdisk}$ (R$_{disk}$)}  & \colhead{T${in}$ (K)} & \colhead{Other/Comments}} 
\startdata
Normal (1) 		& 0.015   &  450    & 0.1-10000   &  0.16  & 1500   &  r$_{crys}$=0.15 AU  \\
Massive (2) 		& 0.052   &  450  &  0.1/0.1-100  &  0.2  & 1500   &  r$_{comp}$=3.0 AU  \\
Normal/Depleted (3A) 	& 0.00013 &  100 & 0.1-10000   &  0.13  & 1500   &  Fits general structure   \\
Normal/Depleted (3B) 	& 0.00013 &  100 & 0.1/0.1-10000 & 0.13    & 1500    & r$_{comp}$=0.1 AU, improves near-IR excess   \\
Settled/Depleted (4) 	& 0.005   &  200  &  6-10000  &  0.02  &  1500   &  Uniform grain composition  \\
Kink (5A) 		& 0.02    &  150  & 0.1/0.1-100   &	0.08	&  1500  &  r$_{comp}$=2.0 AU; Good for silicate \\
Kink (5B) 		& 0.02    &  300  & 10-10000/0.1-100   & 0.08  &  1500  &   r$_{comp}$=5.0 AU; Good for general structure \\
Old CTTS disk (6) 	& 0.0006  & 200   &  0.1/1-10000  &  0.09  & 1500     &  r$_{comp}$=2.0 AU \\
\enddata
\tablecomments{Properties of the RADMC models for the disk prototypes: disk mass and radius,
maximum and minimum grain size, vertical pressure scale height at the outer disk radius
(H$_{rdisk}$) in units of disk radius, inner rim temperature, and other properties 
(radius at which the dust composition changes [r$_{comp}$], radius below which all grains are crystalline [r$_{crys}$]).
The global dependency of the vertical pressure scale height goes as H$_p$/R$\propto$R$^{1/7}$.
The power law exponent of the grain distribution is always assumed to be collisional (a=-3.5),
unless in the cases where a single grain population is assumed for part of the disk. The Schmidt number regulating
the mixing of the two populations after the radius at which grain properties change (r$_{crys}$ or r$_{comp}$) is
always taken to be 2/3 (see Pavlyuchenkov \& Dullemond 2007). }
\end{deluxetable}

\begin{deluxetable}{lcccl}
\tabletypesize{\scriptsize}
\tablenum{13}
\tablecolumns{5} 
\tablewidth{0pc} 
\tablecaption{Correlation table: Spearman rank coefficients  \label{correlation-table}} 
\tablehead{
 \colhead{Parameters}  & \colhead{R} & \colhead{p} & \colhead{R Range} & \colhead{Comments}} 
\startdata
Peak vs. Sp.Type 		& -0.37	& 0.006	& -0.61 --- -0.06 	& Excluding A,B stars		\\
Size vs. Peak			& -0.57	& 9$\times$10$^{-5}$	& 	-0.77 --- -0.26  &		\\
Peak vs. Age			& 0.21	& 0.14	&	-0.11 --- 0.50  &	Excluding G stars  or earlier	\\
\.{M} vs. Peak			& 0.14	& 0.46	&	-0.31 --- 0.54 &		\\
\.{M} vs. Size			& -0.10	& 0.67	&	-0.59 --- 0.44 &		\\
$\alpha$(H-5.8) vs. Peak	& 0.27	& 0.04	&	-0.04 --- 0.53  &		\\
$\alpha$(H-8) vs. Peak		& 0.28	& 0.03	&	-0.02 --- 0.54  &		\\
$\alpha$(H-12) vs. Peak		& 0.48	& 1$\times$10$^{-4}$	&	0.20 --- 0.68  &		\\
\enddata
\tablecomments{Correlation coefficients (R) and false-alarm probability (p) for different
stellar, disk, and silicate properties, including height of the normalized peak, spectral types,
size of the amorphous grains, age, forsterite to enstatite ratio (F/E), and SED slope at
different wavelengths ($\alpha$).}
\end{deluxetable}

\end{document}